\iftrue
\documentclass[aps,prl,twocolumn,
			   groupedaddress,superscriptaddress,
			   amsfonts,amssymb,amsmath,
			   citeautoscript,
			   a4paper]{revtex4-1}
\else
\documentclass[aps,pra,preprint,
			   groupedaddress,superscriptaddress,notitlepage,
			   amsfonts,amssymb,amsmath,
			   citeautoscript,
			   a4paper]{revtex4-1}
\fi

\usepackage[pdftex]{hyperref}
\hypersetup{colorlinks,
			linkcolor={blue!75!black!80!yellow},
			citecolor={blue!75!black!80!yellow},
			urlcolor={blue!75!black!80!yellow},
			pdfstartview=FitH}
\usepackage{graphicx}
\usepackage{mathrsfs}
\usepackage{amsthm}
\usepackage{physics}
\usepackage{xspace}
\usepackage{braket}
\usepackage{xr}
\usepackage{gensymb} 
\usepackage{xcolor,soul}
\usepackage{stmaryrd} 
\usepackage[UKenglish]{babel}
\usepackage{placeins} 
\usepackage{siunitx}
\DeclareSIUnit[number-unit-product=]\percent{\char`\%} 
\usepackage{makecell}
\usepackage{diagbox}
\usepackage{float}
\usepackage{amsmath} 
\usepackage{empheq} 

\usepackage{txfonts}  
\usepackage{txfontsb} 

\newcommand{\appropto}{\mathrel{\vcenter{
			\offinterlineskip\halign{\hfil$##$\cr
				\propto\cr\noalign{\kern2pt}\sim\cr\noalign{\kern-2pt}}}}}

\newcommand{\iu}{\mathrm{i}}

\makeatletter
\newcommand*{\addFileDependency}[1]{
  \typeout{(#1)}
  \@addtofilelist{#1}
  \IfFileExists{#1}{}{\typeout{No file #1.}}
}
\makeatother

\newcommand*{\myexternaldocument}[1]{%
    \externaldocument{#1}%
    \addFileDependency{#1.tex}%
    \addFileDependency{#1.aux}%
}

\myexternaldocument{SM_clean}

\usepackage{scalerel}

\usepackage{textcomp} 
\usepackage{xifthen}
\usepackage{etoolbox}
\newboolean{togglecomments}
\newboolean{togglechanges} 

\setboolean{togglecomments}{true}
\setboolean{togglechanges}{false}

\newcommand{\comment}[2]{%
    \ifbool{togglecomments}%
    {\textcolor{blue!70!black}{\small\textsf{%
    \textsuperscript{\textsc{\textsf{\MakeLowercase{#1}}}}%
    [#2]}}} 
    {}}     
\newcommand{\swap}[2]{\ifbool{togglechanges}
    {#2}  
    {\textcolor{red!70!black}{[#1]}\textrightarrow{}\textcolor{green!50!black}{[#2]}}}
\newcommand{\remove}[1]{\ifbool{togglechanges}
    {}    
    {\textcolor{red!70!black}{#1}}}
\newcommand{\inset}[1]{\ifbool{togglechanges}
    {#1}  
    {\textcolor{green!50!black}{#1}}}
\newcommand{\optional}[1]{\ifbool{togglechanges}
    {#1}  
    {\textcolor{yellow!50!orange!80!gray}{#1}}}
\newcommand{\citeremind}[1]{%
    [\textcolor{blue!75!black!80!yellow}{
        $\blacksquare$%
           \ifthenelse{\isempty{#1}}
               {}
               {\textsuperscript{\textsf{#1}}}%
        }]\xspace}
\newcommand{\todo}[1]{
    \textcolor{orange!80!yellow!95!black}{\textbf{[}%
        \ifthenelse{\isempty{#1}}%
        {\text{$\blacksquare$}}%
        {{\small\textsf{#1}}}%
        \textbf{]}}}
        
\newcommand{\hkuaffil}{\footnotesize Department of Physics and HK Institute of Quantum Science and Technology, The University of Hong Kong, Pokfulam, Hong Kong, China}

\begin{document}

\title{Synthetic non-Abelian Electric Fields and Spin-Orbit Coupling in Photonic Synthetic Dimensions}

\author{Bengy Tsz Tsun Wong}
\affiliation{\hkuaffil}
\author{Zehai Pang}
\affiliation{\hkuaffil}
\author{Yi~Yang}
\email{yiyg@hku.hk}
\affiliation{\hkuaffil}

\begin{abstract}
We theoretically propose a scheme to synthesize photonic non-Abelian electric field and spin-orbit coupling (SOC) in the synthetic frequency dimension based on a polarization-multiplexed time-modulated ring resonator. 
Inside the ring resonator, the cascade of polarization-dependent phase modulation, polarization rotation, and phase retardation enables a photonic realization of minimal SOC composed of equal Rashba and Dresselhaus terms.
The synthetic bands and associated spin textures can be effectively probed by the continuous-wave polarization- and time-resolved band structure measurements. 
Meanwhile, the pulse dynamics in the frequency dimension can be manipulated by both synthetic Abelian and non-Abelian electric fields, leading to Bloch oscillation, Zitterbewegung, and their interplay dynamics. 
Our proposed setup can be realized with free-space optics, fibers, and integrated photonics, and provides pseudospin control knobs for large-scale synthetic lattices based on frequency combs. 
\end{abstract}

\maketitle

Synthetic gauge fields have played a significant role in the study of topology in engineered physical systems~\cite{aidelsburger2018review} like cold atoms, photons, and superconducting circuits.
A crucial part of the study is to synthesize non-Abelian gauge fields that do not commute.
This effort enables the ability the generate and control spin-orbit coupling (SOC)~\cite{zhai2015degenerate,lin2011spin,zhang2015spin}, giving rise to topological insulating phenomena~\cite{goldman2014light,yang20202d}, nontrivial momentum spin texture~\cite{whittaker2021optical,terccas2014non}, Zitterbewegung wavepacket dynamics~\cite{sedov2018zitterbewegung,hasan2022wave,polimeno2021experimental,lovett2023observation,chen2019non}, and non-Abelian Aharonov--Bohm interference~\cite{yang2019synthesis,dalibard2011colloquium,osterloh2005cold,chen2019non,dong2024temporal}.
These gauge fields are expected to be an indispensable ingredient in the development of the quantum simulation of the lattice gauge theory~\cite{zohar2015quantum}
Specifically in photonics, a multitude of methods have been developed for creating these gauge fields based on temporal modulation~\cite{fang2012realizing}, strains~\cite{rechtsman2013strain}, delay lines~\cite{hafezi2011robust}, and so on.
These methods have been successfully applied to creating photonic non-Abelian gauge fields, and efforts are being made for their non-Abelian counterparts~\cite{yang2024non,yan2023non}, which hold implications for topology, braiding theory, light-matter interactions, and non-Hermitian physics.

An intriguing playground for synthesizing gauge fields is the synthetic dimension~\cite{celi2014synthetic,ozawa2019topological,yuan2018synthetic,lustig2021topological,arguello2024synthetic}, where suitable degrees of freedom like frequencies and time are engineered to emulate spatial dimensions. 
In particular, the synthetic frequency dimension of photonic ring resonators has attracted substantial interest for its compatibility of free-space, fiber, and integrated implementation and versatility of linear and nonlinear control~\cite{hey2018advances,ozawa2016synthetic,yuan2016bloch,bell2017spectral,qin2018spectrum,dutt2020single,balvcytis2022synthetic,tusnin2020nonlinear,suh2024photonic,du2022giant,wang2021generating,cheng2021arbitrary,hu2020realization,senanian2023programmable}.
On this platform, synthetic Abelian electric and magnetic fields have been achieved, and recent efforts have been spared to create non-Abelian vector potentials along the photonic frequency dimension~\cite{cheng2023artificial,pang2024synthetic,cheng2024non}. 
However, these non-Abelian vector potentials correspond to non-Abelian magnetic fields, whereas synthesizing non-Abelian electric fields remains elusive in photonic synthetic dimensions.

In this work, we theoretically propose a polarization-multiplexed time-modulated ring resonator setup for synthesizing non-Abelian electric fields in the frequency dimension, enabling a faithful realization of minimally coupled SOC for photons. We show how to probe the resulting spin texture in continuous-wave transmission measurements. In the pulse dynamics, the Zitterbewegung phenomenon can be synthesized and displays rich interplay with Bloch oscillation. 

\emph{Proposed setup and underlying Hamiltonian.}
The non-Abelian electric field and magnetic fields are the $01$-th and the $ij$-th component of the rank-two Yang-Mills field strength tensor~\cite{yang1954conservation}
\begin{align}
& F^{\mu\nu} = \partial^\mu A^\nu - \partial^\nu A^\mu -\mathrm{i} [A^\mu , A^\nu]\,, \label{eq:nonAB_general}
\\
& E^i = F^{0i} = -\partial A^i /\partial t - \nabla_i V -\mathrm{i}[V, A^i]\,, \label{eq:nonAB_E_general}
\\
& B^i = -\frac{1}{2}\epsilon^{ijk}F_{jk}= (\nabla\times A)^i -\mathrm{i}(A\times A)^i\,.
\label{eq:nonAB_B_general}
\end{align}
Here, we have chosen the $\mathrm{diag(-+++)}$ signature of the Minkowski metric, and all field variables $A^{\mu}$ are matrices.
In Eq.~\eqref{eq:nonAB_E_general}, the non-Abelian electric field has an extra last term $\mathrm{i}[V,A^i]$ when the scalar and a vector potential do not commute. In Eq.~\eqref{eq:nonAB_B_general}, a similar non-commutative term arises in the non-Abelian magnetic field under non-commutative vector potentials, which have been recently introduced for both Hermitian and non-Hermitian systems in photonic synthetic dimensions~~\cite{cheng2023artificial,pang2024synthetic,cheng2024non}.
To enable full access and manipulation of the field strength tensor in Eq.~\eqref{eq:nonAB_general}, this study aims to present a method for generating non-Abelian electric fields in photonic synthetic dimensions.

\begin{figure}
    \centering
    \includegraphics[width=\linewidth]{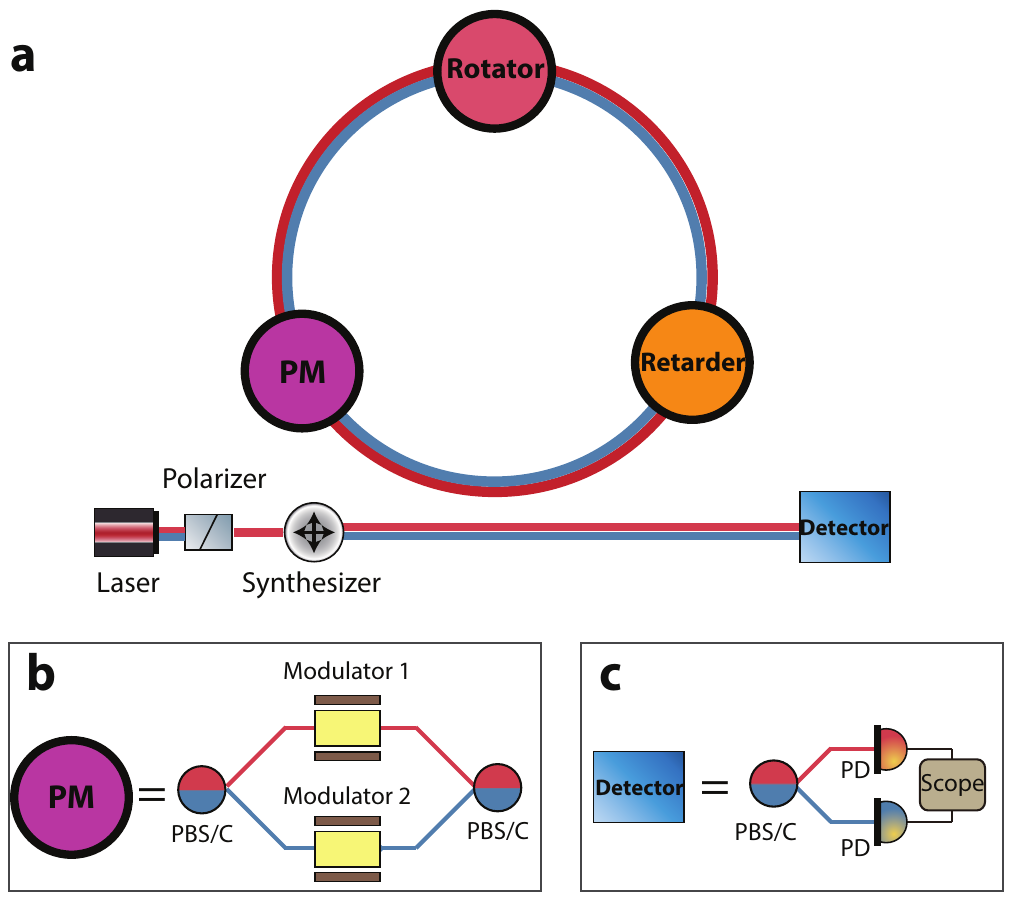}
    \centering
    \caption{\textbf{Setup schematic of a polarization-multiplexed time-modulated ring resonator.} \textbf{a-c.}  The setup contains polarization-dependent phase modulation (PM) enabled by two PBS/Cs and two phase modulators (b), a rotator, and a retarder inside the ring. 
    Outside the ring, a laser gets polarized and can be prepared into arbitrary input states based on the polarization synthesizer. The light intensity can be detected by projection detection in the linear basis (c).  PBS/C: polarization beam splitter/combiner; PD: photodetector.
        }
    \label{fig:setupSOC}
\end{figure}

We consider a system of polarization-multiplexed photonic ring resonator, where the two orthogonal polarizations [horizontal (h) and vertical (v)] serve as the pseudospin. 
The polarization multiplexing is possible with polarization-maintaining fibers and free-space optics.
In the ring resonator, there are three major components: a phase retarder $\mathrm{e}^{\mathrm{i}\phi_z\sigma_z}$ introduces a phase delay between the two orthogonal polarizations, a polarization rotator $\mathrm{e}^{\mathrm{i}\phi_y\sigma_y}$ mixes the two polarizations, and a polarization-dependent phase modulation 
$\mathrm{e}^{\mathrm{i}g \cos(\Omega_{\mathrm{R}} t +\theta\sigma_z)}$ (Fig.~\ref{fig:setupSOC}b) where $g$ is the modulation depth and $\theta$ is the modulation phase opposite for the two polarizations.
Here, $\sigma_y$ and $\sigma_z$ are the Pauli spin matrices of the $y$ and $z$ components, respectively. 
The separation lengths among the three components are denoted by $L_{1,2,3}$, respectively (Fig.~\ref{fig:setupSOC}a).
The ring weakly couples to a waveguide, where, before the light enters the ring, arbitrary polarization can be prepared as the input state using the polarization synthesizer. 
After light exits the ring, they can be tomographically detected, such as in the linear basis (Fig.~\ref{fig:setupSOC}c).

To derive the effective Hamiltonian of this setup, we perform a full electromagnetic derivation by generalizing the method developed in \cite{yuan2016bloch,yuan2016photonic,yuan2021synthetic} for Abelian gauge fields and generalize it for non-Abelian ones.
Inside the ring, light transforms as
\begin{align}\label{eq:transfer}
E (t^+, \mathbf{r}_{\perp},z_0) \!  = \! \mathrm{e}^{\mathrm{i}g \cos(\Omega t +\theta\sigma_z)}\,\mathrm{e}^{\mathrm{i}\phi_y\sigma_y}\mathrm{e}^{\mathrm{i}\phi_z\sigma_z} E (t^-, \mathbf{r}_{\perp}, z_0)
\,,
\end{align}
where $E =(E_\mathrm{h}, E_\mathrm{v})^T$ is a spinor formed by the electric fields in the horizontal and vertical polarizations, $z_0$ is an initial fixed point along the ring,  $\mathbf{r}_\perp=(x,y)$ is the transverse plane perpendicular to the $z$ direction, $t^+$ and $t^-$ are the time before and after the field passes through these optical components.

There potentially exists birefringence between the two polarizations (e.g., as in polarization maintaining fibers). Nevertheless, the birefringence is in the $\sigma_z$ basis and commutes with the modulator and retarder, and thus, it can be incorporated as a calibration to $\phi_z$.
Eq.~\eqref{eq:transfer} requires that the phase delay among the three components is modulo $2\pi$ such that the transfer matrices of these components can group together, which imposes constraints on the modulation frequency $\Omega\approx\Omega_\mathrm{R}\equiv\mathrm{lcm}(\Omega_1,\Omega_2,\Omega_3)$, where lcm denote least common multiple and $\Omega_{i}$ are the associated frequency spectral range (FSR) for length $L_i$ (where $L_1$, $L_2$, and $L_3$ are the optical path length between the retarder and the rotator, the rotator and the phase modulators, and the phase modulators and the retarder, respectively; see SM Sec. S5). 
For example, if $L_1=L_2=L_3$, and the modulation frequency will be three times the FSR of the entire ring resonator.

In the ring resonator, the light field can be expanded as the summation over different resonant modes, $E (t, \mathbf{r}_{\perp}, z) = \sum_{m} E_{m}(t,z) u_{ m}(\mathbf{r}_\perp)\mathrm{e}^{\mathrm{i}\omega_m t}$, where $ E_{ m}(t,z)$ is the modal amplitude of the $m$-th mode, $u_{m}(\mathbf{r}_{\perp})$ is the field profile in the transverse plane, and we assume $u_{m}(\mathbf{r}_{\perp})$ are the same for both polarizations at all $m$, a good approximation for these longitudinal modes when $\omega_m\gg\Omega_{\mathrm{R}}$. Ignoring group velocity dispersion, the propagation of these modes satisfies
$
\left[ \frac{\partial}{\partial z} + \mathrm{i}\beta(\omega_m)\right] E_{ m} + \frac{1}{v_g}\frac{\partial}{\partial t}E_{ m} =0
$
~\cite{haus1984waves}, where $\beta(\omega_m)$ is the propagation constant for mode at $\omega_m$ and $v_g$ is the group velocity (assuming negligible dispersion). 
The general solution follows the ansatz: 
\begin{equation} \label{eq:EfieldFullsolution}
E_{m}(t,z) = A_m \mathrm{e}^{-\mathrm{i} \left(\omega_0 + m \Omega\right)z/v_g}f\left[ (\omega_0 + m\Omega)\left(t- z/v_g \right)  \right] \otimes |\sigma \rangle\,,
\end{equation}
where $f$ is an arbitrary function, $|\sigma\rangle$ is the spinor representing polarization, $A_m$ is the amplitude, $\omega_0$ is the central frequency of the laser, and $\Omega$ is the modulation frequency. 

Due to the circular geometry of the ring resonator, the light fields obey a periodic boundary condition of $E_{ m}(t,z) = E_{ m}(t,z+L)$. Combining the transfer equation Eq.~\eqref{eq:transfer} and the general solution Eq.~\eqref{eq:EfieldFullsolution} leads to the following coupled equation
\begin{align} \label{eq:nonAbelianEfield2}
\begin{aligned}
 E_{m}(t^+ , z_0)= \sum_{p=-\infty}^{\infty}  & \mathrm{i}^{p-m} \mathcal{J}_{p-m}(g)\mathrm{e}^{-\mathrm{i}(p-m)\Delta_{\Omega} t} \\
 & \times \mathrm{e}^{-\mathrm{i}(p-m)\theta \sigma_z } \mathrm{e}^{\mathrm{i}\phi_y \sigma_y }\mathrm{e}^{\mathrm{i}\phi_z \sigma_z } E_{p}(t^- , z_0)\,,
\end{aligned}
\end{align}
where $\mathcal{J}_p$ is the Bessel function of the first kind of the $p$-th order and $\Delta_{\Omega} = \Omega - \Omega_{\mathrm{R}}$ is the detuning frequency. 
We identify the scattering matrix element as $\mathbf{S}_{mp} \equiv \mathrm{i}^{p-m} \mathcal{J}_{p-m}(g)\mathrm{e}^{-\mathrm{i}(p-m)\Delta_{\Omega} t} \left[\mathrm{e}^{-\mathrm{i}(p-m)\theta \sigma_z } \mathrm{e}^{\mathrm{i}\phi_y \sigma_y }\mathrm{e}^{\mathrm{i}\phi_z \sigma_z }\right]$. 
The details of the full derivation of Eq.~\eqref{eq:nonAbelianEfield2} are presented in the supplementary material SM Sec. S1.

Under the first-order approximation on the Bessel function (i.e. only keep $\abs{p-m}\leq1$ terms) for weak modulation $g \rightarrow 0$, we can approximate $\mathcal{J}_0(g) \approx 1$ and $\mathcal{J}_1(g) \approx g/2$. In the limit of $t_\mathrm{R} \rightarrow 0$ (where $t_\mathrm{R}$ is the round-trip time), we can recover a Schrödinger-like equation as follows (see full derivations in SM Sec. S1) :
\begin{align}
\begin{aligned}
&\quad \iu\frac{dE_{m}(t,z_0)}{dt} =-\omega_y \sigma_y E_{m}(t,z_0)-\omega_z \sigma_zE_{m}(t,z_0)\\
&+ J\mathrm{e}^{-\mathrm{i}\Delta_{\Omega} t} \mathrm{e}^{-\mathrm{i}\theta\sigma_z} E_{m+1}(t,z_0) + J\mathrm{e}^{\mathrm{i}\Delta_{\Omega} t} \mathrm{e}^{\mathrm{i}\theta\sigma_z} E_{m-1}(t,z_0) \,,
\end{aligned}
\end{align}
where $\omega_y = \phi_y/t_\mathrm{R} $ and $\omega_z = \phi_z/t_\mathrm{R}$, and $J=-\mathcal{J}_1(g)/t_\mathrm{R}$.
Afer a rotation-wave gauge transformation $|\tilde{\Psi}\rangle =\sum_{m} a_{m} \mathrm{e}^{-\mathrm{i}m\Delta_{\Omega} t} |\psi_{m}\rangle= \sum_m       a_{m} \mathrm{e}^{-\mathrm{i}m\Delta_{\Omega} t}\hat{a}_{m}^\dagger |0\rangle$, a second quantized Hamiltonian can be obtained as
\begin{align} \label{eq:fullhamBloch}
\begin{aligned}
\hat{H} = &\sum_{m}\bigg[ m\Delta_{\Omega} \hat{a}^\dagger_{m}\hat{a}_{m} -\omega_y \hat{a}^\dagger_{m} \sigma_y \hat{a}_{m} -\omega_z \hat{a}^\dagger_{m} \sigma_z \hat{a}_{m}\\
&+J\hat{a}_{m}^\dagger \mathrm{e}^{-\mathrm{i}\theta\sigma_z} \hat{a}_{m+1} + J\hat{a}^\dagger_{m+1}\mathrm{e}^{\mathrm{i}\theta\sigma_z}\hat{a}_{m}   \bigg] \,.
\end{aligned}
\end{align}
The first term is a staggered on-site potential associated to an Abelian electric field~\cite{yuan2016bloch}, the second and third terms are homogeneous non-Abelian on-site scalar potentials, while the remaining correspond to non-Abelian vector potentials. 

In particular, when there is no detuning $\Delta_\Omega =0$, this setup gives rise to a non-Abelian electric field because of the non-commutativity between the scalar and the vector potentials in Eq.~\eqref{eq:nonAB_E_general}
\begin{align} \label{eq:nonAbelianElectricField}
\hat{E}_x = -\mathrm{i}[V, A_x] = \mathrm{i}[\omega_y\sigma_y+\omega_z\sigma_z, (\theta/e) \sigma_z]= (2\omega_y\theta /e)\sigma_x,
\end{align}
where we take $\hbar=1$ in the natural unit. 
To complement our derivation here, we provide an alternative approach to deriving this Hamiltonian using a transfer matrix method (see SM Sec. S2).

\begin{figure}[htbp]
    \centering
    \includegraphics[width=\linewidth]{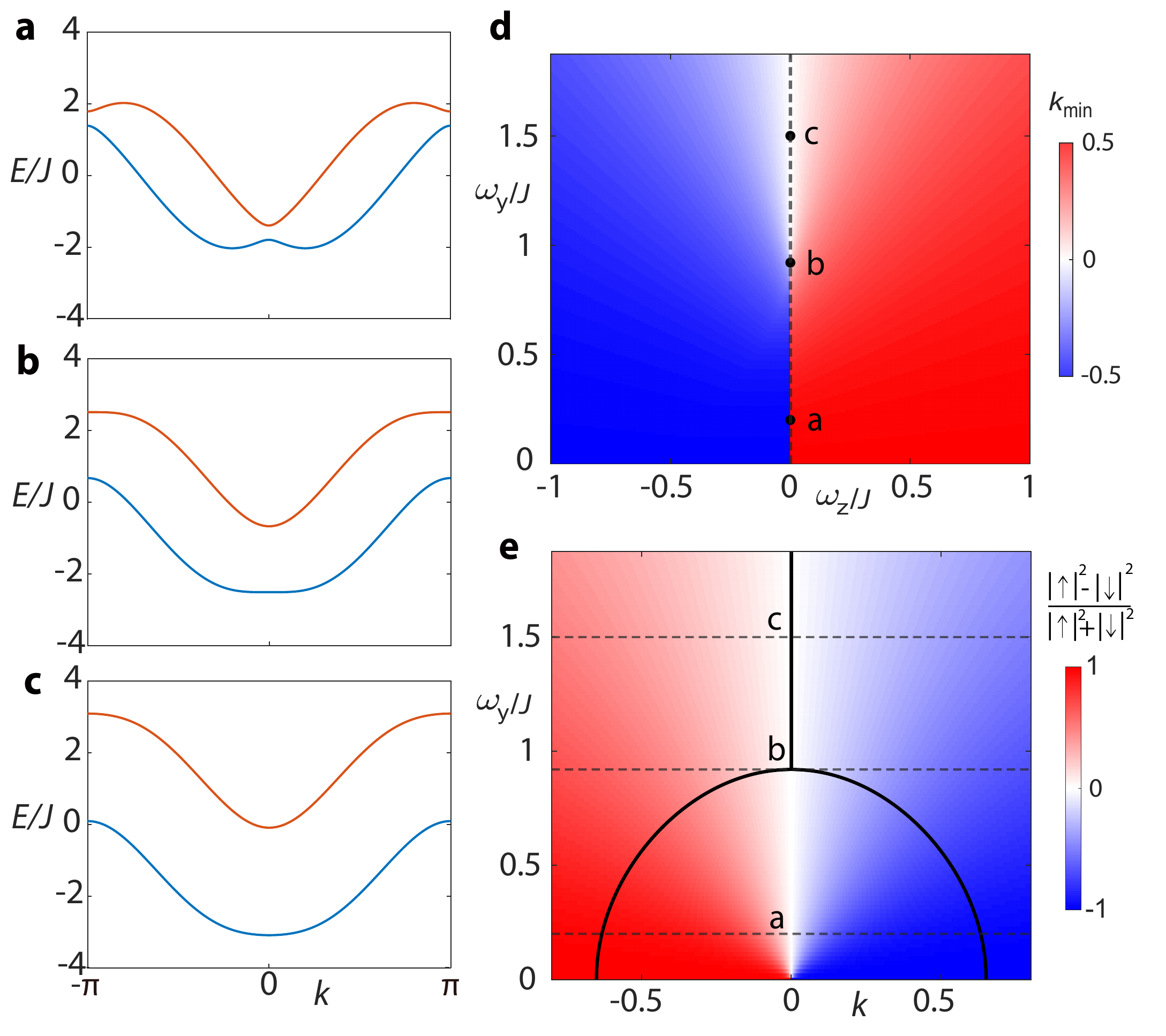}
    \centering
    \caption{\textbf{Photonic band structure and spin texture under synthetic SOC.} 
        \textbf{a-c.} Band structure under different conditions $\omega_y/J < \omega_{y}^{\mathrm{c}}/J$ (a),  $\omega_y/J = \omega_{y}^{\mathrm{c}}/J$ (b), and $\omega_y/J > \omega_{y}^\mathrm{c}/J$ (c). 
        \textbf{d.} Ground-state momenta as a function of $\omega_{y}/J$ and $\omega_{z}/J$. 
        \textbf{e.} Spin texture overlaid with the trajectory of the ground state momenta (black line) when $\omega_z=0$ (i.e., along the dashed line in d). 
        In the whole figure, we fix $\theta = 0.65$ and $\omega_{y}^{\mathrm{c}}/J =2 \sin\theta \tan\theta =0.92$.
        }
    \label{fig:SOCenergyband}
\end{figure}

\emph{Synthetic Spin-Orbit Coupling. }
The proposed experimental setup in Fig.~\ref{fig:setupSOC} faithfully realizes synthetic spin-orbit coupling (SOC) in the photonic synthetic dimension.
Under zero detuning $\Delta_\Omega=0$, the Bloch Hamiltonian reads
\begin{equation} \label{eq:nonAbelianH}
H(k) = 2J\cos(\theta- k\sigma_z) -\omega_y \sigma_y-\omega_z \sigma_z \,.
\end{equation}
which is a one-dimensional lattice under an equal mixing of Rashba and Dresselhaus SOC (see SM Sec. S3).
The eigen-energies of Eq.~\eqref{eq:nonAbelianH} are given by
$
E_{\pm}(k) = 2J\cos\theta\cos k \pm \sqrt{\omega_y^2 + \omega_z^2 +4J^2 \sin^2\theta\sin^2 ka -4J\omega_z \sin\theta\sin k }.
$
We first consider $\omega_z =0$. The local extrema of the lower band $E_{-}(k)$ occurs at 
\begin{equation} \label{eq:minimum}
k=0 \quad \text{and}\quad k = \sin^{-1}\left(\pm\sqrt{\sin^2\theta - \bigg(\frac{\omega_y}{2J}\bigg)^2\cot^2\theta} \right) \,.
\end{equation}

When $\abs{\omega_y} < \omega_{y}^{\mathrm{c}} \equiv 2J \sin\theta \tan\theta$, there are two energy-degenerate minima at different momenta (Fig.~\ref{fig:SOCenergyband}a); when $\omega_y = \pm \omega_{y}^{\mathrm{c}}$, the two minima merge at $k=0$ (Fig.~\ref{fig:SOCenergyband}b); when $\abs{\omega_y} > \omega_{y}^{\mathrm{c}}$, only a single minimum remains (Fig.~\ref{fig:SOCenergyband}c). 
The parameter choices of $\omega_y$ and $\omega_z$ in Fig.~\ref{fig:SOCenergyband}a-c together with the associated ground-state momenta are shown in Fig.~\ref{fig:SOCenergyband}d.
Along the $\omega_z=0$ line cut, we plot the ground-state momenta as a function of $\omega_y$ in Fig.~\ref{fig:SOCenergyband}e, where the background is overlaid with the contrast $\eta \equiv (I_{\mathrm{h}} -  I_{\mathrm{v}})/( I_{\mathrm{h}} +  I_{\mathrm{v}})$ where $I_{\mathrm{h,v}}$ are intensity measured in the horizontal and vertical basis, respectively.
Notably, the ground state momenta obey $k(\omega_y) \sim  (\omega_{y}^{\mathrm{c}}- \omega_y )^{1/2}$.
When $\omega_z\neq0$, the spectra become asymmetric with regard to $k=0$ (Fig.~\ref{fig:SOCenergyband}d). 
The sign of $\omega_z$ determines the asymmetry. If $\omega_z <0$, it is left-tilted (Fig.~\ref{fig:SOCenergyband}d) and right-tilted otherwise. 

\begin{figure}[htbp]
    \centering
    \includegraphics[width=\linewidth]{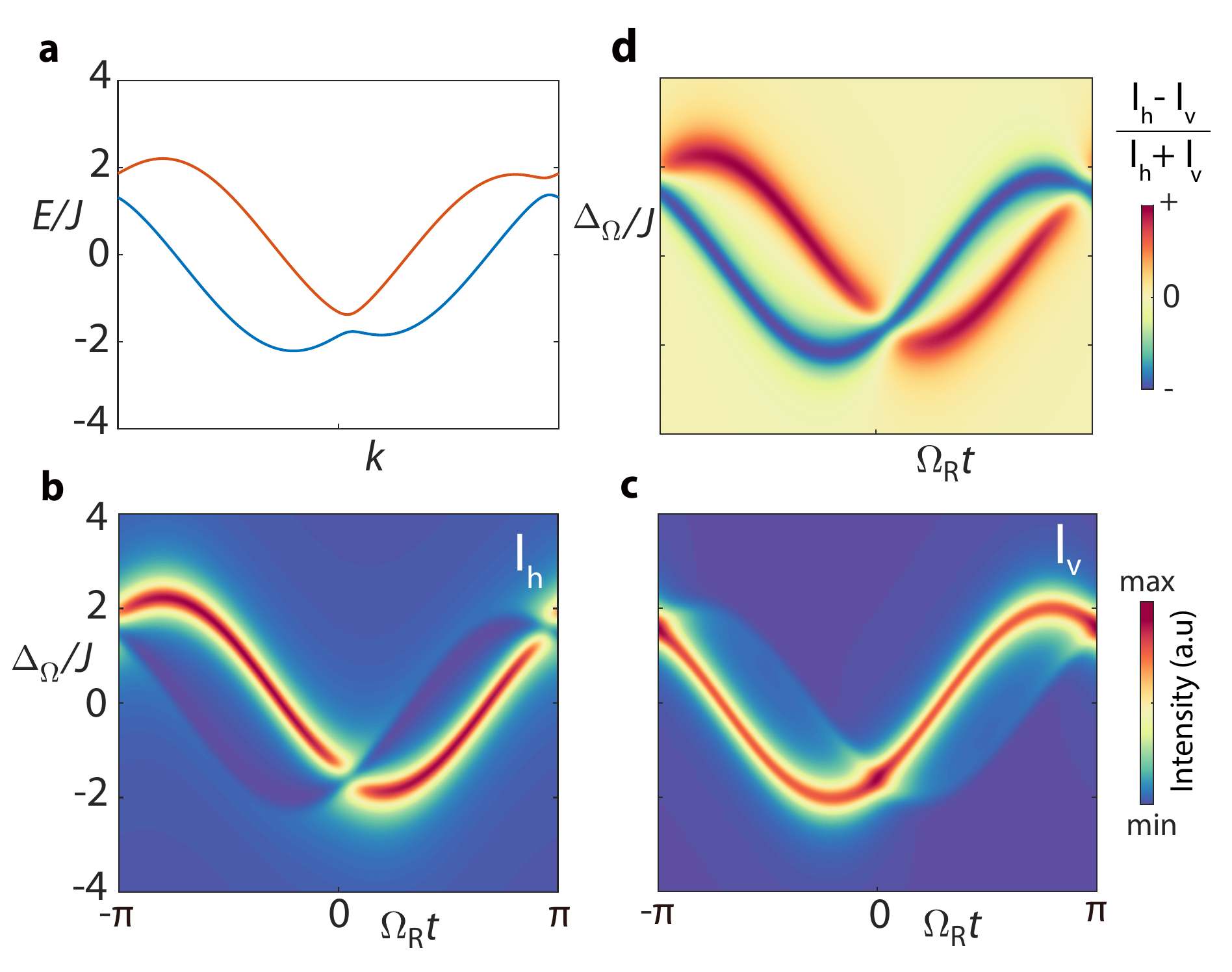}
    \centering
    \caption{\textbf{Continuous-wave measurement of synthetic bands and pseudospin texture. } 
        \textbf{a.} An example band structure with simultaneously nonzero $\omega_y$ and $\omega_z$. 
        \textbf{b-c.} Projection detection of intensity in the horizontal (b) and vertical (c) polarization according to Eq.~\eqref{eq:projectedIntensity}.
        \textbf{d.} Intensity contrast as a function of $I_{\mathrm{h}}$ and $I_{\mathrm{v}}$ confirming the synthetic SOC.
        Parameter choices are $\theta=0.65$, $\omega_y/J = 0.2$, $\omega_z /J= -0.19$ and $\gamma_{\mathrm{in}}/J = 0.3$. We pick the input state as $\ket{\psi_{\mathrm{\mathrm{i}n}}} = \frac{1}{\sqrt{2}}(1, 1)^T$. 
        }
    \label{fig:signalProjection}
\end{figure}

\emph{Band structure spectroscopy.}
To experimentally confirm the SOC, the spin texture of the bands can be probed using polarization-resolved band structure spectroscopy along the synthetic dimension (Fig.~\ref{fig:setupSOC}c)~\cite{dutt2019experimental,pang2024synthetic}.
A continuous-wave laser beam with detuning $\Delta_{\Omega}$ and arbitrary polarization is injected to the ring resonator through the polarizer and synthesizer (see Fig.~\ref{fig:setupSOC}b). 
We plot the calculated projected intensity in the linear basis $I_{\mathrm{h}}(\Delta_{\Omega},k)$ and $I_{\mathrm{v}}(\Delta_{\Omega},k)$ as a function of laser detuning frequency $\Delta_{\Omega}$ and quasi-momentum $k$, which can be directly measured by the two 
photodetectors illustrated in Fig.~\ref{fig:setupSOC}c. In this synthetic-dimension experiment, the quasi-momentum $k$ corresponds to time $t=k/\Omega_\mathrm{R}$~\cite{wang2021generating}. The projected intensities in the linear basis (Fig.~\ref{fig:signalProjection}b-c) for the horizontal and vertical polarization are 
\begin{align} \label{eq:projectedIntensity}
I_\alpha  = |\psi_\alpha|^2 =\bigg\vert \braket{ \psi_\alpha | \psi_{\mathrm{\mathrm{i}n}}} + \mathrm{i}\gamma_{\mathrm{\mathrm{i}n}} \sum_{j=1,2} \frac{ \braket{\psi_\alpha | \psi_j } \braket{ \psi_j |\psi_{\mathrm{\mathrm{i}n}}}} { E_j -\Delta_{\Omega} + \iu\gamma_{\mathrm{\mathrm{i}n}} } \bigg\vert^2 \,,
\end{align}
where $\alpha=\left\{\mathrm{h},\mathrm{v}\right\}$, $\ket{\psi_{1,2}}$ are the eigenvectors of the Bloch Hamiltonian in Eq.~\eqref{eq:nonAbelianH}, and $\ket{\psi_{\mathrm{\mathrm{i}n}}}$ is the input state. 
These intensities permit the calculation of their contrast $\eta$ (Fig.~\ref{fig:signalProjection}d), which shows that the two bands are predominated by opposite spins on opposite sides of the band structure, indicating the onset of SOC.

\begin{figure}
    \centering
    \includegraphics[width=\linewidth]{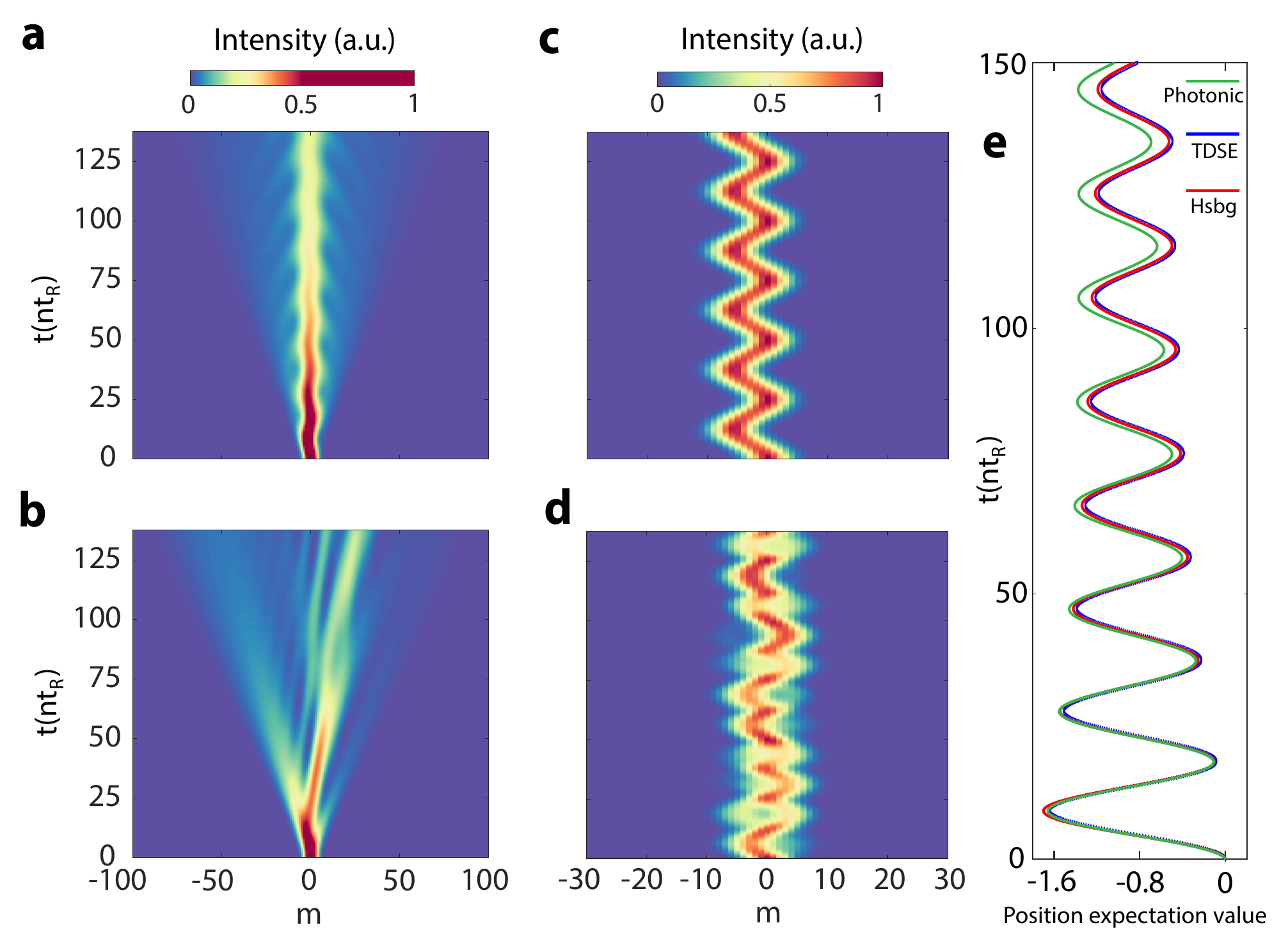}
    \centering
    \caption{\textbf{Pulse dynamics in the synthetic frequency dimension showing Zitterbewegung (ZB), Bloch oscillation, and their interplay.}
         \textbf{a.} Dynamics without frequency detuning when $\omega_y>\omega_y^\mathrm{c}$ (c.f. Fig.~\ref{fig:SOCenergyband}c where the bands feature single minima) exhibits ZB oscillation. Here $\omega_y /J= 0.4$ and  $\omega_z /J= 0$, and $\omega_y > \omega_y^\mathrm{c}$.
         \textbf{b.} Dynamics without frequency detuning when $\omega_y<\omega_y^\mathrm{c}$ (c.f. Fig.~\ref{fig:signalProjection}a) where the bands feature tilted double minima. Here $\omega_y/J = 0.1$ and $\omega_z /J= -0.13$. 
         The color bar is saturated at 0.5 for both (a) and (b) for better visualization. 
         \textbf{c.} Bloch oscillation under pure Abelian gauge fields. Here $\omega_y/J =\omega_z/J =0$.
         \textbf{d.} Complicated dynamic pattern resulting from the interplay of ZB and Bloch oscillation with both Abelian and non-Abelian electric field present. Here $\omega_y/J = 0.5$ and $\omega_z /J= -0.13$.
         Frequency detuning is $\Delta_\Omega = 0.01 \Omega_{\mathrm{R}}$ for both c and d. 
         %
         In a, b, and c, $\theta = 0.4$\,rad, while in d, we choose $\theta=\pi/2$\,rad.
         \textbf{e.} Position expectation in the photonic simulation (green; associated to Fig.~\ref{fig:FullEMSimulation_vertical}a) benchmarked with quantum mechanical solutions [blue: direct numerical solution to time-dependent Schrodinger equation (TDSE); red: expectation value evaluated by the semi-analytical Heisenberg calculations (Hsbg)].
         In all the calculations, we choose the input state as $\ket{\psi_{\mathrm{\mathrm{i}n}}} = \frac{1}{\sqrt{2}}(1,  1)^T$.
         }    \label{fig:FullEMSimulation_vertical}
\end{figure}

\emph{Dynamics under non-Abelian electric field.}
The pseudospin dynamics in the frequency dimension under the synthetic non-Abelian electric field can be measured under pulsed excitation.
Under this configuration, after $n$ round trips, electric fields transform as
\begin{align} \label{eq:Efieldtransform}
E_{m} (t+nt_\mathrm{R},z) = \left[{\left(\mathbf{P}\otimes \sigma_0\right)\mathbf{S}}\right]^n E_{m} (t,z),
\end{align}
where $n$ is the number of round trips, $\mathbf{S}$ is the scattering matrix defined previously, $\sigma_0$ is the $2\times 2$ identity matrix. $\mathbf{P}$ is the free propagator given by $\mathbf{P}_{\alpha\beta} = e^{-\mathrm{i}\Theta(\alpha)}\delta_{\alpha\beta}$, where $\alpha,\beta = -N,-(N-1),\cdots, N-1, N$, $\Theta(m) = (\omega_0 + m\Omega)L/v_g$, and $\omega_0$ is the central frequency. 
We include our discussion on the general configurations with arbitrary delays among the optical components in SM Sec. S5.

In our analysis, we take $A_m =\frac{1}{\sqrt{L}}\exp(-(\omega_m - \omega_0)^2 / \chi \Omega_{\mathrm{R}}^2)$ as a Gaussian profile and $\chi$ parameterizes its width. 
We track the total intensity of each mode $m$ (summing both polarizations) by $\mathcal{I}_m = \int_0^L (|E_{m\,\mathrm{h}}(t,z)|^2 +  |E_{m\,\mathrm{v}}(t,z)|^2 ) dz \,.$
To elucidate the interplay between Abelian and non-Abelian electric fields, we dichotomize our study into two parts. 
First, with zero detuning $\Delta_\Omega=0$, the propagator $\mathbf{P}$ reduces to identity, and only the non-Abelian electric field is present in the system. Fig.~\ref{fig:FullEMSimulation_vertical}a and  Fig.~\ref{fig:FullEMSimulation_vertical}b showcase the dynamics without detuning when $\Delta_\Omega=0$. Fig.~\ref{fig:FullEMSimulation_vertical}a illustrates the dynamics when $\omega_z /J =0$ and $\omega_y /J > \omega_y^{\mathrm{c}}/J$, corresponding to the case of energy dispersion with one single minimum only (Fig.~\ref{fig:SOCenergyband}c). The evolution of center of mass can be calculated as $\sum_{m}m\mathcal{I}_m/\sum_{m}\mathcal{I}_m$, as shown by the green curve in Fig.~\ref{fig:FullEMSimulation_vertical}e, which demonstrates a Zitterbewegung (ZB) feature [explicit expression obtained from quantum mechanical calculation shown in Eq.~\eqref{eq:xbasisHsbgmaintext}].
In contrast, Fig.~\ref{fig:FullEMSimulation_vertical}b shows the dynamics when $\omega_z /J \neq 0$ and two minima occur for the energy bands (Fig.~\ref{fig:SOCenergyband}a), resulting in two main distinguished and sided trajectories. 
After that, we turn on detuning $\Delta_\Omega$ to introduce the staggered Abelian gauge fields [first term in Eq.~\eqref{eq:fullhamBloch}]. When $\omega_y=\omega_z=0$, the non-Abelian electric field vanishes, and the associated dynamics exhibit typical Bloch oscillation (Fig.~\ref{fig:FullEMSimulation_vertical}c)~\cite{yuan2016bloch}.
The angular frequency of the Bloch oscillation is $\omega_\mathrm{B} = \Delta_{\Omega}$.
In Fig.~\ref{fig:FullEMSimulation_vertical}d, when both Abelian and non-Abelian electric fields are present, the dynamics display a complicated trajectory due to the interference between Bloch oscillation and ZB contributions.

Finally, we perform two quantum mechanical benchmarks---a direct numerical solution to the time-dependent Schrodinger equation (TDSE) and the semi-analytical center-of-mass calculation in the Heisenberg picture---to confirm the validity of the photonic simulation.
In the TDSE calculation (details in SM Sec. S6), the wave function of the time-evolved state can be obtained from
$
    \psi(x,t) = \int_{-\infty}^{\infty} dp \,\mathrm{e}^{-\mathrm{i}H(p)t}\mathrm{e}^{\mathrm{i}px} \ip{p}{\psi(0)} \,,
$
where $H(p)$ is the Hamiltonian matrix in Eq.~\eqref{eq:nonAbelianH} and $\ip{p}{\psi(0)} =e^{-\gamma p^2}| \sigma \rangle$ with the relation of $\gamma /\chi= 1/4$ from the Fourier transform of a Gaussian. The position expectation value can then be evaluated by $\braket{x(t)} = \int dx\,x|\psi(x,t)|^2$. 
Meanwhile, in the Heisenberg center-of-mass calculation (details in SM. Sec. S7), we obtain the semi-analytical solutions for the position, velocity, and acceleration operators and evaluate their expectation values under a low-momentum approximation for the Yang-Mills Hamiltonian. The trajectory displays a typical Zitterbewegung feature under $\sigma_x$ input state, whose explicit trajectory is given by (derivation shown in SM. Sec. S7)
\begin{align} \label{eq:xbasisHsbgmaintext}
 \langle \hat{x} (t) \rangle_{\sigma_x} = - \frac{\omega_y}{m}\sqrt{\frac{2\gamma}{\pi}}\int_{-\infty}^{\infty}  \frac{1-\cos\big(2\theta G(p,\omega_y,\theta)t\big)}{2\theta G^2 (p,\omega_y,\theta)} \mathrm{e}^{-2\gamma p^2} dp \,.
\end{align}
where $G(p,\omega_y,\theta)\equiv \sqrt{(p/m)^2 + (\omega_y/\theta)^2 }$ with effective mass $m=-1/2J$. 
Evidently, the integrand's oscillatory behavior gets inherited by the position expectation value, leading to a ZB trajectory in Fig.~\ref{fig:FullEMSimulation_vertical}a.
The photonic simulation achieves excellent agreement with the two quantum mechanical benchmarks (see Fig.~\ref{fig:FullEMSimulation_vertical}e).

\emph{Conclusions.}
To sum up, we theoretically propose the synthesis of non-Abelian electric field and spin-orbit coupling in photonic synthetic dimensions based on a polarization-multiplexed time-modulated ring resonator.
The underlying band structure and spin texture can be spectroscopically probed under continuous-wave excitation.
Under pulse excitation, frequency-domain Zitterbewegung appears and exhibits rich interplay with Bloch oscillation.
The proposed setup could be realized on various photonic platforms, such as free-space optics, fibers, and integrated photonics.
Looking forward, the combinatorial use of both second-order and third-order optical nonlinearity on this setup may help create photonic many-body effects with spin-orbit coupling.

\emph{Acknowledgments}. We thank Prof. Shizhong Zhang for the helpful discussion and proofreading of the manuscript.


\begin{thebibliography}{50}%
\makeatletter
\providecommand \@ifxundefined [1]{%
 \@ifx{#1\undefined}
}%
\providecommand \@ifnum [1]{%
 \ifnum #1\expandafter \@firstoftwo
 \else \expandafter \@secondoftwo
 \fi
}%
\providecommand \@ifx [1]{%
 \ifx #1\expandafter \@firstoftwo
 \else \expandafter \@secondoftwo
 \fi
}%
\providecommand \natexlab [1]{#1}%
\providecommand \enquote  [1]{``#1''}%
\providecommand \bibnamefont  [1]{#1}%
\providecommand \bibfnamefont [1]{#1}%
\providecommand \citenamefont [1]{#1}%
\providecommand \href@noop [0]{\@secondoftwo}%
\providecommand \href [0]{\begingroup \@sanitize@url \@href}%
\providecommand \@href[1]{\@@startlink{#1}\@@href}%
\providecommand \@@href[1]{\endgroup#1\@@endlink}%
\providecommand \@sanitize@url [0]{\catcode `\\12\catcode `\$12\catcode `\&12\catcode `\#12\catcode `\^12\catcode `\_12\catcode `\%12\relax}%
\providecommand \@@startlink[1]{}%
\providecommand \@@endlink[0]{}%
\providecommand \url  [0]{\begingroup\@sanitize@url \@url }%
\providecommand \@url [1]{\endgroup\@href {#1}{\urlprefix }}%
\providecommand \urlprefix  [0]{URL }%
\providecommand \Eprint [0]{\href }%
\providecommand \doibase [0]{http://dx.doi.org/}%
\providecommand \selectlanguage [0]{\@gobble}%
\providecommand \bibinfo  [0]{\@secondoftwo}%
\providecommand \bibfield  [0]{\@secondoftwo}%
\providecommand \translation [1]{[#1]}%
\providecommand \BibitemOpen [0]{}%
\providecommand \bibitemStop [0]{}%
\providecommand \bibitemNoStop [0]{.\EOS\space}%
\providecommand \EOS [0]{\spacefactor3000\relax}%
\providecommand \BibitemShut  [1]{\csname bibitem#1\endcsname}%
\let\auto@bib@innerbib\@empty
\bibitem [{\citenamefont {Aidelsburger}\ \emph {et~al.}(2018)\citenamefont {Aidelsburger}, \citenamefont {Nascimbene},\ and\ \citenamefont {Goldman}}]{aidelsburger2018review}%
  \BibitemOpen
  \bibfield  {author} {\bibinfo {author} {\bibfnamefont {M.}~\bibnamefont {Aidelsburger}}, \bibinfo {author} {\bibfnamefont {S.}~\bibnamefont {Nascimbene}}, \ and\ \bibinfo {author} {\bibfnamefont {N.}~\bibnamefont {Goldman}},\ }\href@noop {} {\bibfield  {journal} {\bibinfo  {journal} {Comptes Rendus Physique}\ }\textbf {\bibinfo {volume} {19}},\ \bibinfo {pages} {394} (\bibinfo {year} {2018})}\BibitemShut {NoStop}%
\bibitem [{\citenamefont {Zhai}(2015)}]{zhai2015degenerate}%
  \BibitemOpen
  \bibfield  {author} {\bibinfo {author} {\bibfnamefont {H.}~\bibnamefont {Zhai}},\ }\href@noop {} {\bibfield  {journal} {\bibinfo  {journal} {Reports on Progress in Physics}\ }\textbf {\bibinfo {volume} {78}},\ \bibinfo {pages} {026001} (\bibinfo {year} {2015})}\BibitemShut {NoStop}%
\bibitem [{\citenamefont {Lin}\ \emph {et~al.}(2011)\citenamefont {Lin}, \citenamefont {Jim{\'e}nez-Garc{\'\i}a},\ and\ \citenamefont {Spielman}}]{lin2011spin}%
  \BibitemOpen
  \bibfield  {author} {\bibinfo {author} {\bibfnamefont {Y.-J.}\ \bibnamefont {Lin}}, \bibinfo {author} {\bibfnamefont {K.}~\bibnamefont {Jim{\'e}nez-Garc{\'\i}a}}, \ and\ \bibinfo {author} {\bibfnamefont {I.~B.}\ \bibnamefont {Spielman}},\ }\href@noop {} {\bibfield  {journal} {\bibinfo  {journal} {Nature}\ }\textbf {\bibinfo {volume} {471}},\ \bibinfo {pages} {83} (\bibinfo {year} {2011})}\BibitemShut {NoStop}%
\bibitem [{\citenamefont {Zhang}\ \emph {et~al.}(2015)\citenamefont {Zhang}, \citenamefont {Cole}, \citenamefont {Paramekanti},\ and\ \citenamefont {Trivedi}}]{zhang2015spin}%
  \BibitemOpen
  \bibfield  {author} {\bibinfo {author} {\bibfnamefont {S.}~\bibnamefont {Zhang}}, \bibinfo {author} {\bibfnamefont {W.~S.}\ \bibnamefont {Cole}}, \bibinfo {author} {\bibfnamefont {A.}~\bibnamefont {Paramekanti}}, \ and\ \bibinfo {author} {\bibfnamefont {N.}~\bibnamefont {Trivedi}},\ }\href@noop {} {\bibfield  {journal} {\bibinfo  {journal} {Annual Review of Cold Atoms and Molecules}\ ,\ \bibinfo {pages} {135}} (\bibinfo {year} {2015})}\BibitemShut {NoStop}%
\bibitem [{\citenamefont {Goldman}\ \emph {et~al.}(2014)\citenamefont {Goldman}, \citenamefont {Juzeli{\=u}nas}, \citenamefont {{\"O}hberg},\ and\ \citenamefont {Spielman}}]{goldman2014light}%
  \BibitemOpen
  \bibfield  {author} {\bibinfo {author} {\bibfnamefont {N.}~\bibnamefont {Goldman}}, \bibinfo {author} {\bibfnamefont {G.}~\bibnamefont {Juzeli{\=u}nas}}, \bibinfo {author} {\bibfnamefont {P.}~\bibnamefont {{\"O}hberg}}, \ and\ \bibinfo {author} {\bibfnamefont {I.~B.}\ \bibnamefont {Spielman}},\ }\href@noop {} {\bibfield  {journal} {\bibinfo  {journal} {Reports on Progress in Physics}\ }\textbf {\bibinfo {volume} {77}},\ \bibinfo {pages} {126401} (\bibinfo {year} {2014})}\BibitemShut {NoStop}%
\bibitem [{\citenamefont {Yang}\ \emph {et~al.}(2020)\citenamefont {Yang}, \citenamefont {Zhen}, \citenamefont {Joannopoulos},\ and\ \citenamefont {Solja{\v{c}}i{\'{c}}}}]{yang20202d}%
  \BibitemOpen
  \bibfield  {author} {\bibinfo {author} {\bibfnamefont {Y.}~\bibnamefont {Yang}}, \bibinfo {author} {\bibfnamefont {B.}~\bibnamefont {Zhen}}, \bibinfo {author} {\bibfnamefont {J.~D.}\ \bibnamefont {Joannopoulos}}, \ and\ \bibinfo {author} {\bibfnamefont {M.}~\bibnamefont {Solja{\v{c}}i{\'{c}}}},\ }\href@noop {} {\bibfield  {journal} {\bibinfo  {journal} {Light: Science {\&} Applications}\ }\textbf {\bibinfo {volume} {9}},\ \bibinfo {pages} {177} (\bibinfo {year} {2020})}\BibitemShut {NoStop}%
\bibitem [{\citenamefont {Whittaker}\ \emph {et~al.}(2021)\citenamefont {Whittaker}, \citenamefont {Dowling}, \citenamefont {Nalitov}, \citenamefont {Yulin}, \citenamefont {Royall}, \citenamefont {Clarke}, \citenamefont {Skolnick}, \citenamefont {Shelykh},\ and\ \citenamefont {Krizhanovskii}}]{whittaker2021optical}%
  \BibitemOpen
  \bibfield  {author} {\bibinfo {author} {\bibfnamefont {C.}~\bibnamefont {Whittaker}}, \bibinfo {author} {\bibfnamefont {T.}~\bibnamefont {Dowling}}, \bibinfo {author} {\bibfnamefont {A.}~\bibnamefont {Nalitov}}, \bibinfo {author} {\bibfnamefont {A.}~\bibnamefont {Yulin}}, \bibinfo {author} {\bibfnamefont {B.}~\bibnamefont {Royall}}, \bibinfo {author} {\bibfnamefont {E.}~\bibnamefont {Clarke}}, \bibinfo {author} {\bibfnamefont {M.}~\bibnamefont {Skolnick}}, \bibinfo {author} {\bibfnamefont {I.}~\bibnamefont {Shelykh}}, \ and\ \bibinfo {author} {\bibfnamefont {D.}~\bibnamefont {Krizhanovskii}},\ }\href@noop {} {\bibfield  {journal} {\bibinfo  {journal} {Nature Photonics}\ }\textbf {\bibinfo {volume} {15}},\ \bibinfo {pages} {193} (\bibinfo {year} {2021})}\BibitemShut {NoStop}%
\bibitem [{\citenamefont {Ter{\c{c}}as}\ \emph {et~al.}(2014)\citenamefont {Ter{\c{c}}as}, \citenamefont {Flayac}, \citenamefont {Solnyshkov},\ and\ \citenamefont {Malpuech}}]{terccas2014non}%
  \BibitemOpen
  \bibfield  {author} {\bibinfo {author} {\bibfnamefont {H.}~\bibnamefont {Ter{\c{c}}as}}, \bibinfo {author} {\bibfnamefont {H.}~\bibnamefont {Flayac}}, \bibinfo {author} {\bibfnamefont {D.}~\bibnamefont {Solnyshkov}}, \ and\ \bibinfo {author} {\bibfnamefont {G.}~\bibnamefont {Malpuech}},\ }\href@noop {} {\bibfield  {journal} {\bibinfo  {journal} {Physical review letters}\ }\textbf {\bibinfo {volume} {112}},\ \bibinfo {pages} {066402} (\bibinfo {year} {2014})}\BibitemShut {NoStop}%
\bibitem [{\citenamefont {Sedov}\ \emph {et~al.}(2018)\citenamefont {Sedov}, \citenamefont {Rubo},\ and\ \citenamefont {Kavokin}}]{sedov2018zitterbewegung}%
  \BibitemOpen
  \bibfield  {author} {\bibinfo {author} {\bibfnamefont {E.}~\bibnamefont {Sedov}}, \bibinfo {author} {\bibfnamefont {Y.}~\bibnamefont {Rubo}}, \ and\ \bibinfo {author} {\bibfnamefont {A.}~\bibnamefont {Kavokin}},\ }\href@noop {} {\bibfield  {journal} {\bibinfo  {journal} {Physical Review B}\ }\textbf {\bibinfo {volume} {97}},\ \bibinfo {pages} {245312} (\bibinfo {year} {2018})}\BibitemShut {NoStop}%
\bibitem [{\citenamefont {Hasan}\ \emph {et~al.}(2022)\citenamefont {Hasan}, \citenamefont {Madasu}, \citenamefont {Rathod}, \citenamefont {Kwong}, \citenamefont {Miniatura}, \citenamefont {Chevy},\ and\ \citenamefont {Wilkowski}}]{hasan2022wave}%
  \BibitemOpen
  \bibfield  {author} {\bibinfo {author} {\bibfnamefont {M.}~\bibnamefont {Hasan}}, \bibinfo {author} {\bibfnamefont {C.~S.}\ \bibnamefont {Madasu}}, \bibinfo {author} {\bibfnamefont {K.~D.}\ \bibnamefont {Rathod}}, \bibinfo {author} {\bibfnamefont {C.~C.}\ \bibnamefont {Kwong}}, \bibinfo {author} {\bibfnamefont {C.}~\bibnamefont {Miniatura}}, \bibinfo {author} {\bibfnamefont {F.}~\bibnamefont {Chevy}}, \ and\ \bibinfo {author} {\bibfnamefont {D.}~\bibnamefont {Wilkowski}},\ }\href@noop {} {\bibfield  {journal} {\bibinfo  {journal} {Physical Review Letters}\ }\textbf {\bibinfo {volume} {129}},\ \bibinfo {pages} {130402} (\bibinfo {year} {2022})}\BibitemShut {NoStop}%
\bibitem [{\citenamefont {Polimeno}\ \emph {et~al.}(2021)\citenamefont {Polimeno}, \citenamefont {Fieramosca}, \citenamefont {Lerario}, \citenamefont {De~Marco}, \citenamefont {De~Giorgi}, \citenamefont {Ballarini}, \citenamefont {Dominici}, \citenamefont {Ardizzone}, \citenamefont {Pugliese}, \citenamefont {Prontera} \emph {et~al.}}]{polimeno2021experimental}%
  \BibitemOpen
  \bibfield  {author} {\bibinfo {author} {\bibfnamefont {L.}~\bibnamefont {Polimeno}}, \bibinfo {author} {\bibfnamefont {A.}~\bibnamefont {Fieramosca}}, \bibinfo {author} {\bibfnamefont {G.}~\bibnamefont {Lerario}}, \bibinfo {author} {\bibfnamefont {L.}~\bibnamefont {De~Marco}}, \bibinfo {author} {\bibfnamefont {M.}~\bibnamefont {De~Giorgi}}, \bibinfo {author} {\bibfnamefont {D.}~\bibnamefont {Ballarini}}, \bibinfo {author} {\bibfnamefont {L.}~\bibnamefont {Dominici}}, \bibinfo {author} {\bibfnamefont {V.}~\bibnamefont {Ardizzone}}, \bibinfo {author} {\bibfnamefont {M.}~\bibnamefont {Pugliese}}, \bibinfo {author} {\bibfnamefont {C.}~\bibnamefont {Prontera}},  \emph {et~al.},\ }\href@noop {} {\bibfield  {journal} {\bibinfo  {journal} {Optica}\ }\textbf {\bibinfo {volume} {8}},\ \bibinfo {pages} {1442} (\bibinfo {year} {2021})}\BibitemShut {NoStop}%
\bibitem [{\citenamefont {Lovett}\ \emph {et~al.}(2023)\citenamefont {Lovett}, \citenamefont {Walker}, \citenamefont {Osipov}, \citenamefont {Yulin}, \citenamefont {Naik}, \citenamefont {Whittaker}, \citenamefont {Shelykh}, \citenamefont {Skolnick},\ and\ \citenamefont {Krizhanovskii}}]{lovett2023observation}%
  \BibitemOpen
  \bibfield  {author} {\bibinfo {author} {\bibfnamefont {S.}~\bibnamefont {Lovett}}, \bibinfo {author} {\bibfnamefont {P.~M.}\ \bibnamefont {Walker}}, \bibinfo {author} {\bibfnamefont {A.}~\bibnamefont {Osipov}}, \bibinfo {author} {\bibfnamefont {A.}~\bibnamefont {Yulin}}, \bibinfo {author} {\bibfnamefont {P.~U.}\ \bibnamefont {Naik}}, \bibinfo {author} {\bibfnamefont {C.~E.}\ \bibnamefont {Whittaker}}, \bibinfo {author} {\bibfnamefont {I.~A.}\ \bibnamefont {Shelykh}}, \bibinfo {author} {\bibfnamefont {M.~S.}\ \bibnamefont {Skolnick}}, \ and\ \bibinfo {author} {\bibfnamefont {D.~N.}\ \bibnamefont {Krizhanovskii}},\ }\href@noop {} {\bibfield  {journal} {\bibinfo  {journal} {Light: Science \& Applications}\ }\textbf {\bibinfo {volume} {12}},\ \bibinfo {pages} {126} (\bibinfo {year} {2023})}\BibitemShut {NoStop}%
\bibitem [{\citenamefont {Chen}\ \emph {et~al.}(2019)\citenamefont {Chen}, \citenamefont {Zhang}, \citenamefont {Xiong}, \citenamefont {Hang}, \citenamefont {Li}, \citenamefont {Shen},\ and\ \citenamefont {Chan}}]{chen2019non}%
  \BibitemOpen
  \bibfield  {author} {\bibinfo {author} {\bibfnamefont {Y.}~\bibnamefont {Chen}}, \bibinfo {author} {\bibfnamefont {R.-Y.}\ \bibnamefont {Zhang}}, \bibinfo {author} {\bibfnamefont {Z.}~\bibnamefont {Xiong}}, \bibinfo {author} {\bibfnamefont {Z.~H.}\ \bibnamefont {Hang}}, \bibinfo {author} {\bibfnamefont {J.}~\bibnamefont {Li}}, \bibinfo {author} {\bibfnamefont {J.~Q.}\ \bibnamefont {Shen}}, \ and\ \bibinfo {author} {\bibfnamefont {C.~T.}\ \bibnamefont {Chan}},\ }\href@noop {} {\bibfield  {journal} {\bibinfo  {journal} {Nature communications}\ }\textbf {\bibinfo {volume} {10}},\ \bibinfo {pages} {3125} (\bibinfo {year} {2019})}\BibitemShut {NoStop}%
\bibitem [{\citenamefont {Yang}\ \emph {et~al.}(2019)\citenamefont {Yang}, \citenamefont {Peng}, \citenamefont {Zhu}, \citenamefont {Buljan}, \citenamefont {Joannopoulos}, \citenamefont {Zhen},\ and\ \citenamefont {Solja{\v{c}}i{\'c}}}]{yang2019synthesis}%
  \BibitemOpen
  \bibfield  {author} {\bibinfo {author} {\bibfnamefont {Y.}~\bibnamefont {Yang}}, \bibinfo {author} {\bibfnamefont {C.}~\bibnamefont {Peng}}, \bibinfo {author} {\bibfnamefont {D.}~\bibnamefont {Zhu}}, \bibinfo {author} {\bibfnamefont {H.}~\bibnamefont {Buljan}}, \bibinfo {author} {\bibfnamefont {J.~D.}\ \bibnamefont {Joannopoulos}}, \bibinfo {author} {\bibfnamefont {B.}~\bibnamefont {Zhen}}, \ and\ \bibinfo {author} {\bibfnamefont {M.}~\bibnamefont {Solja{\v{c}}i{\'c}}},\ }\href@noop {} {\bibfield  {journal} {\bibinfo  {journal} {Science}\ }\textbf {\bibinfo {volume} {365}},\ \bibinfo {pages} {1021} (\bibinfo {year} {2019})}\BibitemShut {NoStop}%
\bibitem [{\citenamefont {Dalibard}\ \emph {et~al.}(2011)\citenamefont {Dalibard}, \citenamefont {Gerbier}, \citenamefont {Juzeli{\=u}nas},\ and\ \citenamefont {{\"O}hberg}}]{dalibard2011colloquium}%
  \BibitemOpen
  \bibfield  {author} {\bibinfo {author} {\bibfnamefont {J.}~\bibnamefont {Dalibard}}, \bibinfo {author} {\bibfnamefont {F.}~\bibnamefont {Gerbier}}, \bibinfo {author} {\bibfnamefont {G.}~\bibnamefont {Juzeli{\=u}nas}}, \ and\ \bibinfo {author} {\bibfnamefont {P.}~\bibnamefont {{\"O}hberg}},\ }\href@noop {} {\bibfield  {journal} {\bibinfo  {journal} {Rev. Mod. Phys.}\ }\textbf {\bibinfo {volume} {83}},\ \bibinfo {pages} {1523} (\bibinfo {year} {2011})}\BibitemShut {NoStop}%
\bibitem [{\citenamefont {Osterloh}\ \emph {et~al.}(2005)\citenamefont {Osterloh}, \citenamefont {Baig}, \citenamefont {Santos}, \citenamefont {Zoller},\ and\ \citenamefont {Lewenstein}}]{osterloh2005cold}%
  \BibitemOpen
  \bibfield  {author} {\bibinfo {author} {\bibfnamefont {K.}~\bibnamefont {Osterloh}}, \bibinfo {author} {\bibfnamefont {M.}~\bibnamefont {Baig}}, \bibinfo {author} {\bibfnamefont {L.}~\bibnamefont {Santos}}, \bibinfo {author} {\bibfnamefont {P.}~\bibnamefont {Zoller}}, \ and\ \bibinfo {author} {\bibfnamefont {M.}~\bibnamefont {Lewenstein}},\ }\href@noop {} {\bibfield  {journal} {\bibinfo  {journal} {Phys. Rev. Lett.}\ }\textbf {\bibinfo {volume} {95}},\ \bibinfo {pages} {010403} (\bibinfo {year} {2005})}\BibitemShut {NoStop}%
\bibitem [{\citenamefont {Dong}\ \emph {et~al.}(2024)\citenamefont {Dong}, \citenamefont {Wu}, \citenamefont {Yang}, \citenamefont {Yu}, \citenamefont {Chen},\ and\ \citenamefont {Yuan}}]{dong2024temporal}%
  \BibitemOpen
  \bibfield  {author} {\bibinfo {author} {\bibfnamefont {Z.}~\bibnamefont {Dong}}, \bibinfo {author} {\bibfnamefont {X.}~\bibnamefont {Wu}}, \bibinfo {author} {\bibfnamefont {Y.}~\bibnamefont {Yang}}, \bibinfo {author} {\bibfnamefont {P.}~\bibnamefont {Yu}}, \bibinfo {author} {\bibfnamefont {X.}~\bibnamefont {Chen}}, \ and\ \bibinfo {author} {\bibfnamefont {L.}~\bibnamefont {Yuan}},\ }\href@noop {} {\bibfield  {journal} {\bibinfo  {journal} {Nature Communications}\ }\textbf {\bibinfo {volume} {15}},\ \bibinfo {pages} {7392} (\bibinfo {year} {2024})}\BibitemShut {NoStop}%
\bibitem [{\citenamefont {Zohar}\ \emph {et~al.}(2015)\citenamefont {Zohar}, \citenamefont {Cirac},\ and\ \citenamefont {Reznik}}]{zohar2015quantum}%
  \BibitemOpen
  \bibfield  {author} {\bibinfo {author} {\bibfnamefont {E.}~\bibnamefont {Zohar}}, \bibinfo {author} {\bibfnamefont {J.~I.}\ \bibnamefont {Cirac}}, \ and\ \bibinfo {author} {\bibfnamefont {B.}~\bibnamefont {Reznik}},\ }\href@noop {} {\bibfield  {journal} {\bibinfo  {journal} {Reports on Progress in Physics}\ }\textbf {\bibinfo {volume} {79}},\ \bibinfo {pages} {014401} (\bibinfo {year} {2015})}\BibitemShut {NoStop}%
\bibitem [{\citenamefont {Fang}\ \emph {et~al.}(2012)\citenamefont {Fang}, \citenamefont {Yu},\ and\ \citenamefont {Fan}}]{fang2012realizing}%
  \BibitemOpen
  \bibfield  {author} {\bibinfo {author} {\bibfnamefont {K.}~\bibnamefont {Fang}}, \bibinfo {author} {\bibfnamefont {Z.}~\bibnamefont {Yu}}, \ and\ \bibinfo {author} {\bibfnamefont {S.}~\bibnamefont {Fan}},\ }\href@noop {} {\bibfield  {journal} {\bibinfo  {journal} {Nat. Photon.}\ }\textbf {\bibinfo {volume} {6}},\ \bibinfo {pages} {782} (\bibinfo {year} {2012})}\BibitemShut {NoStop}%
\bibitem [{\citenamefont {Rechtsman}\ \emph {et~al.}(2013)\citenamefont {Rechtsman}, \citenamefont {Zeuner}, \citenamefont {T{\"u}nnermann}, \citenamefont {Nolte}, \citenamefont {Segev},\ and\ \citenamefont {Szameit}}]{rechtsman2013strain}%
  \BibitemOpen
  \bibfield  {author} {\bibinfo {author} {\bibfnamefont {M.~C.}\ \bibnamefont {Rechtsman}}, \bibinfo {author} {\bibfnamefont {J.~M.}\ \bibnamefont {Zeuner}}, \bibinfo {author} {\bibfnamefont {A.}~\bibnamefont {T{\"u}nnermann}}, \bibinfo {author} {\bibfnamefont {S.}~\bibnamefont {Nolte}}, \bibinfo {author} {\bibfnamefont {M.}~\bibnamefont {Segev}}, \ and\ \bibinfo {author} {\bibfnamefont {A.}~\bibnamefont {Szameit}},\ }\href@noop {} {\bibfield  {journal} {\bibinfo  {journal} {Nat. Photon.}\ }\textbf {\bibinfo {volume} {7}},\ \bibinfo {pages} {153} (\bibinfo {year} {2013})}\BibitemShut {NoStop}%
\bibitem [{\citenamefont {Hafezi}\ \emph {et~al.}(2011)\citenamefont {Hafezi}, \citenamefont {Demler}, \citenamefont {Lukin},\ and\ \citenamefont {Taylor}}]{hafezi2011robust}%
  \BibitemOpen
  \bibfield  {author} {\bibinfo {author} {\bibfnamefont {M.}~\bibnamefont {Hafezi}}, \bibinfo {author} {\bibfnamefont {E.~A.}\ \bibnamefont {Demler}}, \bibinfo {author} {\bibfnamefont {M.~D.}\ \bibnamefont {Lukin}}, \ and\ \bibinfo {author} {\bibfnamefont {J.~M.}\ \bibnamefont {Taylor}},\ }\href@noop {} {\bibfield  {journal} {\bibinfo  {journal} {Nat. Phys.}\ }\textbf {\bibinfo {volume} {7}},\ \bibinfo {pages} {907} (\bibinfo {year} {2011})}\BibitemShut {NoStop}%
\bibitem [{\citenamefont {Yang}\ \emph {et~al.}(2024)\citenamefont {Yang}, \citenamefont {Yang}, \citenamefont {Ma}, \citenamefont {Li}, \citenamefont {Zhang},\ and\ \citenamefont {Chan}}]{yang2024non}%
  \BibitemOpen
  \bibfield  {author} {\bibinfo {author} {\bibfnamefont {Y.}~\bibnamefont {Yang}}, \bibinfo {author} {\bibfnamefont {B.}~\bibnamefont {Yang}}, \bibinfo {author} {\bibfnamefont {G.}~\bibnamefont {Ma}}, \bibinfo {author} {\bibfnamefont {J.}~\bibnamefont {Li}}, \bibinfo {author} {\bibfnamefont {S.}~\bibnamefont {Zhang}}, \ and\ \bibinfo {author} {\bibfnamefont {C.}~\bibnamefont {Chan}},\ }\href@noop {} {\bibfield  {journal} {\bibinfo  {journal} {Science}\ }\textbf {\bibinfo {volume} {383}},\ \bibinfo {pages} {eadf9621} (\bibinfo {year} {2024})}\BibitemShut {NoStop}%
\bibitem [{\citenamefont {Yan}\ \emph {et~al.}(2023)\citenamefont {Yan}, \citenamefont {Wang}, \citenamefont {Wang}, \citenamefont {Ma}, \citenamefont {Lu}, \citenamefont {Ma}, \citenamefont {Hu},\ and\ \citenamefont {Gong}}]{yan2023non}%
  \BibitemOpen
  \bibfield  {author} {\bibinfo {author} {\bibfnamefont {Q.}~\bibnamefont {Yan}}, \bibinfo {author} {\bibfnamefont {Z.}~\bibnamefont {Wang}}, \bibinfo {author} {\bibfnamefont {D.}~\bibnamefont {Wang}}, \bibinfo {author} {\bibfnamefont {R.}~\bibnamefont {Ma}}, \bibinfo {author} {\bibfnamefont {C.}~\bibnamefont {Lu}}, \bibinfo {author} {\bibfnamefont {G.}~\bibnamefont {Ma}}, \bibinfo {author} {\bibfnamefont {X.}~\bibnamefont {Hu}}, \ and\ \bibinfo {author} {\bibfnamefont {Q.}~\bibnamefont {Gong}},\ }\href@noop {} {\bibfield  {journal} {\bibinfo  {journal} {Advances in Optics and Photonics}\ }\textbf {\bibinfo {volume} {15}},\ \bibinfo {pages} {907} (\bibinfo {year} {2023})}\BibitemShut {NoStop}%
\bibitem [{\citenamefont {Celi}\ \emph {et~al.}(2014)\citenamefont {Celi}, \citenamefont {Massignan}, \citenamefont {Ruseckas}, \citenamefont {Goldman}, \citenamefont {Spielman}, \citenamefont {Juzeli{\=u}nas},\ and\ \citenamefont {Lewenstein}}]{celi2014synthetic}%
  \BibitemOpen
  \bibfield  {author} {\bibinfo {author} {\bibfnamefont {A.}~\bibnamefont {Celi}}, \bibinfo {author} {\bibfnamefont {P.}~\bibnamefont {Massignan}}, \bibinfo {author} {\bibfnamefont {J.}~\bibnamefont {Ruseckas}}, \bibinfo {author} {\bibfnamefont {N.}~\bibnamefont {Goldman}}, \bibinfo {author} {\bibfnamefont {I.~B.}\ \bibnamefont {Spielman}}, \bibinfo {author} {\bibfnamefont {G.}~\bibnamefont {Juzeli{\=u}nas}}, \ and\ \bibinfo {author} {\bibfnamefont {M.}~\bibnamefont {Lewenstein}},\ }\href@noop {} {\bibfield  {journal} {\bibinfo  {journal} {Physical review letters}\ }\textbf {\bibinfo {volume} {112}},\ \bibinfo {pages} {043001} (\bibinfo {year} {2014})}\BibitemShut {NoStop}%
\bibitem [{\citenamefont {Ozawa}\ and\ \citenamefont {Price}(2019)}]{ozawa2019topological}%
  \BibitemOpen
  \bibfield  {author} {\bibinfo {author} {\bibfnamefont {T.}~\bibnamefont {Ozawa}}\ and\ \bibinfo {author} {\bibfnamefont {H.~M.}\ \bibnamefont {Price}},\ }\href@noop {} {\bibfield  {journal} {\bibinfo  {journal} {Nature Reviews Physics}\ }\textbf {\bibinfo {volume} {1}},\ \bibinfo {pages} {349} (\bibinfo {year} {2019})}\BibitemShut {NoStop}%
\bibitem [{\citenamefont {Yuan}\ \emph {et~al.}(2018)\citenamefont {Yuan}, \citenamefont {Lin}, \citenamefont {Xiao},\ and\ \citenamefont {Fan}}]{yuan2018synthetic}%
  \BibitemOpen
  \bibfield  {author} {\bibinfo {author} {\bibfnamefont {L.}~\bibnamefont {Yuan}}, \bibinfo {author} {\bibfnamefont {Q.}~\bibnamefont {Lin}}, \bibinfo {author} {\bibfnamefont {M.}~\bibnamefont {Xiao}}, \ and\ \bibinfo {author} {\bibfnamefont {S.}~\bibnamefont {Fan}},\ }\href@noop {} {\bibfield  {journal} {\bibinfo  {journal} {Optica}\ }\textbf {\bibinfo {volume} {5}},\ \bibinfo {pages} {1396} (\bibinfo {year} {2018})}\BibitemShut {NoStop}%
\bibitem [{\citenamefont {Lustig}\ and\ \citenamefont {Segev}(2021)}]{lustig2021topological}%
  \BibitemOpen
  \bibfield  {author} {\bibinfo {author} {\bibfnamefont {E.}~\bibnamefont {Lustig}}\ and\ \bibinfo {author} {\bibfnamefont {M.}~\bibnamefont {Segev}},\ }\href@noop {} {\bibfield  {journal} {\bibinfo  {journal} {Advances in Optics and Photonics}\ }\textbf {\bibinfo {volume} {13}},\ \bibinfo {pages} {426} (\bibinfo {year} {2021})}\BibitemShut {NoStop}%
\bibitem [{\citenamefont {Arg{\"u}ello-Luengo}\ \emph {et~al.}(2024)\citenamefont {Arg{\"u}ello-Luengo}, \citenamefont {Bhattacharya}, \citenamefont {Celi}, \citenamefont {Chhajlany}, \citenamefont {Grass}, \citenamefont {P{\l}odzie{\'n}}, \citenamefont {Rakshit}, \citenamefont {Salamon}, \citenamefont {Stornati}, \citenamefont {Tarruell} \emph {et~al.}}]{arguello2024synthetic}%
  \BibitemOpen
  \bibfield  {author} {\bibinfo {author} {\bibfnamefont {J.}~\bibnamefont {Arg{\"u}ello-Luengo}}, \bibinfo {author} {\bibfnamefont {U.}~\bibnamefont {Bhattacharya}}, \bibinfo {author} {\bibfnamefont {A.}~\bibnamefont {Celi}}, \bibinfo {author} {\bibfnamefont {R.~W.}\ \bibnamefont {Chhajlany}}, \bibinfo {author} {\bibfnamefont {T.}~\bibnamefont {Grass}}, \bibinfo {author} {\bibfnamefont {M.}~\bibnamefont {P{\l}odzie{\'n}}}, \bibinfo {author} {\bibfnamefont {D.}~\bibnamefont {Rakshit}}, \bibinfo {author} {\bibfnamefont {T.}~\bibnamefont {Salamon}}, \bibinfo {author} {\bibfnamefont {P.}~\bibnamefont {Stornati}}, \bibinfo {author} {\bibfnamefont {L.}~\bibnamefont {Tarruell}},  \emph {et~al.},\ }\href@noop {} {\bibfield  {journal} {\bibinfo  {journal} {Communications Physics}\ }\textbf {\bibinfo {volume} {7}},\ \bibinfo {pages} {143} (\bibinfo {year} {2024})}\BibitemShut {NoStop}%
\bibitem [{\citenamefont {Hey}\ and\ \citenamefont {Li}(2018)}]{hey2018advances}%
  \BibitemOpen
  \bibfield  {author} {\bibinfo {author} {\bibfnamefont {D.}~\bibnamefont {Hey}}\ and\ \bibinfo {author} {\bibfnamefont {E.}~\bibnamefont {Li}},\ }\href@noop {} {\bibfield  {journal} {\bibinfo  {journal} {Royal Society Open Science}\ }\textbf {\bibinfo {volume} {5}},\ \bibinfo {pages} {172447} (\bibinfo {year} {2018})}\BibitemShut {NoStop}%
\bibitem [{\citenamefont {Ozawa}\ \emph {et~al.}(2016)\citenamefont {Ozawa}, \citenamefont {Price}, \citenamefont {Goldman}, \citenamefont {Zilberberg},\ and\ \citenamefont {Carusotto}}]{ozawa2016synthetic}%
  \BibitemOpen
  \bibfield  {author} {\bibinfo {author} {\bibfnamefont {T.}~\bibnamefont {Ozawa}}, \bibinfo {author} {\bibfnamefont {H.~M.}\ \bibnamefont {Price}}, \bibinfo {author} {\bibfnamefont {N.}~\bibnamefont {Goldman}}, \bibinfo {author} {\bibfnamefont {O.}~\bibnamefont {Zilberberg}}, \ and\ \bibinfo {author} {\bibfnamefont {I.}~\bibnamefont {Carusotto}},\ }\href@noop {} {\bibfield  {journal} {\bibinfo  {journal} {Physical Review A}\ }\textbf {\bibinfo {volume} {93}},\ \bibinfo {pages} {043827} (\bibinfo {year} {2016})}\BibitemShut {NoStop}%
\bibitem [{\citenamefont {Yuan}\ and\ \citenamefont {Fan}(2016)}]{yuan2016bloch}%
  \BibitemOpen
  \bibfield  {author} {\bibinfo {author} {\bibfnamefont {L.}~\bibnamefont {Yuan}}\ and\ \bibinfo {author} {\bibfnamefont {S.}~\bibnamefont {Fan}},\ }\href@noop {} {\bibfield  {journal} {\bibinfo  {journal} {Optica}\ }\textbf {\bibinfo {volume} {3}},\ \bibinfo {pages} {1014} (\bibinfo {year} {2016})}\BibitemShut {NoStop}%
\bibitem [{\citenamefont {Bell}\ \emph {et~al.}(2017)\citenamefont {Bell}, \citenamefont {Wang}, \citenamefont {Solntsev}, \citenamefont {Neshev}, \citenamefont {Sukhorukov},\ and\ \citenamefont {Eggleton}}]{bell2017spectral}%
  \BibitemOpen
  \bibfield  {author} {\bibinfo {author} {\bibfnamefont {B.~A.}\ \bibnamefont {Bell}}, \bibinfo {author} {\bibfnamefont {K.}~\bibnamefont {Wang}}, \bibinfo {author} {\bibfnamefont {A.~S.}\ \bibnamefont {Solntsev}}, \bibinfo {author} {\bibfnamefont {D.~N.}\ \bibnamefont {Neshev}}, \bibinfo {author} {\bibfnamefont {A.~A.}\ \bibnamefont {Sukhorukov}}, \ and\ \bibinfo {author} {\bibfnamefont {B.~J.}\ \bibnamefont {Eggleton}},\ }\href@noop {} {\bibfield  {journal} {\bibinfo  {journal} {Optica}\ }\textbf {\bibinfo {volume} {4}},\ \bibinfo {pages} {1433} (\bibinfo {year} {2017})}\BibitemShut {NoStop}%
\bibitem [{\citenamefont {Qin}\ \emph {et~al.}(2018)\citenamefont {Qin}, \citenamefont {Zhou}, \citenamefont {Peng}, \citenamefont {Sounas}, \citenamefont {Zhu}, \citenamefont {Wang}, \citenamefont {Dong}, \citenamefont {Zhang}, \citenamefont {Al{\`u}},\ and\ \citenamefont {Lu}}]{qin2018spectrum}%
  \BibitemOpen
  \bibfield  {author} {\bibinfo {author} {\bibfnamefont {C.}~\bibnamefont {Qin}}, \bibinfo {author} {\bibfnamefont {F.}~\bibnamefont {Zhou}}, \bibinfo {author} {\bibfnamefont {Y.}~\bibnamefont {Peng}}, \bibinfo {author} {\bibfnamefont {D.}~\bibnamefont {Sounas}}, \bibinfo {author} {\bibfnamefont {X.}~\bibnamefont {Zhu}}, \bibinfo {author} {\bibfnamefont {B.}~\bibnamefont {Wang}}, \bibinfo {author} {\bibfnamefont {J.}~\bibnamefont {Dong}}, \bibinfo {author} {\bibfnamefont {X.}~\bibnamefont {Zhang}}, \bibinfo {author} {\bibfnamefont {A.}~\bibnamefont {Al{\`u}}}, \ and\ \bibinfo {author} {\bibfnamefont {P.}~\bibnamefont {Lu}},\ }\href@noop {} {\bibfield  {journal} {\bibinfo  {journal} {Physical Review Letters}\ }\textbf {\bibinfo {volume} {120}},\ \bibinfo {pages} {133901} (\bibinfo {year} {2018})}\BibitemShut {NoStop}%
\bibitem [{\citenamefont {Dutt}\ \emph {et~al.}(2020)\citenamefont {Dutt}, \citenamefont {Lin}, \citenamefont {Yuan}, \citenamefont {Minkov}, \citenamefont {Xiao},\ and\ \citenamefont {Fan}}]{dutt2020single}%
  \BibitemOpen
  \bibfield  {author} {\bibinfo {author} {\bibfnamefont {A.}~\bibnamefont {Dutt}}, \bibinfo {author} {\bibfnamefont {Q.}~\bibnamefont {Lin}}, \bibinfo {author} {\bibfnamefont {L.}~\bibnamefont {Yuan}}, \bibinfo {author} {\bibfnamefont {M.}~\bibnamefont {Minkov}}, \bibinfo {author} {\bibfnamefont {M.}~\bibnamefont {Xiao}}, \ and\ \bibinfo {author} {\bibfnamefont {S.}~\bibnamefont {Fan}},\ }\href@noop {} {\bibfield  {journal} {\bibinfo  {journal} {Science}\ }\textbf {\bibinfo {volume} {367}},\ \bibinfo {pages} {59} (\bibinfo {year} {2020})}\BibitemShut {NoStop}%
\bibitem [{\citenamefont {Bal{\v{c}}ytis}\ \emph {et~al.}(2022)\citenamefont {Bal{\v{c}}ytis}, \citenamefont {Ozawa}, \citenamefont {Ota}, \citenamefont {Iwamoto}, \citenamefont {Maeda},\ and\ \citenamefont {Baba}}]{balvcytis2022synthetic}%
  \BibitemOpen
  \bibfield  {author} {\bibinfo {author} {\bibfnamefont {A.}~\bibnamefont {Bal{\v{c}}ytis}}, \bibinfo {author} {\bibfnamefont {T.}~\bibnamefont {Ozawa}}, \bibinfo {author} {\bibfnamefont {Y.}~\bibnamefont {Ota}}, \bibinfo {author} {\bibfnamefont {S.}~\bibnamefont {Iwamoto}}, \bibinfo {author} {\bibfnamefont {J.}~\bibnamefont {Maeda}}, \ and\ \bibinfo {author} {\bibfnamefont {T.}~\bibnamefont {Baba}},\ }\href@noop {} {\bibfield  {journal} {\bibinfo  {journal} {Science advances}\ }\textbf {\bibinfo {volume} {8}},\ \bibinfo {pages} {eabk0468} (\bibinfo {year} {2022})}\BibitemShut {NoStop}%
\bibitem [{\citenamefont {Tusnin}\ \emph {et~al.}(2020)\citenamefont {Tusnin}, \citenamefont {Tikan},\ and\ \citenamefont {Kippenberg}}]{tusnin2020nonlinear}%
  \BibitemOpen
  \bibfield  {author} {\bibinfo {author} {\bibfnamefont {A.~K.}\ \bibnamefont {Tusnin}}, \bibinfo {author} {\bibfnamefont {A.~M.}\ \bibnamefont {Tikan}}, \ and\ \bibinfo {author} {\bibfnamefont {T.~J.}\ \bibnamefont {Kippenberg}},\ }\href@noop {} {\bibfield  {journal} {\bibinfo  {journal} {Physical Review A}\ }\textbf {\bibinfo {volume} {102}},\ \bibinfo {pages} {023518} (\bibinfo {year} {2020})}\BibitemShut {NoStop}%
\bibitem [{\citenamefont {Suh}\ \emph {et~al.}(2024)\citenamefont {Suh}, \citenamefont {Kim}, \citenamefont {Park}, \citenamefont {Fan}, \citenamefont {Park},\ and\ \citenamefont {Yu}}]{suh2024photonic}%
  \BibitemOpen
  \bibfield  {author} {\bibinfo {author} {\bibfnamefont {J.}~\bibnamefont {Suh}}, \bibinfo {author} {\bibfnamefont {G.}~\bibnamefont {Kim}}, \bibinfo {author} {\bibfnamefont {H.}~\bibnamefont {Park}}, \bibinfo {author} {\bibfnamefont {S.}~\bibnamefont {Fan}}, \bibinfo {author} {\bibfnamefont {N.}~\bibnamefont {Park}}, \ and\ \bibinfo {author} {\bibfnamefont {S.}~\bibnamefont {Yu}},\ }\href@noop {} {\bibfield  {journal} {\bibinfo  {journal} {Physical Review Letters}\ }\textbf {\bibinfo {volume} {132}},\ \bibinfo {pages} {033803} (\bibinfo {year} {2024})}\BibitemShut {NoStop}%
\bibitem [{\citenamefont {Du}\ \emph {et~al.}(2022)\citenamefont {Du}, \citenamefont {Zhang}, \citenamefont {Wu}, \citenamefont {Kockum},\ and\ \citenamefont {Li}}]{du2022giant}%
  \BibitemOpen
  \bibfield  {author} {\bibinfo {author} {\bibfnamefont {L.}~\bibnamefont {Du}}, \bibinfo {author} {\bibfnamefont {Y.}~\bibnamefont {Zhang}}, \bibinfo {author} {\bibfnamefont {J.-H.}\ \bibnamefont {Wu}}, \bibinfo {author} {\bibfnamefont {A.~F.}\ \bibnamefont {Kockum}}, \ and\ \bibinfo {author} {\bibfnamefont {Y.}~\bibnamefont {Li}},\ }\href@noop {} {\bibfield  {journal} {\bibinfo  {journal} {Physical Review Letters}\ }\textbf {\bibinfo {volume} {128}},\ \bibinfo {pages} {223602} (\bibinfo {year} {2022})}\BibitemShut {NoStop}%
\bibitem [{\citenamefont {Wang}\ \emph {et~al.}(2021)\citenamefont {Wang}, \citenamefont {Dutt}, \citenamefont {Yang}, \citenamefont {Wojcik}, \citenamefont {Vu{\v{c}}kovi{\'c}},\ and\ \citenamefont {Fan}}]{wang2021generating}%
  \BibitemOpen
  \bibfield  {author} {\bibinfo {author} {\bibfnamefont {K.}~\bibnamefont {Wang}}, \bibinfo {author} {\bibfnamefont {A.}~\bibnamefont {Dutt}}, \bibinfo {author} {\bibfnamefont {K.~Y.}\ \bibnamefont {Yang}}, \bibinfo {author} {\bibfnamefont {C.~C.}\ \bibnamefont {Wojcik}}, \bibinfo {author} {\bibfnamefont {J.}~\bibnamefont {Vu{\v{c}}kovi{\'c}}}, \ and\ \bibinfo {author} {\bibfnamefont {S.}~\bibnamefont {Fan}},\ }\href@noop {} {\bibfield  {journal} {\bibinfo  {journal} {Science}\ }\textbf {\bibinfo {volume} {371}},\ \bibinfo {pages} {1240} (\bibinfo {year} {2021})}\BibitemShut {NoStop}%
\bibitem [{\citenamefont {Cheng}\ \emph {et~al.}(2021)\citenamefont {Cheng}, \citenamefont {Peng}, \citenamefont {Wang}, \citenamefont {Chen}, \citenamefont {Yuan},\ and\ \citenamefont {Fan}}]{cheng2021arbitrary}%
  \BibitemOpen
  \bibfield  {author} {\bibinfo {author} {\bibfnamefont {D.}~\bibnamefont {Cheng}}, \bibinfo {author} {\bibfnamefont {B.}~\bibnamefont {Peng}}, \bibinfo {author} {\bibfnamefont {D.-W.}\ \bibnamefont {Wang}}, \bibinfo {author} {\bibfnamefont {X.}~\bibnamefont {Chen}}, \bibinfo {author} {\bibfnamefont {L.}~\bibnamefont {Yuan}}, \ and\ \bibinfo {author} {\bibfnamefont {S.}~\bibnamefont {Fan}},\ }\href@noop {} {\bibfield  {journal} {\bibinfo  {journal} {Physical Review Research}\ }\textbf {\bibinfo {volume} {3}},\ \bibinfo {pages} {033069} (\bibinfo {year} {2021})}\BibitemShut {NoStop}%
\bibitem [{\citenamefont {Hu}\ \emph {et~al.}(2020)\citenamefont {Hu}, \citenamefont {Reimer}, \citenamefont {Shams-Ansari}, \citenamefont {Zhang},\ and\ \citenamefont {Loncar}}]{hu2020realization}%
  \BibitemOpen
  \bibfield  {author} {\bibinfo {author} {\bibfnamefont {Y.}~\bibnamefont {Hu}}, \bibinfo {author} {\bibfnamefont {C.}~\bibnamefont {Reimer}}, \bibinfo {author} {\bibfnamefont {A.}~\bibnamefont {Shams-Ansari}}, \bibinfo {author} {\bibfnamefont {M.}~\bibnamefont {Zhang}}, \ and\ \bibinfo {author} {\bibfnamefont {M.}~\bibnamefont {Loncar}},\ }\href@noop {} {\bibfield  {journal} {\bibinfo  {journal} {Optica}\ }\textbf {\bibinfo {volume} {7}},\ \bibinfo {pages} {1189} (\bibinfo {year} {2020})}\BibitemShut {NoStop}%
\bibitem [{\citenamefont {Senanian}\ \emph {et~al.}(2023)\citenamefont {Senanian}, \citenamefont {Wright}, \citenamefont {Wade}, \citenamefont {Doyle},\ and\ \citenamefont {McMahon}}]{senanian2023programmable}%
  \BibitemOpen
  \bibfield  {author} {\bibinfo {author} {\bibfnamefont {A.}~\bibnamefont {Senanian}}, \bibinfo {author} {\bibfnamefont {L.~G.}\ \bibnamefont {Wright}}, \bibinfo {author} {\bibfnamefont {P.~F.}\ \bibnamefont {Wade}}, \bibinfo {author} {\bibfnamefont {H.~K.}\ \bibnamefont {Doyle}}, \ and\ \bibinfo {author} {\bibfnamefont {P.~L.}\ \bibnamefont {McMahon}},\ }\href@noop {} {\bibfield  {journal} {\bibinfo  {journal} {Nature Physics}\ }\textbf {\bibinfo {volume} {19}},\ \bibinfo {pages} {1333} (\bibinfo {year} {2023})}\BibitemShut {NoStop}%
\bibitem [{\citenamefont {Cheng}\ \emph {et~al.}(2023)\citenamefont {Cheng}, \citenamefont {Wang},\ and\ \citenamefont {Fan}}]{cheng2023artificial}%
  \BibitemOpen
  \bibfield  {author} {\bibinfo {author} {\bibfnamefont {D.}~\bibnamefont {Cheng}}, \bibinfo {author} {\bibfnamefont {K.}~\bibnamefont {Wang}}, \ and\ \bibinfo {author} {\bibfnamefont {S.}~\bibnamefont {Fan}},\ }\href@noop {} {\bibfield  {journal} {\bibinfo  {journal} {Physical Review Letters}\ }\textbf {\bibinfo {volume} {130}},\ \bibinfo {pages} {083601} (\bibinfo {year} {2023})}\BibitemShut {NoStop}%
\bibitem [{\citenamefont {Pang}\ \emph {et~al.}(2024)\citenamefont {Pang}, \citenamefont {Wong}, \citenamefont {Hu},\ and\ \citenamefont {Yang}}]{pang2024synthetic}%
  \BibitemOpen
  \bibfield  {author} {\bibinfo {author} {\bibfnamefont {Z.}~\bibnamefont {Pang}}, \bibinfo {author} {\bibfnamefont {B.~T.~T.}\ \bibnamefont {Wong}}, \bibinfo {author} {\bibfnamefont {J.}~\bibnamefont {Hu}}, \ and\ \bibinfo {author} {\bibfnamefont {Y.}~\bibnamefont {Yang}},\ }\href@noop {} {\bibfield  {journal} {\bibinfo  {journal} {Physical Review Letters}\ }\textbf {\bibinfo {volume} {132}},\ \bibinfo {pages} {043804} (\bibinfo {year} {2024})}\BibitemShut {NoStop}%
\bibitem [{\citenamefont {Cheng}\ \emph {et~al.}(2024)\citenamefont {Cheng}, \citenamefont {Wang}, \citenamefont {Roques-Carmes}, \citenamefont {Lustig}, \citenamefont {Long}, \citenamefont {Wang},\ and\ \citenamefont {Fan}}]{cheng2024non}%
  \BibitemOpen
  \bibfield  {author} {\bibinfo {author} {\bibfnamefont {D.}~\bibnamefont {Cheng}}, \bibinfo {author} {\bibfnamefont {K.}~\bibnamefont {Wang}}, \bibinfo {author} {\bibfnamefont {C.}~\bibnamefont {Roques-Carmes}}, \bibinfo {author} {\bibfnamefont {E.}~\bibnamefont {Lustig}}, \bibinfo {author} {\bibfnamefont {O.~Y.}\ \bibnamefont {Long}}, \bibinfo {author} {\bibfnamefont {H.}~\bibnamefont {Wang}}, \ and\ \bibinfo {author} {\bibfnamefont {S.}~\bibnamefont {Fan}},\ }\href@noop {} {\bibfield  {journal} {\bibinfo  {journal} {arXiv preprint arXiv:2406.00321}\ } (\bibinfo {year} {2024})}\BibitemShut {NoStop}%
\bibitem [{\citenamefont {Yang}\ and\ \citenamefont {Mills}(1954)}]{yang1954conservation}%
  \BibitemOpen
  \bibfield  {author} {\bibinfo {author} {\bibfnamefont {C.-N.}\ \bibnamefont {Yang}}\ and\ \bibinfo {author} {\bibfnamefont {R.~L.}\ \bibnamefont {Mills}},\ }\href@noop {} {\bibfield  {journal} {\bibinfo  {journal} {Physical Review}\ }\textbf {\bibinfo {volume} {96}},\ \bibinfo {pages} {191} (\bibinfo {year} {1954})}\BibitemShut {NoStop}%
\bibitem [{\citenamefont {Yuan}\ \emph {et~al.}(2016)\citenamefont {Yuan}, \citenamefont {Shi},\ and\ \citenamefont {Fan}}]{yuan2016photonic}%
  \BibitemOpen
  \bibfield  {author} {\bibinfo {author} {\bibfnamefont {L.}~\bibnamefont {Yuan}}, \bibinfo {author} {\bibfnamefont {Y.}~\bibnamefont {Shi}}, \ and\ \bibinfo {author} {\bibfnamefont {S.}~\bibnamefont {Fan}},\ }\href@noop {} {\bibfield  {journal} {\bibinfo  {journal} {Optics letters}\ }\textbf {\bibinfo {volume} {41}},\ \bibinfo {pages} {741} (\bibinfo {year} {2016})}\BibitemShut {NoStop}%
\bibitem [{\citenamefont {Yuan}\ \emph {et~al.}(2021)\citenamefont {Yuan}, \citenamefont {Dutt},\ and\ \citenamefont {Fan}}]{yuan2021synthetic}%
  \BibitemOpen
  \bibfield  {author} {\bibinfo {author} {\bibfnamefont {L.}~\bibnamefont {Yuan}}, \bibinfo {author} {\bibfnamefont {A.}~\bibnamefont {Dutt}}, \ and\ \bibinfo {author} {\bibfnamefont {S.}~\bibnamefont {Fan}},\ }\href@noop {} {\bibfield  {journal} {\bibinfo  {journal} {APL Photonics}\ }\textbf {\bibinfo {volume} {6}} (\bibinfo {year} {2021})}\BibitemShut {NoStop}%
\bibitem [{\citenamefont {Haus}(1984)}]{haus1984waves}%
  \BibitemOpen
  \bibfield  {author} {\bibinfo {author} {\bibfnamefont {H.~A.}\ \bibnamefont {Haus}},\ }\href@noop {} {\bibfield  {journal} {\bibinfo  {journal} {Waves and fields in optoelectronics}\ } (\bibinfo {year} {1984})}\BibitemShut {NoStop}%
\bibitem [{\citenamefont {Dutt}\ \emph {et~al.}(2019)\citenamefont {Dutt}, \citenamefont {Minkov}, \citenamefont {Lin}, \citenamefont {Yuan}, \citenamefont {Miller},\ and\ \citenamefont {Fan}}]{dutt2019experimental}%
  \BibitemOpen
  \bibfield  {author} {\bibinfo {author} {\bibfnamefont {A.}~\bibnamefont {Dutt}}, \bibinfo {author} {\bibfnamefont {M.}~\bibnamefont {Minkov}}, \bibinfo {author} {\bibfnamefont {Q.}~\bibnamefont {Lin}}, \bibinfo {author} {\bibfnamefont {L.}~\bibnamefont {Yuan}}, \bibinfo {author} {\bibfnamefont {D.~A.}\ \bibnamefont {Miller}}, \ and\ \bibinfo {author} {\bibfnamefont {S.}~\bibnamefont {Fan}},\ }\href@noop {} {\bibfield  {journal} {\bibinfo  {journal} {Nature Communications}\ }\textbf {\bibinfo {volume} {10}},\ \bibinfo {pages} {3122} (\bibinfo {year} {2019})}\BibitemShut {NoStop}%
\end{thebibliography}

\providecommand{\noopsort}[1]{}\providecommand{\singleletter}[1]{#1}

\end{document}


\title{
Supplementary Materials\\
Synthetic non-Abelian Electric Fields and Spin-Orbit Coupling in Photonic Synthetic Dimensions
    }

\author{Bengy Tsz Tsun Wong}
\affiliation{\hkuaffil}
\author{Zehai Pang}
\affiliation{\hkuaffil}
\author{Yi~Yang}
\email{yiyg@hku.hk}
\affiliation{\hkuaffil}

\maketitle

\setlength{\parindent}{0em}
\setlength{\parskip}{.5em}

\noindent{\textbf{\textsf{CONTENTS}}}\\ 
\twocolumngrid
\begingroup 
    \let\bfseries\relax 
    \deactivateaddvspace 
    \deactivatetocsubsections 
    \tableofcontents
\endgroup
\onecolumngrid

\hyphenation{eigen-index}

\section{Hamiltonian derivation by explicit wave propagation} \label{FullEM}

Throughout the derivation in the supplementary, we adopt the notation of $\left\{\alpha,\beta\right\} = \left\{\mathrm{h},\mathrm{v}\right\}$ denoting the horizontal and vertical polarizations and Einstein summation on repeated indices are inferred throughout the supplementary. Conversely, we adopt the matrix notation in the main text.

The following derivation is inspired by and generalizes previous work on synthetic Abelian gauge fields in synthetic frequency dimension~\cite{yuan2016bloch,yuan2016photonic,yuan2021synthetic}. 
Inside the ring resonator, we consider the following transformation between horizontal and vertical polarization at position $(\mathbf{r}_\perp,z_0)$ and time $t$
\begin{align}
\begin{pmatrix}
E_{\mathrm{h}} (t^+, \mathbf{r}_\perp,z_0) \\
E_{\mathrm{v}} (t^+, \mathbf{r}_\perp,z_0)
\end{pmatrix} =\mathrm{e}^{\mathrm{i}g\cos(\Omega t + \theta \sigma_z) }\mathrm{e}^{\mathrm{i}\phi_y\sigma_y}\mathrm{e}^{\mathrm{i}\phi_z\sigma_z}
\begin{pmatrix}
E_{\mathrm{h}} (t^-, \mathbf{r}_\perp,z_0) \\
E_{\mathrm{v}} (t^-, \mathbf{r}_\perp,z_0)
\end{pmatrix} \,.
\end{align}
where the composite transfer matrix 
\begin{align}
\begin{aligned}
\pmb{\mathrm{T}} &= \mathrm{e}^{\mathrm{i}g\cos(\Omega t + \theta \sigma_z) }\mathrm{e}^{\mathrm{i}\phi_y\sigma_y} \mathrm{e}^{\mathrm{i}\phi_z\sigma_z}=\begin{pmatrix}
\mathrm{e}^{\mathrm{i}g\cos(\Omega t + \theta)} & 0 \\
0 & \mathrm{e}^{\mathrm{i}g \cos(\Omega t - \theta)}
\end{pmatrix}
\begin{pmatrix}
\cos \phi_y & \sin\phi_y \\
-\sin\phi_y & \cos\phi_y
\end{pmatrix}
\begin{pmatrix}
\mathrm{e}^{\mathrm{i}\phi_z} & 0 \\
0 & \mathrm{e}^{-\mathrm{i}\phi_z}
\end{pmatrix}\\
&= \begin{pmatrix}
\mathrm{e}^{\mathrm{i}g\cos(\Omega t+\theta )}\mathrm{e}^{\mathrm{i}\phi_z}\cos\phi_y & \mathrm{e}^{\mathrm{i}g\cos(\Omega t + \theta)}\mathrm{e}^{-\mathrm{i}\phi_z}\sin\phi_y \\
-\mathrm{e}^{\mathrm{i}g \cos(\Omega t-\theta )}\mathrm{e}^{\mathrm{i}\phi_z}\sin\phi_y & \mathrm{e}^{\mathrm{i}g \cos(\Omega t-\theta )}\mathrm{e}^{-\mathrm{i}\phi_z}\cos\phi_y
\end{pmatrix}\,.
\end{aligned}
\end{align}
Then it follows that we have
\begin{align} \label{eq:twocases}
\begin{cases}
E_{\mathrm{h}}(t^+ ,\mathbf{r}_\perp,z_0) = \mathrm{e}^{\mathrm{i}g\cos(\Omega t + \theta)} \Big( \mathrm{e}^{\mathrm{i}\phi_z}   \cos\phi_y E_{\mathrm{h}} (t^-, \mathbf{r}_\perp, z_0) + \mathrm{e}^{-\mathrm{i}\phi_z}\sin\phi_y  E_{\mathrm{v}} (t^-, \mathbf{r}_\perp, z_0)    \Big) \\
E_{\mathrm{v}}(t^+ ,\mathbf{r}_\perp,z_0) = \mathrm{e}^{\mathrm{i}g\cos(\Omega t - \theta)} \Big(    -\mathrm{e}^{\mathrm{i}\phi_z}\sin\phi_y E_{\mathrm{h}} (t^-, \mathbf{r}_\perp, z_0) + \mathrm{e}^{-\mathrm{i}\phi_z}\cos\phi_y  E_{\mathrm{v}} (t^-, \mathbf{r}_\perp, z_0)    \Big) \,.
\end{cases}
\end{align}
Note that here we idealize the different optical components into a single component and neglect the propagation time and phase delay among them, whose consequences and associated constraints are analyzed in Section~\ref{dynamics}.

We can expand the electric field into different eigenmodes 
\begin{align} \label{eq:Eexpansion}
E_\alpha (t^\pm , \mathbf{r}_\perp, z_0)= \sum_m E_{\alpha m} (t^\pm , z_0) u_{\alpha m}(\mathbf{r}_\perp) \mathrm{e}^{\mathrm{i}\omega_m t}\,,
\end{align}
where $\alpha = \{\mathrm{h},\mathrm{v}\}$. Furthermore, using the Jacobi–Anger expansion, we have
\begin{equation} \label{eq:Anger}
\mathrm{e}^{\mathrm{i}g \cos\theta} =\sum_{n=-\infty}^{\infty} \mathrm{i}^n \mathcal{J}_n(g) \mathrm{e}^{\mathrm{i}n\theta} =\sum_{n=-\infty}^{\infty} \iu ^n \mathcal{J}_n(g) \mathrm{e}^{-\mathrm{i}n\theta} \,.
\end{equation}

Substituting Eqs~\ref{eq:Anger} and \ref{eq:Eexpansion} into \ref{eq:twocases}, we have
\begin{align}
\begin{aligned}
E_{\mathrm{h}}(t^+ , \mathbf{r}_\perp,z_0) =\sum_{q=-\infty}^\infty \mathrm{i}^q \mathcal{J}_q(g) \mathrm{e}^{-\mathrm{i}q(\Omega t + \theta)}\bigg[ \mathrm{e}^{\mathrm{i}\phi_z}\cos\phi_y\sum_m E_{{\mathrm{h}}\,m}(t^-,z_0)u_{\mathrm{h}\,m}(\mathbf{r}_\perp) \mathrm{e}^{\mathrm{i}\omega_m t} + \mathrm{e}^{-\mathrm{i}\phi_z}\sin\phi_y\sum_m E_{{\mathrm{v}}\,m}(t^-,z_0)u_{\mathrm{v}\,m}(\mathbf{r}_\perp) \mathrm{e}^{\mathrm{i}\omega_m t}  \bigg] \,.
\end{aligned}
\end{align}
By definiting $\Delta_{\Omega} \equiv \Omega - \Omega_{\mathrm{R}}$ the detunning frequency and $\omega_m = \omega_0 + m\Omega_{\mathrm{R}}$ the resonant frequencies, we have
\begin{align}
\begin{aligned}
&\quad E_{\mathrm{h}}(t^+ , \mathbf{r}_\perp,z_0)\\
& = \sum_{q=-\infty}^\infty \mathrm{i}^q \mathcal{J}_q(g) \mathrm{e}^{-\mathrm{i}q\theta} \bigg[\mathrm{e}^{\mathrm{i}\phi_z} \cos\phi_y\sum_m E_{{\mathrm{h}}\,m}(t^-,z_0)u_{{\mathrm{h}}\,m}(\mathbf{r}_\perp) + \mathrm{e}^{-\mathrm{i}\phi_z} \sin\phi_y\sum_m E_{{\mathrm{v}}\,m}(t^-,z_0)u_{{\mathrm{v}}\,m}(\mathbf{r}_\perp)  \bigg]\mathrm{e}^{\mathrm{i}(\omega_0 + (m-q)\Omega_{\mathrm{R}})t}\mathrm{e}^{-\mathrm{i}q\Delta_{\Omega} t}\,.
\end{aligned}
\end{align}
By taking $m^\prime = m -q$ and further assuming the transverse amplitude profiles identical for both horizontal polarization and vertical polarization $u_{{\mathrm{h}}\,m}(\mathbf{r}_\perp)$ = $u_{{\mathrm{v}}\,m}(\mathbf{r}_\perp)= u(\mathbf{r}_\perp)$, we get
\begin{align}
\begin{aligned}
&\quad E_{\mathrm{h}} (t^+ , \mathbf{r}_\perp,z_0)\\
& =  \sum_{m^\prime=-\infty}^\infty  u(\mathbf{r}_\perp) \mathrm{e}^{\mathrm{i}\omega_{m^\prime}t}  \sum_{m}   i^{m-m^\prime} \mathcal{J}_{m-m^\prime}(g)   \big(\mathrm{e}^{\mathrm{i}\phi_z}\cos\phi_y E_{{\mathrm{h}}\,m}(t^-,z_0)  + \mathrm{e}^{-\mathrm{i}\phi_z}\sin\phi_y E_{{\mathrm{v}}\,m}(t^-,z_0) \big)\mathrm{e}^{-\mathrm{i}(m-m^\prime)\theta} \mathrm{e}^{-\mathrm{i}(m-m^\prime) \Delta_{\Omega} t} \,.
\end{aligned}
\end{align}
By swapping the indices $m$ and $m^\prime$ and matching coefficients, we obtain for the horizontal polarization
\begin{align}
E_{{\mathrm{h}}\,m}(t^+ ,z_0)
= \sum_{m^\prime}   \mathrm{i}^{m^\prime-m} \mathcal{J}_{m^\prime-m}(g)   \big(\mathrm{e}^{\mathrm{i}\phi_z}\cos\phi_y E_{{\mathrm{h}}\,m^\prime}(t^-,z_0) + \mathrm{e}^{-\mathrm{i}\phi_z}\sin\phi_y E_{{\mathrm{v}}\,m^\prime}(t^-,z_0) \big)\mathrm{e}^{-\mathrm{i}(m^\prime-m)\theta} \mathrm{e}^{-\mathrm{i}(m^\prime-m) \Delta_{\Omega} t} \,.
\end{align}
Now redefine $p \equiv m^\prime -m$, we obtain
\begin{align} \label{eq:EhPlus}
E_{{\mathrm{h}}\,m}(t^+ ,z_0) = \sum_{p=-\infty-m}^{\infty-m} \mathrm{i}^p \mathcal{J}_{p}(g)\mathrm{e}^{-\mathrm{i}p\theta}\mathrm{e}^{-\mathrm{i}p\Delta_{\Omega} t} \big(\mathrm{e}^{\mathrm{i}\phi_z}\cos\phi_y E_{{\mathrm{h}}\,m+p}(t^-,z_0)  + \mathrm{e}^{-\mathrm{i}\phi_z}\sin\phi_y E_{{\mathrm{v}}\,m+p}(t^-,z_0) \big).
\end{align}
Similarly for the vertical polarization, we have
\begin{align}
 E_{{\mathrm{v}}\,m}(t^+ ,z_0)
= \sum_{m^\prime}  \mathrm{i}^{m^\prime-m} \mathcal{J}_{m^\prime-m}(g)   \big(-\mathrm{e}^{\mathrm{i}\phi_z}\sin\phi_y E_{{\mathrm{h}}\,m^\prime}(t^-,z_0) + \mathrm{e}^{-\mathrm{i}\phi_z}\cos\phi_y E_{{\mathrm{v}}\,m^\prime}(t^-,z_0) \big)\mathrm{e}^{\mathrm{i}(m^\prime-m)\theta} \mathrm{e}^{\mathrm{i}(m^\prime-m) \Delta_{\Omega} t}
\end{align}
and
\begin{align} \label{eq:EvPlus}
E_{{\mathrm{v}}\,m}(t^+ ,z_0) = \sum_{p=-\infty-m}^{\infty-m} \mathrm{i}^p \mathcal{J}_{p}(g)\mathrm{e}^{\mathrm{i}p\theta}\mathrm{e}^{-\mathrm{i}p\Delta_{\Omega} t} \big(-\mathrm{e}^{\mathrm{i}\phi_z}\sin\phi_y E_{{\mathrm{h}}\,m+p}(t^-,z_0)  +\mathrm{e}^{-\mathrm{i}\phi_z} \cos\phi_y E_{{\mathrm{v}}\,m+p}(t^-,z_0) \big).
\end{align} 
We can compactly combine Eqs.~\ref{eq:EhPlus} and \ref{eq:EvPlus} into a matrix form and carry out the summation expansion as follows,
\begin{align}
\begin{aligned}
\begin{pmatrix}
E_{{\mathrm{h}}\,m}(t^+,z_0)\\
E_{{\mathrm{v}}\,m}(t^+,z_0)
\end{pmatrix} &= \mathcal{J}_0(g)
\begin{pmatrix}
\mathrm{e}^{\mathrm{i}\phi_z}\cos\phi_y & \mathrm{e}^{-\mathrm{i}\phi_z} \sin\phi_y \\
-\mathrm{e}^{\mathrm{i}\phi_z}\sin\phi_y & \mathrm{e}^{-\mathrm{i}\phi_z}\cos\phi_y
\end{pmatrix}
\begin{pmatrix}
E_{{\mathrm{h}}\,m}(t^-,z_0)\\
E_{{\mathrm{v}}\,m}(t^-,z_0)
\end{pmatrix} \\
&+\sum_p \mathrm{i}^p \mathcal{J}_p(g) \bigg[ 
\begin{pmatrix}
\mathrm{e}^{-\mathrm{i}p\theta}\mathrm{e}^{\mathrm{i}\phi_z}\cos\phi_y & \mathrm{e}^{-\mathrm{i}p\theta}\mathrm{e}^{-\mathrm{i}\phi_z}\sin\phi_y \\
-\mathrm{e}^{\mathrm{i}p\theta}\mathrm{e}^{\mathrm{i}\phi_z}\sin\phi_y & \mathrm{e}^{-\mathrm{i}\phi_z}\mathrm{e}^{\mathrm{i}p\theta}\sin\cos_y \\
\end{pmatrix}
\begin{pmatrix}
E_{{\mathrm{h}}\,m+p}(t^-,z_0)\\
E_{{\mathrm{v}}\,m+p}(t^-,z_0)
\end{pmatrix} \mathrm{e}^{-\mathrm{i}p\Delta_{\Omega} t}\\
&\quad\quad\quad\quad\quad\quad +
\begin{pmatrix}
\mathrm{e}^{\mathrm{i}p\theta}\mathrm{e}^{\mathrm{i}\phi_z}\cos\phi_y & \mathrm{e}^{\mathrm{i}p\theta}\mathrm{e}^{-\mathrm{i}\phi_z}\sin\phi_y \\
-\mathrm{e}^{-\mathrm{i}p\theta}\mathrm{e}^{\mathrm{i}\phi_z}\sin\phi_y & \mathrm{e}^{-\mathrm{i}p\theta}\mathrm{e}^{-\mathrm{i}\phi_z}\sin\phi_y \\
\end{pmatrix}
\begin{pmatrix}
E_{{\mathrm{h}}\,m-p}(t^-,z_0)\\
E_{{\mathrm{v}}\,m-p}(t^-,z_0)
\end{pmatrix} \mathrm{e}^{\mathrm{i}p\Delta_{\Omega} t}
\bigg] \\
&= \mathcal{J}_0(g) \begin{pmatrix}
\cos\phi_y & \sin\phi_y \\
-\sin\phi_y & \cos\phi_y
\end{pmatrix}
\begin{pmatrix}
\mathrm{e}^{\mathrm{i}\phi_z} & 0 \\
0 & \mathrm{e}^{-\mathrm{i}\phi_z}
\end{pmatrix}
\begin{pmatrix}
E_{{\mathrm{h}}\,m}(t^-,z_0)\\
E_{{\mathrm{v}}\,m}(t^-,z_0)
\end{pmatrix}\\
&+\sum_p \mathrm{i}^p \mathcal{J}_p(g) \bigg[ 
\begin{pmatrix}
\mathrm{e}^{-\mathrm{i}p\theta} & 0\\
0 & \mathrm{e}^{\mathrm{i}p\theta}
\end{pmatrix}
\begin{pmatrix}
\cos\phi_y &\sin\phi_y \\
-\sin\phi_y & \sin\cos_y \\
\end{pmatrix}
\begin{pmatrix}
\mathrm{e}^{\mathrm{i}\phi_z} & 0 \\
0 & \mathrm{e}^{-\mathrm{i}\phi_z}
\end{pmatrix}
\begin{pmatrix}
E_{{\mathrm{h}}\,m+p}(t^-,z_0)\\
E_{{\mathrm{v}}\,m+p}(t^-,z_0)
\end{pmatrix} \mathrm{e}^{-\mathrm{i}p\Delta_{\Omega} t}\\
&\quad\quad\quad\quad\quad\quad +
\begin{pmatrix}
\mathrm{e}^{\mathrm{i}p\theta} & 0\\
0 & \mathrm{e}^{-\mathrm{i}p\theta}
\end{pmatrix}
\begin{pmatrix}
\cos\phi_y & \sin\phi_y \\
-\sin\phi_y & \cos_y \\
\end{pmatrix}
\begin{pmatrix}
\mathrm{e}^{\mathrm{i}\phi_z} & 0 \\
0 & \mathrm{e}^{-\mathrm{i}\phi_z}
\end{pmatrix}
\begin{pmatrix}
E_{{\mathrm{h}}\,m-p}(t^-,z_0)\\
E_{{\mathrm{v}}\,m-p}(t^-,z_0)
\end{pmatrix} \mathrm{e}^{\mathrm{i}p\Delta_{\Omega} t}
\bigg] \,.
\end{aligned}
\end{align}
Therefore, compactly we have
\begin{align}
\begin{aligned}
\begin{pmatrix}
E_{{\mathrm{h}}\,m}(t^+,z_0)\\
E_{{\mathrm{v}}\,m}(t^+,z_0)
\end{pmatrix}  &= \mathcal{J}_0(g)\mathrm{e}^{\mathrm{i}\phi_y\sigma_y}
\begin{pmatrix}
E_{{\mathrm{h}}\,m}(t^-,z_0)\\
E_{{\mathrm{v}}\,m}(t^-,z_0)
\end{pmatrix}\\
&+\sum_p \mathrm{i}^p \mathcal{J}_p(g)\bigg[\mathrm{e}^{-\mathrm{i}p\theta\sigma_z}\mathrm{e}^{\mathrm{i}\phi_y\sigma_y} \mathrm{e}^{\mathrm{i}\phi_z\sigma_z} \begin{pmatrix}
E_{{\mathrm{h}}\,m+p}(t^-,z_0)\\
E_{{\mathrm{v}}\,m+p}(t^-,z_0)
\end{pmatrix}\mathrm{e}^{-\mathrm{i}p\Delta_{\Omega} t^-} + \mathrm{e}^{\mathrm{i}p\theta\sigma_z}\mathrm{e}^{\mathrm{i}\phi_y\sigma_y} \mathrm{e}^{\mathrm{i}\phi_z\sigma_z}  
\begin{pmatrix}
E_{{\mathrm{h}}\,m-p}(t^-,z_0)\\
E_{{\mathrm{v}}\,m-p}(t^-,z_0)
\end{pmatrix} \mathrm{e}^{\mathrm{i}p\Delta_{\Omega} t} 
\bigg] \,.
\end{aligned}
\end{align}
We can write this as
\begin{align} \label{eq:compacttwoE}
\begin{pmatrix}
E_{{\mathrm{h}}\,m}(t^+,z_0)\\
E_{{\mathrm{v}}\,m}(t^+,z_0)
\end{pmatrix}
=
\lim_{N\rightarrow\infty}\sum_{p=-N-m}^{N-m} \mathrm{i}^p \mathcal{J}_{p}(g) \mathrm{e}^{-\mathrm{i}p\theta\sigma_z}\mathrm{e}^{\mathrm{i}\phi_y \sigma_y} \mathrm{e}^{\mathrm{i}\phi_z\sigma_z} \begin{pmatrix}
E_{{\mathrm{h}}\,m+p}(t^-,z_0)\\
E_{{\mathrm{v}}\,m+p}(t^-,z_0)
\end{pmatrix}\mathrm{e}^{-\mathrm{i}p\Delta_{\Omega} t} \,,
\end{align}
For convenience, we can write equation \ref{eq:compacttwoE} in the standard tensor form,
\begin{align} \label{eq:tensorform}
E_{\alpha m}(t^+ , z_0)= \lim_{N\rightarrow\infty}\sum_{p=-N-m}^{N-m} \sum_\beta \mathrm{i}^p \mathcal{J}_p(g) \left(\mathrm{e}^{-\mathrm{i}p\theta \sigma_z } \mathrm{e}^{\mathrm{i}\phi_y \phi_y } \mathrm{e}^{\mathrm{i}\phi_z\sigma_z}\right)_{\alpha\beta} E_{\beta\, m+p}(t^- , z_0)\mathrm{e}^{-\mathrm{i}p\Delta_{\Omega} t}\,.
\end{align}
Next, by shifting $p\rightarrow p-m$, then we finally get
\begin{align} 
E_{\alpha m}(t^+ , z_0)= \sum_{p=-\infty}^{\infty} \sum_\beta \mathrm{i}^{p-m} \mathcal{J}_{p-m}(g)\mathrm{e}^{-\mathrm{i}(p-m)\Delta_{\Omega} t} \left(\mathrm{e}^{-\mathrm{i}(p-m)\theta \sigma_z } \mathrm{e}^{\mathrm{i}\phi_y \phi_y } \mathrm{e}^{\mathrm{i}\phi_z\sigma_z} \right)_{\alpha\beta} E_{\beta\,p}(t^- , z_0)\,.
\end{align}
Thus we identify the Scattering matrix elements as
\begin{equation} \label{eq:SMatrix}
S_{mp,\alpha\beta} \equiv \mathrm{i}^{p-m} \mathcal{J}_{p-m}(g)\mathrm{e}^{-\mathrm{i}(p-m)\Delta_{\Omega} t} \left(\mathrm{e}^{-\mathrm{i}(p-m)\theta \sigma_z } \mathrm{e}^{\mathrm{i}\phi_y \phi_y } \mathrm{e}^{\mathrm{i}\phi_z\sigma_z} \right)_{\alpha\beta}  \,.
\end{equation}

We adopt the approximation of weak modulation such that
$\mathcal{J}_0(g) \approx 1$, $\mathcal{J}_1(g) \approx g/2$, only the coupling between the nearest neighbors need to be considered, i.e. $ p-m = \pm 1$. It follows that 
\begin{align}
\begin{aligned}
&\quad\begin{pmatrix}
E_{\mathrm{h}\,m}(t^+,z_0)\\
E_{\mathrm{v}\,m}(t^+,z_0)
\end{pmatrix} -\mathrm{e}^{\mathrm{i}\phi_y\sigma_y} \mathrm{e}^{\mathrm{i}\phi_z\sigma_z}
\begin{pmatrix}
E_{\mathrm{h}\,m}(t^-,z_0)\\
E_{\mathrm{v}\,m}(t^-,z_0)
\end{pmatrix}\\ 
&=\frac{\mathrm{i}g}{2} \bigg[ \mathrm{e}^{-\mathrm{i}\theta\sigma_z}\mathrm{e}^{\mathrm{i}\phi_y\sigma_y} \mathrm{e}^{\mathrm{i}\phi_z\sigma_z}\begin{pmatrix}
E_{\mathrm{h}\,\,m+1}(t^-,z_0)\\
E_{\mathrm{v}\,\,m+1}(t^-,z_0)
\end{pmatrix} \mathrm{e}^{-\mathrm{i}\Delta_{\Omega} t} +  \mathrm{e}^{\mathrm{i}\theta\sigma_z}\mathrm{e}^{\mathrm{i}\phi_y\sigma_y} \mathrm{e}^{\mathrm{i}\phi_z\sigma_z} 
\begin{pmatrix}
E_{\mathrm{h}\,m-1}(t^-,z_0)\\
E_{\mathrm{v}\,m-1}(t^-,z_0)
\end{pmatrix} \mathrm{e}^{\mathrm{i}\Delta_{\Omega} t}
\bigg] \,.
\end{aligned}
\end{align}
Then we use the first order approximation of $\mathrm{e}^{\mathrm{i}\phi_y\sigma_y} \mathrm{e}^{\mathrm{i}\phi_z\sigma_z} \approx 1+\mathrm{i}\omega_y t_\mathrm{R} \sigma_y +\mathrm{i}\omega_z t_\mathrm{R} \sigma_z $, which requires that $\phi_y$ and $\phi_z$ to be small but does not limit the magnitude of $\omega_y$ and $\omega_z$ provided that $t_\mathrm{R}$ is small enough. 
It follows that
\begin{equation}
\begin{aligned}
&\quad\begin{pmatrix}
E_{\mathrm{h}\,m}(t^+,z_0)\\
E_{\mathrm{v}\,m}(t^+,z_0)
\end{pmatrix} -
\begin{pmatrix}
E_{\mathrm{h}\,m}(t^-,z_0)\\
E_{\mathrm{v}\,m}(t^-,z_0)
\end{pmatrix}\\
&= (\mathrm{i}\omega_y t_\mathrm{R} \sigma_y + \mathrm{i}\omega_z t_\mathrm{R} \sigma_z)\begin{pmatrix}
E_{\mathrm{h}\,m}(t^-,z_0)\\
E_{\mathrm{v}\,m}(t^-,z_0)
\end{pmatrix}
+\mathrm{i}\frac{g}{2}
\bigg[
\mathrm{e}^{-\mathrm{i}\theta\sigma_z}(1+\mathrm{i}\omega_y t_\mathrm{R} \sigma_y +  \mathrm{i}\omega_z t_\mathrm{R}\sigma_z)   \begin{pmatrix}
E_{\mathrm{h}\,m+1}(t^-,z_0)\\
E_{\mathrm{v}\,m+1}(t^-,z_0)
\end{pmatrix} \mathrm{e}^{-\mathrm{i}\Delta_{\Omega} t}
\\
&\quad\quad\quad\quad\quad\quad\quad\quad\quad\quad\quad\quad\quad\quad \mathrm{e}^{\mathrm{i}\theta\sigma_z}(1+\mathrm{i}\omega_y t_\mathrm{R} \sigma_y + \mathrm{i}\omega_z t_\mathrm{R}\sigma_z) \begin{pmatrix}
E_{\mathrm{h}\,m-1}(t^-,z_0)\\
E_{\mathrm{v}\,m-1}(t^-,z_0)
\end{pmatrix} \mathrm{e}^{\mathrm{i}\Delta_{\Omega} t} \bigg] \,.
\end{aligned}
\end{equation}
Define $\Delta E_{\alpha m}(t)= E_{\alpha m} (t + t_\mathrm{R})-E_{\alpha m}(t)$ and divide $t_\mathrm{R}$ both sides, we obtain
\begin{equation}
\begin{aligned} \mathrm{i}
\begin{pmatrix}
\frac{\Delta E_{\mathrm{h}\,m}(t,z_0)}{t_\mathrm{R}}\\
\frac{\Delta E_{\mathrm{v}\,m}(t,z_0)}{t_\mathrm{R}}
\end{pmatrix} & = (-\omega_y  \sigma_y - \omega_z\sigma_z) \begin{pmatrix}
E_{\mathrm{h}\,m}(t,z_0)\\
E_{\mathrm{v}\,m}(t,z_0)
\end{pmatrix}
-\frac{g}{2t_\mathrm{R}}
\bigg[
\mathrm{e}^{-\mathrm{i}\theta\sigma_z}(1+\mathrm{i}\omega_y t_\mathrm{R} \sigma_y + \mathrm{i}\omega_z t_\mathrm{R}\sigma_z)   \begin{pmatrix}
E_{\mathrm{h}\,m+1}(t,z_0)\\
E_{\mathrm{h}\,m+1}(t,z_0)
\end{pmatrix} \mathrm{e}^{-\mathrm{i}\Delta_{\Omega} t}
\\
&\quad\quad\quad\quad\quad\quad\quad\quad\quad\quad\quad\quad\quad\quad \mathrm{e}^{\mathrm{i}\theta\sigma_z}(1+\mathrm{i}\omega_y t_\mathrm{R} \sigma_y + \mathrm{i}\omega_z t_\mathrm{R}\sigma_z) \begin{pmatrix}
E_{\mathrm{h}\,m-1}(t,z_0)\\
E_{\mathrm{v}\,m-1}(t,z_0)
\end{pmatrix} \mathrm{e}^{\mathrm{i}\Delta_{\Omega} t} \bigg] \,.
\end{aligned}
\end{equation}
Define $J= -g/2t_\mathrm{R}$ and take the limit of $t_\mathrm{R}\rightarrow 0$, we get
\begin{equation}
\begin{aligned} \mathrm{i}
\begin{pmatrix}
\frac{dE_{\mathrm{h}\,m}(t,z_0)}{dt}\\
\frac{dE_{\mathrm{v}\,m}(t,z_0)}{dt}
\end{pmatrix} & = (-\omega_y  \sigma_y -\omega_z\sigma_z) \begin{pmatrix}
E_{\mathrm{h}\,m}(t,z_0)\\
E_{\mathrm{v}\,m}(t,z_0)
\end{pmatrix}
+J
\bigg[
\mathrm{e}^{-\mathrm{i}\theta\sigma_z}   \begin{pmatrix}
E_{\mathrm{h}\,m+1}(t,z_0)\\
E_{\mathrm{v}\,m+1}(t,z_0)
\end{pmatrix} \mathrm{e}^{-\mathrm{i}\Delta_{\Omega} t} +
\mathrm{e}^{\mathrm{i}\theta\sigma_z} \begin{pmatrix}
E_{\mathrm{h}\,m-1}(t,z_0)\\
E_{\mathrm{v}\,m-1}(t,z_0)
\end{pmatrix} \mathrm{e}^{\mathrm{i}\Delta_{\Omega} t} \bigg] \,.
\end{aligned}
\end{equation}
This can be compactly written as
\begin{equation}
\mathrm{i}\frac{dE_{\alpha m}(t,z_0)}{dt} = -\omega_y (\sigma_y)_{\alpha\beta}E_{\beta m}(t,z_0)  -\omega_z (\sigma_z)_{\alpha\beta}E_{\beta m}(t,z_0)+ J\mathrm{e}^{-\mathrm{i}\Delta_{\Omega} t}(\mathrm{e}^{-\mathrm{i}\theta\sigma_z})_{\alpha\beta}E_{\beta\,m+1}(t,z_0) + J\mathrm{e}^{\mathrm{i}\Delta_{\Omega} t} (\mathrm{e}^{\mathrm{i}\theta\sigma_z})_{\alpha\beta}E_{\beta\,m-1}(t,z_0) \,.
\end{equation}
This corresponds to the second quantized Hamiltonian
\begin{equation}
\hat{H}=\sum_m \sum_{\alpha\beta}\bigg[-\omega_y \hat{a}^\dagger_{m\alpha}(\sigma_y)_{\alpha\beta}\hat{a}_{m\beta}-\omega_z \hat{a}^\dagger_{m\alpha}(\sigma_z)_{\alpha\beta}\hat{a}_{m\beta}+J\mathrm{e}^{-\mathrm{i}\Delta_{\Omega} t}\hat{a}_{m\alpha}^\dagger (\mathrm{e}^{-\mathrm{i}\theta\sigma_z})_{\alpha\beta}\hat{a}_{m+1\,\beta} + J\mathrm{e}^{\mathrm{i}\Delta_{\Omega} t}\hat{a}^\dagger_{m+1\,\alpha}(\mathrm{e}^{\mathrm{i}\theta\sigma_z})_{\alpha\beta}\hat{a}_{m\beta}   \bigg] \,.
\end{equation}
Under the gauge transformation from rotation wave approximation, we can transform the Hamiltonian above as
\begin{equation}
\hat{H}=\sum_{m}\sum_{\alpha} m\Delta \hat{a}^\dagger_{m\alpha}\hat{a}_{m\alpha}+ \sum_m \sum_{\alpha\beta}\bigg[-\omega_y \hat{a}^\dagger_{m\alpha}(\sigma_y)_{\alpha\beta}\hat{a}_{m\beta}-\omega_z \hat{a}^\dagger_{m\alpha}(\sigma_z)_{\alpha\beta}\hat{a}_{m\beta} +J\hat{a}_{m\alpha}^\dagger (\mathrm{e}^{-\mathrm{i}\theta\sigma_z})_{\alpha\beta}\hat{a}_{m+1\,\beta} + J\hat{a}^\dagger_{m+1\,\alpha}(\mathrm{e}^{\mathrm{i}\theta\sigma_z})_{\alpha\beta}\hat{a}_{m\beta}   \bigg] \,.
\end{equation}
When there is no detuning ($\Omega=\Omega_{\mathrm{R}}$ and $\Delta_{\Omega} =0$), 
\begin{equation} \label{eq:requiredHam}
\hat{H}=\sum_m \sum_{\alpha\beta}\bigg[-\omega_y \hat{a}^\dagger_{m\alpha}(\sigma_y)_{\alpha\beta}\hat{a}_{m\beta}-\omega_z \hat{a}^\dagger_{m\alpha}(\sigma_z)_{\alpha\beta}\hat{a}_{m\beta}+J\hat{a}_{m\alpha}^\dagger(\mathrm{e}^{-\mathrm{i}\theta\sigma_z})_{\alpha\beta}\hat{a}_{m+1\,\beta} + J\hat{a}^\dagger_{m+1\,\alpha}(\mathrm{e}^{\mathrm{i}\theta\sigma_z})_{\alpha\beta}\hat{a}_{m\beta}   \bigg]\,.
\end{equation}

\section{Hamiltonian derivation by transfer matrix method} \label{FullQM}
In this section, we provide an alternative derivation of the effective Hamiltonian using quantum mechanical method. We ansatz the transfer matrix can be written down as a product of optical elements as follows:
\begin{align}
\pmb{\mathrm{T}} = \mathrm{e}^{\mathrm{i}\omega t_{\mathrm{R}}-\frac{\gamma_{\mathrm{in}}}{2}t_{\mathrm{R}}} \mathrm{e}^{-\mathrm{i}A\cos(\Omega_{\mathrm{R}} t + \theta \sigma_z)  }\mathrm{e}^{\mathrm{i}\phi_y \sigma_y}\mathrm{e}^{\mathrm{i}\phi_z \sigma_z} \,,
\end{align}
We assume that both of the amplitude of phase modulation $A$ and the coupling to waveguide $\gamma_{\mathrm{in}}$ are small, so that we approximate, up to first order in Taylor expansion as
\begin{equation} \label{eq:approx1}
\mathrm{e}^{-\mathrm{i}A\cos(\Omega_{\mathrm{R}} t + \theta\sigma_z) -\frac{\gamma_{\mathrm{in}}}{2}t_{\mathrm{R}}} \approx 1-\mathrm{i}A\cos(\Omega_{\mathrm{R}} t + \theta\sigma_z) -\frac{\gamma_{\mathrm{in}}t_{\mathrm{R}} }{2} \,.
\end{equation}
By implementing the transfer matrix to the Hamiltonian equation in~\cite{wang2021topological}
\begin{align} 
\pmb{\mathrm{H}}_c = \frac{\mathrm{i}}{t_{\mathrm{R}}} \bigg[ \bigg( \frac{\gamma_{\mathrm{in}}t_{\mathrm{R}}}{2} -1\bigg)\mathbf{\sigma}_0 + \mathrm{e}^{-\mathrm{i}\omega t_{\mathrm{R}}} \pmb{\mathrm{T}} \bigg] \,,
\end{align}
it follows that
\begin{align}
\pmb{\mathrm{H}}_c =\frac{1}{t_{\mathrm{R}}}\begin{pmatrix}
\mathrm{i}\Gamma(\mathrm{e}^{\mathrm{i}\phi_z }\cos\phi_y -1) +A\cos(\Omega_\mathrm{R} t + \theta) \mathrm{e}^{\mathrm{i}\phi_z}\cos\phi_y & \mathrm{i}\Gamma\mathrm{e}^{-\mathrm{i}\phi_z}\sin\phi_y + A\cos (\Omega_{\mathrm{R}} t+\theta) \mathrm{e}^{-\mathrm{i}\phi_z}\sin \phi_y \\
-\mathrm{i}\Gamma\mathrm{e}^{\mathrm{i}\phi_z}\sin\phi_y - A\cos(\Omega_{\mathrm{R}} t - \theta)\mathrm{e}^{\mathrm{i}\phi_z}\sin\phi_y & \mathrm{i}\Gamma(\mathrm{e}^{\mathrm{i}\phi_z}\cos\phi_y -1) +A\cos(\Omega_\mathrm{R} t - \theta) \mathrm{e}^{-\mathrm{i}\phi_z}\cos\phi_y
\end{pmatrix} \,,
\end{align}
where we have defined $\Gamma = 1- \frac{\gamma_{\mathrm{in}} t_\mathrm{R}}{2}$ and $t_\mathrm{R}$ is the round-trip time. The amplitude must satisfy the $\tau-$dependent Schrodinger equation~\cite{wang2021topological}:
\begin{align} \label{eq:120}
\mathrm{i}\frac{\partial}{\partial\tau}\ket{\psi(\tau ,t)} = \pmb{\mathrm{H}}_c (t)\ket{\psi(\tau ,t)}
\end{align}
where
\begin{align} \label{eq:121}
|\psi(\tau,t)\rangle = \sum_{n}\mathrm{e}^{-\mathrm{i}\omega_n t} \begin{pmatrix}
a_{n\mathrm{h}} (\tau)\\
a_{n\mathrm{v}} (\tau) 
\end{pmatrix} \,.
\end{align} 
Using the fact that $a_{n\pm 1\, \mathrm{h/v}} =\mathrm{e}^{\mp \mathrm{i}\Omega_\mathrm{R} t}a_{n\,\mathrm{h/v}}$, then we can get the first equation of motion of the $\mathrm{h}$ state as follows:
\begin{align} \label{eq:eomh}
\begin{aligned}
\mathrm{i}\frac{da_{n\mathrm{h}}}{d\tau} &= \frac{\mathrm{i}\Gamma}{t_\mathrm{R}}(\mathrm{e}^{\mathrm{i}\phi_z }\cos\phi_y -1) a_{n\mathrm{h}} + \frac{A}{2t_\mathrm{R}}\mathrm{e}^{\mathrm{i}\theta}\mathrm{e}^{\mathrm{i}\phi_z }\cos\phi_y a_{n-1\,\mathrm{h}} + \frac{A}{2t_\mathrm{R}} \mathrm{e}^{-\mathrm{i}\theta}\mathrm{e}^{\mathrm{i}\phi_z }\cos\phi_y a_{n+1\,\mathrm{h}} \\
&\quad + \frac{\mathrm{i}\Gamma}{t_\mathrm{R}}\mathrm{e}^{-\mathrm{i}\phi_z }\sin\phi_y a_{n\mathrm{v}} +\frac{A}{2t_\mathrm{R}}\mathrm{e}^{\mathrm{i}\theta}\mathrm{e}^{-\mathrm{i}\phi_z }\sin\phi_y a_{n-1\,\mathrm{v}} +\frac{A}{2t_\mathrm{R}}\mathrm{e}^{-\mathrm{i}\theta}\mathrm{e}^{-\mathrm{i}\phi_z }\sin\phi_y
a_{n+1\,\mathrm{v}} \,.
\end{aligned}
\end{align}
Similarly, we have the second equation of motion for the $\mathrm{v}$ state as follows:
\begin{align} \label{eq:nonAbelianEOM2}
\begin{aligned}
\mathrm{i}\frac{da_{n\mathrm{v}}}{d\tau} &= \frac{\mathrm{i}\Gamma}{t_\mathrm{R}}(\mathrm{e}^{-\mathrm{i}\phi_z}\cos\phi_y -1) a_{n\mathrm{v}} + \frac{A}{2t_\mathrm{R}}\mathrm{e}^{-\mathrm{i}\theta}\mathrm{e}^{-\mathrm{i}\phi_z}\cos\phi_y a_{n-1\,\mathrm{v}} + \frac{A}{2t_\mathrm{R}}\mathrm{e}^{\mathrm{i}\theta}\mathrm{e}^{-\mathrm{i}\phi_z} \cos\phi_y a_{n+1\,\mathrm{v}} \\
&- \frac{\mathrm{i}\Gamma}{t_\mathrm{R}} \mathrm{e}^{\mathrm{i}\phi_z}\sin\phi_y a_{n\mathrm{h}} -\frac{A}{2t_\mathrm{R}}\mathrm{e}^{-\mathrm{i}\theta}\mathrm{e}^{\mathrm{i}\phi_z} \sin\phi_y a_{n-1 \,h} -\frac{A}{2t_\mathrm{R}}\mathrm{e}^{\mathrm{i}\theta}\mathrm{e}^{\mathrm{i}\phi_z} \sin\phi_y
a_{n+1\,\mathrm{h}}
\end{aligned}\,.
\end{align}
Then we can recover the explicit second-quantized Hamiltonian from the above two equations of motion:
\begin{align} \label{eq:HAML}
\begin{aligned}
\hat{H} &=\sum_n\bigg[\frac{\mathrm{i}\Gamma}{t_\mathrm{R}}(\mathrm{e}^{\mathrm{i}\phi_z}\cos\phi_y -1) \hat{a}^\dagger_{n\mathrm{h}}\hat{a}_{n\mathrm{h}}  +\frac{\mathrm{i}\Gamma}{t_\mathrm{R}}(\mathrm{e}^{-\mathrm{i}\phi_z}\cos\phi_y -1)\hat{a}^\dagger_{n\mathrm{v}}\hat{a}_{n\mathrm{v}} -\frac{\mathrm{i}\Gamma}{t_\mathrm{R}}\sin\phi_y (\mathrm{e}^{\mathrm{i}\phi_z}\hat{a}^\dagger_{n\mathrm{v}}\hat{a}_{n\mathrm{h}} -\mathrm{e}^{-\mathrm{i}\phi_z}\hat{a}^\dagger_{n\mathrm{h}}\hat{a}_{n\mathrm{v}})\\
&\quad +\frac{A}{2t_\mathrm{R}} \mathrm{e}^{\mathrm{i}\phi_z}\cos\phi_y (\mathrm{e}^{\mathrm{i}\theta} \hat{a}^\dagger_{n+1\,\mathrm{h}}\hat{a}_{n\mathrm{h}} + \mathrm{e}^{-\mathrm{i}\theta} \hat{a}^\dagger_{n\mathrm{h}}\hat{a}_{n+1\,\mathrm{h}})  +\frac{A}{2t_\mathrm{R}}\mathrm{e}^{-\mathrm{i}\phi_z} \cos\phi_y (\mathrm{e}^{-\mathrm{i}\theta} \hat{a}^\dagger_{n+1\,\mathrm{v}}\hat{a}_{n\mathrm{v}} + \mathrm{e}^{\mathrm{i}\theta} \hat{a}^\dagger_{n\mathrm{v}}\hat{a}_{n+1\,\mathrm{v}} )\\
&\quad +\frac{A}{2t_\mathrm{R}}\mathrm{e}^{-\mathrm{i}\phi_z}\sin\phi_y (\mathrm{e}^{\mathrm{i}\theta}\hat{a}^\dagger_{n+1\,\mathrm{h}}\hat{a}_{n\mathrm{v}}+\mathrm{e}^{-\mathrm{i}\theta}\hat{a}^\dagger_{n\mathrm{h}}\hat{a}_{n+1\,\mathrm{v}} )-\frac{A}{2t_\mathrm{R}}\mathrm{e}^{\mathrm{i}\phi_z}\sin\phi_y (\mathrm{e}^{-\mathrm{i}\theta}\hat{a}^\dagger_{n+1\,\mathrm{v}}\hat{a}_{n\mathrm{h}} +\mathrm{e}^{\mathrm{i}\theta}\hat{a}^\dagger_{n\mathrm{v}} \hat{a}_{n+1\,\mathrm{h}} ) 
\bigg]
\end{aligned}
\end{align}
which finally gives 
\begin{align}
\begin{aligned}
\hat{H} &=\sum_n \sum_{\alpha\beta} \hat{a}^\dagger_{n\alpha}\mathrm{i}\bigg(\frac{1}{t_\mathrm{R}} - \frac{\gamma_{\mathrm{in}}}{2}\bigg)(\mathrm{e}^{\mathrm{i}\phi_y\sigma_y}\mathrm{e}^{\mathrm{i}\phi_z\sigma_z} - \pmb{\mathrm{I}})_{\alpha\beta}\hat{a}_{n\beta}\\
&+J\sum_{n}\sum_{\alpha\beta} \bigg(\hat{a}^\dagger_{n+1 \,\alpha}(\mathrm{e}^{\mathrm{i}\theta\sigma_z}\mathrm{e}^{\mathrm{i}\phi_y\sigma_y}\mathrm{e}^{\mathrm{i}\phi_z\sigma_z})_{\alpha\beta}\hat{a}_{n\beta} +\hat{a}^\dagger_{n \alpha}(\mathrm{e}^{-\mathrm{i}\theta\sigma_z}\mathrm{e}^{\mathrm{i}\phi_y\sigma_y}\mathrm{e}^{\mathrm{i}\phi_z\sigma_z})_{\alpha\beta}\hat{a}_{n+1\beta}   \bigg) \,,
\end{aligned}
\end{align}
where we have defined $J=A/2t_{\mathrm{R}}$ in the limit of $t_\mathrm{R} \rightarrow 0$ and also $A$ is small enough.  This Hamiltonian is non-hermitian. Now again we take $\phi_z = \omega_z t_\mathrm{R} $ and $\phi_y = \omega_y t_\mathrm{R} $. Under the first-order approximation for small $t_{\mathrm{R}}$, we obtain the following effective Hermitian Hamiltonian
\begin{align}
\hat{H}=\sum_m \sum_{\alpha\beta}\bigg[-\omega_y \hat{a}^\dagger_{m\alpha}(\sigma_y)_{\alpha\beta}\hat{a}_{m\beta}-\omega_z \hat{a}^\dagger_{m\alpha}(\sigma_z)_{\alpha\beta}\hat{a}_{m\beta}+J\hat{a}_{m\alpha}^\dagger(\mathrm{e}^{-\mathrm{i}\theta\sigma_z})_{\alpha\beta}\hat{a}_{m+1\,\beta} + J\hat{a}^\dagger_{m+1\,\alpha}(\mathrm{e}^{\mathrm{i}\theta\sigma_z})_{\alpha\beta}\hat{a}_{m\beta}   \bigg]\,,
\end{align}
which is exactly same as \ref{eq:requiredHam}.

\section{Equal mixing of Rashaba and Dresselhaus SOC} \label{RDSOC}
In this section, we will show that our SOC belongs to an equal mixing of Rashaba and Dresselhaus type. We will first recall the case in Raman-induced SOC. The Hamiltonian of  Raman-induced SOC is given by~\cite{zhai2015degenerate}
\begin{equation}
H_0 = \frac{(k_0 + k_x \sigma_z)^2}{2m} + \frac{\delta}{2}\sigma_z + \frac{\Omega}{2} \sigma_x\,.
\end{equation}
Upon a spin-rotation along $y$-axis by $\pi/2$, $\sigma_x \rightarrow \sigma_z$ and $\sigma_z \rightarrow -\sigma_x$. 
Thus, $H_0$ can be expressed by
\begin{equation}
H_0^\prime =UH_0 U^\dagger= \frac{(k_0 + k_x \sigma_x)^2}{2m} - \frac{\delta}{2}\sigma_x + \frac{\Omega}{2} \sigma_z\,,
\label{smeq:raman_soc}
\end{equation} 
where
\begin{equation}
U =\frac{1}{\sqrt{2}} \begin{pmatrix}
1 & 1 \\
-1 & 1 \\
\end{pmatrix} = e^{i\frac{\pi}{4}\sigma_y}\,.
\end{equation}
Therefore, it corresponds to an equal weight of mixing of Rashba $(k_x \sigma_x + k_y\sigma_y)$ SOC and Dresselhaus $(k_x \sigma_x - k_y\sigma_y)$ SOC.

Now returning to our case
\begin{equation}
H = 2J\cos(\theta - k\sigma_z ) -\omega_y \sigma_y -\omega_z \sigma_z\,.
\end{equation}
Under the transformation of $V$ (a spin rotation of $\pi/2$ along the $y$-direction followed by a spin rotation of $-\pi/2$ along the $x$-direction), we have
\begin{equation}
H^\prime = VHV^{-1} = 2J\cos(\theta+k\sigma_x ) -\omega_y \sigma_z +\omega_z \sigma_x\,,
\end{equation}
which is a lattice version of Eq.~\eqref{smeq:raman_soc}.

\section{Signal extraction} \label{signal}
We perform signal extraction and projection detection of the transmission intensity in the photon linear basis for both Abelian SOC and non-Abelian SOC (denoted as $I_{\mathrm{h}}$ and $I_{\mathrm{v}}$, respectively), which allows direct experimental validation.
Inspired by~\cite{wang2021topological,pang2024synthetic}, we will derive the required signal detection in the linear basis to confirm the spin texture our Hermitian SOC. The transmission signal can be written as
\begin{align}
s=2\gamma_{\mathrm{in}} \mathrm{Re} \matrixel{\psi_{\mathrm{in}}}{\pmb{\mathrm{G}}_k }{\psi_{\mathrm{in}}}  -\gamma_{\mathrm{in}}^2 \matrixel{ \psi_{\mathrm{in}}}{\pmb{\mathrm{G}}^\dagger_k\pmb{\mathrm{G}}_k}{ \psi_{\mathrm{in}} } \,.
\end{align}
where $\gamma_{\mathrm{in}}$ is the small coupling between the ring resonator and the waveguide, $\pmb{\mathrm{G}}_k$ is the Green's function and $\ket{\psi_{\mathrm{in}}}$ is the input state. 
The Hamiltonian can be expanded by the eigen-energy basis $\ket{\psi_1}$ and $\ket{\psi_2 }$
of the Hamiltonian,
\begin{align}
H_k = E_1 \ketbra{\psi_1 }{ \psi_1 } + E_2 \ketbra{\psi_2 }{ \psi_2 } \,.
\end{align}
In addition, the Green's function can be expanded in the basis of $\ket{\psi_1}$ and $\ket{\psi_2 }$ as follows:
\begin{align}
\pmb{\mathrm{G}}_k = \frac{\mathrm{i}}{\Delta_{\Omega} - E_1 + \mathrm{i}\gamma_{\mathrm{in}}} \ketbra{\psi_1 }{ \psi_1 }+ \frac{\mathrm{i}}{\Delta_{\Omega} - E_2 + \mathrm{i}\gamma_{\mathrm{in}}} \ketbra{\psi_2 }{ \psi_2 }\,
\end{align}
The matrix element of the Green's function is then given by
\begin{align}
\begin{aligned}
\langle \psi_{\mathrm{in}}|\pmb{\mathrm{G}}_k |\psi_{\mathrm{in}} \rangle &= \frac{\mathrm{i}}{\Delta_{\Omega}-E_1 +\mathrm{i}\gamma_{\mathrm{in}}} \braket{ \psi_{\mathrm{in}} }{\psi_1 } \braket{ \psi_1}{\psi_{\mathrm{in}}} +\frac{\mathrm{i}}{\Delta_{\Omega}-E_2+\mathrm{i}\gamma_{\mathrm{in}}} \braket{ \psi_{\mathrm{in}} }{ \psi_2 } \braket{ \psi_2}{\psi_{\mathrm{in}}} \\
&=\mathrm{i}\bigg(\frac{1}{\Delta_{\Omega}-E_1+\mathrm{i}\gamma_{\mathrm{in}}} |\braket{ \psi_{\mathrm{in}}}{\psi_1} |^2 + \frac{1}{\Delta_{\Omega}-E_2+\mathrm{i}\gamma_{\mathrm{in}}} | \braket{ \psi_{\mathrm{in}}}{\psi_2} |^2 \bigg) \,.
\end{aligned} \,.
\end{align}
Thus,
\begin{align}
s^{(1)} = 2\gamma_{\mathrm{in}} \mathrm{Re}\matrixel{ \psi_{in}}{\pmb{\mathrm{G}}_k }{\psi_{\mathrm{in}}} = 2\sum_{j=1,2} \frac{\gamma_{\mathrm{in}}^2}{(E_j - \Delta_{\Omega})^2 +\gamma_{\mathrm{in}}^2}\big|\braket{ \psi_{\mathrm{in}} }{ \psi_j }\big|^2 \,.
\end{align}
Next, we move on to the second order term. Notice that
\begin{align}
\pmb{\mathrm{G}}_k^\dagger = \frac{-\mathrm{i}}{\Delta_{\Omega} - E_1 -\mathrm{i}\gamma_{\mathrm{in}}} \ketbra{\psi_1 }{  \psi_1 } + \frac{-\mathrm{i}}{\Delta_{\Omega} - E_2 -\mathrm{i}\gamma_{\mathrm{in}}}\ketbra{\psi_2 }{  \psi_2 } \,,
\end{align}
where we have used the fact that $E_i^* = E_i$ as for Hermitian system. Also, notice that $\braket{ \psi_1}{ \psi_2 } = \braket{ \psi_2 }{\psi_1 }= 0$, so there is no cross term. Then we obtain
\begin{align}
\matrixel{\psi_{\mathrm{in}}}{\pmb{\mathrm{G}}^\dagger_k\pmb{\mathrm{G}}_k }{\psi_{\mathrm{in}} } = \frac{1}{(E_1 -\Delta_{\Omega})^2 +\gamma_{\mathrm{in}}^2} | \braket{ \psi_{in}}{\psi_1 }|^2 + \frac{1}{(E_2 -\Delta_{\Omega})^2 +\gamma_{\mathrm{in}}^2} | \braket{ \psi_{in}}{\psi_2 }|^2 \,.
\end{align}
Therefore,
\begin{align}
s^{(2)} =-\gamma_{\mathrm{in}}^2 \langle \psi_{\mathrm{in}}|\pmb{\mathrm{G}}^\dagger_k\pmb{\mathrm{G}}_k |\psi_{\mathrm{in}} \rangle = -\sum_{j=1,2} \frac{\gamma_{\mathrm{in}}^2}{(E_j - \Delta_{\Omega})^2 +\gamma_{\mathrm{in}}^2}  |\langle \psi_{\mathrm{in}} | \psi_j \rangle |^2 \,. 
\end{align}
All together, this gives
\begin{align} \label{eq:signalformula}
s=s^{(1)} + s^{(2)} =\sum_{j=1,2} \frac{\gamma_{\mathrm{in}}^2}{(E_j - \Delta_{\Omega})^2 +\gamma_{\mathrm{in}}^2} \big|\braket{\psi_{\mathrm{in}} }{ \psi_j } \big|^2  \,,
\end{align}
which is the summation of the detected intensity of the two polarizations.

Next, we will derive the explicit formula for projection detection in the linear basis (horizontal and vertical). First, consider the completeness relation,
\begin{align} \label{eq:completeness}
\ketbra{\psi_\mathrm{h} }{ \psi_\mathrm{h} } +\ketbra{\psi_\mathrm{v} }{ \psi_\mathrm{v} } = 1 \,.
\end{align}
The direct transmission is given by
\begin{align}
\xi = \matrixel{ \psi_{\mathrm{in}} }{ \pmb{\mathrm{S}}_k^\dagger \pmb{\mathrm{S}}}{ \psi_{\mathrm{in}} } \,,
\end{align}
where the scattering matrix can be defined as~\cite{wang2021topological}
\begin{align}
\pmb{\mathrm{S}}_k = 1-\gamma_{\mathrm{in}}\pmb{\mathrm{G}}_k \,.
\end{align}
Then by inserting the completeness relation in~\ref{eq:completeness}, we obtain the direct transmission as follows:
\begin{align}
\begin{aligned}
\xi &= \bra{\psi_{\mathrm{in}} } \pmb{\mathrm{S}}_k^\dagger \ketbra{\psi_{\mathrm{h}}}{ \psi_\mathrm{h}}  \pmb{\mathrm{S}} \ket{ \psi_{\mathrm{in}}} + \bra{\psi_{\mathrm{in}} } \pmb{\mathrm{S}}_k^\dagger \ketbra{\psi_\mathrm{v}}{ \psi_v}  \pmb{\mathrm{S}} \ket{ \psi_{\mathrm{in}}}\\
&=|\matrixel{\psi_\mathrm{h}}{ \pmb{\mathrm{S}}}{ \psi_{\mathrm{in}}}|^2 + |\matrixel{\psi_v}{ \pmb{\mathrm{S}}}{ \psi_{\mathrm{in}}}|^2\\
&= I_{\mathrm{h}} + I_{\mathrm{v}} \,.
\end{aligned}
\end{align}

By explicit computation,
\begin{align} \label{eq:Ih}
I_{\mathrm{h}} =|\matrixel{ \psi_\mathrm{h}}{ \pmb{\mathrm{S}}}{\psi_{\mathrm{\mathrm{in}}}} |^2  = |\psi_\mathrm{h}|^2 =\Bigg\vert \braket{\psi_\mathrm{h}}{ \psi_{\mathrm{in}}} + \mathrm{i}\gamma_{\mathrm{in}} \sum_{j=1,2}\frac{\braket{ \psi_\mathrm{h}}{ \psi_j}\braket{ \psi_j }{\psi_{\mathrm{in}}}}{E_j -\Delta_{\Omega} + \mathrm{i}\gamma_{\mathrm{in}}}  \Bigg\vert^2
\end{align}
and
\begin{align} \label{eq:Iv}
I_{\mathrm{v}} =|\matrixel{ \psi_\mathrm{v}}{ \pmb{\mathrm{S}}}{\psi_{\mathrm{in}}} |^2  = |\psi_\mathrm{v}|^2 =\Bigg\vert \braket{\psi_\mathrm{v}}{ \psi_{in}} + \mathrm{i}\gamma_{\mathrm{in}} \sum_{j=1,2}\frac{\braket{ \psi_\mathrm{v}}{ \psi_j}\braket{ \psi_j }{\psi_{\mathrm{in}}}}{E_j -\Delta_{\Omega} + \mathrm{i}\gamma_{\mathrm{in}}}  \Bigg\vert^2 \,,
\end{align}
which jointly permit the definition of contrast $\eta = (I_h - I_v)/(I_h + I_v)$.

\section{Photonic calculation of dynamics under non-Abelian electric fields} \label{dynamics}

We derive the scattering matrix and further consider the delay lines among the modulator, retarder, and rotator.
Along the passage in the ring resonator, the transformation of the input electric field reads

\begin{align} \label{eq:fulltransform}
E_{\alpha m}(t + nt_\mathrm{R},z)=\left[ (\mathbf{P}_3\otimes \mathbf{\sigma}_0)
\mathrm{i}^{p-m} \mathcal{J}_{p-m}(g)\mathrm{e}^{-\mathrm{i}(p-m)\Delta_{\Omega} t}  \mathrm{e}^{-\mathrm{i}(p-m)\theta \sigma_z } (\mathbf{P}_2\otimes \mathbf{\sigma}_0) \mathrm{e}^{\mathrm{i}\phi_y \sigma_y }(\mathbf{P}_1\otimes \mathbf{\sigma}_0) \mathrm{e}^{\mathrm{i}\phi_z \sigma_z }\right]^n_{\alpha\beta} E_{\beta p} (t,z) \,,
\end{align}
where $\mathbf{P}_{j\,\alpha\beta} = e^{-\mathrm{i}\Theta_j (\alpha)}\delta_{\alpha\beta}$, $\alpha$ and $\beta$ are integers denoting the resonant modes, $\Theta_j(\alpha) = \Big(\frac{\omega_0 + \alpha\Omega}{v_g}\Big)L_j$ is the phase delay associated with $L_j$ for $j=1,2,3$ (see Fig.1 in the main text).

By choosing the modulation frequency $\Omega=\Omega_\mathrm{lcm}\equiv\mathrm{lcm}(\Omega_1,\Omega_2,\Omega_3)$ (where $\Omega_j$ is the FSR associated with $L_j$) with $\Omega_{\mathrm{lcm}}$ the least common multiplier of individual modulation frequency $\Omega_1, \Omega_2 $ and $\Omega_3$. By doing so, we can also define an effective length $l$ via $\Omega_{\mathrm{lcm}} = 2\pi v_g /l$. 
With the choice of $\Omega = \Omega_{\mathrm{lcm}}$, the phase matrix becomes
\begin{align}
\mathbf{P}_{j\,\alpha\beta} = \exp\bigg[-\mathrm{i}
\bigg(\frac{\omega_0 + \alpha\Omega_{\mathrm{lcm}}}{v_g}\bigg) L_j\bigg]\delta_{\alpha\beta}= \exp\bigg(
\frac{-\mathrm{i}\omega_0 L_j}{v_g}\bigg) \exp\bigg( \frac{-2\mathrm{i}\alpha\pi L_j}{l} \bigg)\delta_{\alpha\beta}\,.
\end{align}
The first exponential term will become unity as we take the modulus square of the output electric field afterwards. The second phase term is unity since $L_j /l$ is an integer. Thus, under the choice of 
$\Omega=\Omega_{\mathrm{lcm}}$, we can factorize the phase matrix out in Eq.\eqref{eq:fulltransform} as follows
\begin{align}
 \begin{aligned}
 E_{\alpha m}(t+nt_\mathrm{R},z)&=\left[ (\mathbf{P}_3\otimes \mathbf{\sigma}_0)(\mathbf{P}_2\otimes\mathbf{\sigma}_0)(\mathbf{P}_1\otimes \mathbf{\sigma}_0)
\mathrm{i}^{p-m} \mathcal{J}_{p-m}(g)\mathrm{e}^{-\mathrm{i}(p-m)\Delta_{\Omega} t}  \mathrm{e}^{-\mathrm{i}(p-m)\theta \sigma_z }  \mathrm{e}^{\mathrm{i}\phi_y \sigma_y } \mathrm{e}^{\mathrm{i}\phi_z \sigma_z }\right]^n_{\alpha\beta} E_{\beta p}(t,z) \\
&=\left[ (\mathbf{P}_3 \mathbf{P}_2 \mathbf{P}_1 \otimes\mathbf{\sigma}_0)
\mathrm{i}^{p-m} \mathcal{J}_{p-m}(g)\mathrm{e}^{-\mathrm{i}(p-m)\Delta_{\Omega} t}  \mathrm{e}^{-\mathrm{i}(p-m)\theta \sigma_z }  \mathrm{e}^{\mathrm{i}\phi_y \sigma_y } \mathrm{e}^{\mathrm{i}\phi_z \sigma_z }\right]^n_{\alpha\beta} E_{\beta p}(t,z) \\
&=\left[ (\mathbf{P}\otimes \mathbf{\sigma}_0)
\mathrm{i}^{p-m} \mathcal{J}_{p-m}(g)\mathrm{e}^{-\mathrm{i}(p-m)\Delta_{\Omega} t}  \mathrm{e}^{-\mathrm{i}(p-m)\theta \sigma_z }  \mathrm{e}^{\mathrm{i}\phi_y \sigma_y } \mathrm{e}^{\mathrm{i}\phi_z \sigma_z }\right]^n_{\alpha\beta} E_{\beta p}(t,z)\,,
 \end{aligned}
\end{align}
where we recover the main text Eq.~\ref{eq:Efieldtransform}, 
\begin{align} 
E_{\alpha m} (t+nt_\mathrm{R},z) = [(\mathbf{P}\otimes \mathbf{\sigma}_0)\mathbf{S})^n]_{mp,\alpha\beta} E_{\beta p} (t,z)\,,
\end{align}
where the $\mathbf{S}$-matrix is given by the formula in Eq.~\eqref{eq:SMatrix} and we have defined the composite phase factor as $\textbf{P}=\mathbf{P}_3 \mathbf{P}_2 \mathbf{P}_1 $. In actual implementation, the $\mathbf{S}$-matrix can be computed by the following method,
\begin{align} \label{eq:SMatrixIM}
\begin{aligned}
S&=\sum_{m=1}^N \mathcal{J}_0(g)|m\rangle\langle m|\otimes \mathrm{e}^{\mathrm{i}\phi_y\sigma_y} \mathrm{e}^{\mathrm{i}\phi_z\sigma_z}\\
&+\sum_{m=1}^{N-m} \sum_{n=1}^{N} \Big(\mathrm{i}^m \mathcal{J}_m(g)|m\rangle \langle m+n|\otimes   \mathrm{e}^{-\mathrm{i}m\Delta_{\Omega} t} \mathrm{e}^{-\mathrm{i}m\theta\sigma_z}\mathrm{e}^{\mathrm{i}\phi_y\sigma_y} \mathrm{e}^{\mathrm{i}\phi_z\sigma_z} + \mathrm{i}^{-m} \mathcal{J}_{-m}(g)|m+n\rangle\langle m|\otimes \mathrm{e}^{\mathrm{i}m\Delta_{\Omega} t}\mathrm{e}^{\mathrm{i}m\theta\sigma_z}\mathrm{e}^{\mathrm{i}\phi_y\sigma_y} \mathrm{e}^{\mathrm{i}\phi_z\sigma_z}\Big) \,.
\end{aligned}
\end{align}

\section{Numerical time-dependent Schrodinger equation calculation under non-Abelian electric fields} \label{QM}

\begin{figure} 
    \centering
    \includegraphics[width=\linewidth]{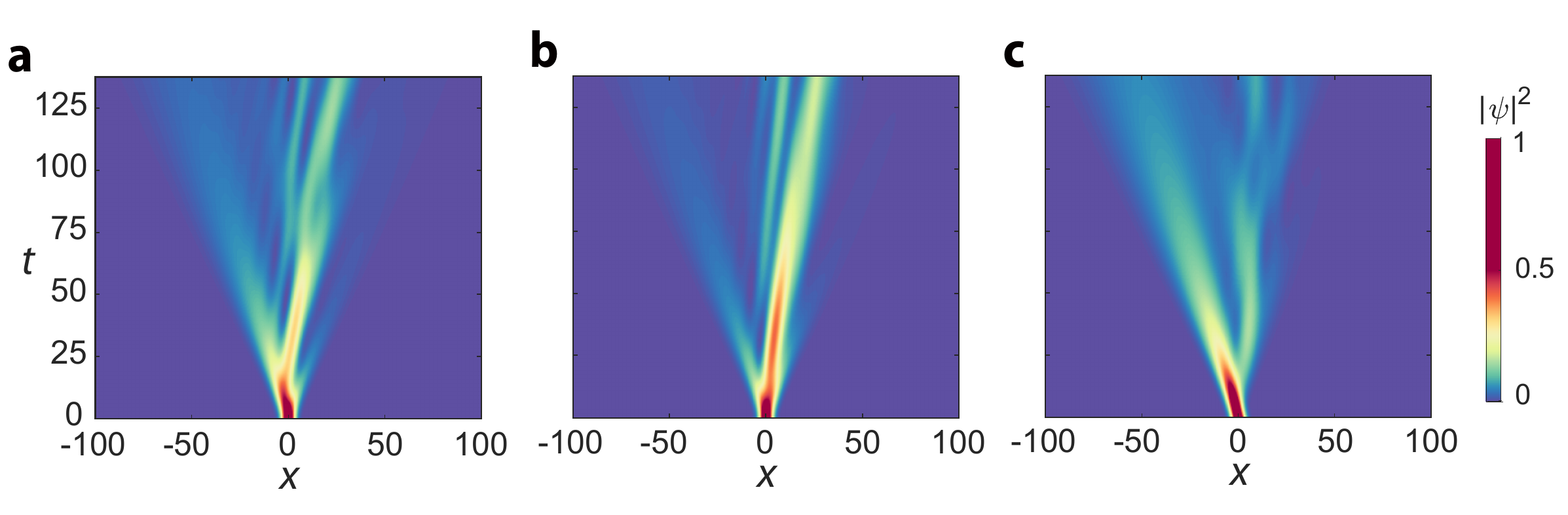}
    \centering
     \caption{
     \textbf{Numerically calculated trajectory with TDSE.}
     \textbf{a-c.} Trajectories for input states in $\sigma_x$ basis (a), $\sigma_y$ basis (b), and $\sigma_z$ basis (c).
     The parameter choices are the same as those of full EM simulation in Fig. 4b in the main text: $\theta=0.4\,\mathrm{rad},\omega_y/J = 0.1\,\,\text{and}\,\,\omega_z/J = -0.13\,.$ 
     Excellent agreement is achieved between the photonic calculation and numerical TDSE (c.f. Fig. 4b and Fig.~\ref{fig:psisquare_Allbasis}a).
      }
    \label{fig:psisquare_Allbasis}
\end{figure}

In this section, we will study the dynamics of synthetic non-Abelian electric fields supported by the framework of the ring resonator through the numerical solution to the time-dependent Schrodinger equation (TDSE). Consider the time-evolved state in the Schrödinger picture, $\ket {\psi(t)} = e^{-\mathrm{i}\hat{H}(\hat{p})t}\ket {\psi(0)}$, by inserting the completeness relation in momentum state and take the bra $\bra{x}$ with $\ket {\psi(t)}$, this gives the position wave function as follows:
\begin{align}
\begin{aligned}
    \psi(x,t) = \ip{x}{\psi(t)} &= \int_{-\infty}^\infty dp \matrixel{x}{e^{-\mathrm{i}\hat{H}(\hat{p})t}}{p} \ip{p}{\psi(0)}\\
    &= \int_{-\infty}^\infty dp \matrixel{x}{e^{-\mathrm{i}\hat{H}(p)t}}{p} \ip{p}{\psi(0)}\\
    &=\int_{-\infty}^\infty dp \,e^{-\mathrm{i}\hat{H}(p)t} \ip{x}{p}\ip{p}{\psi(0)}\\
    &= \frac{1}{\sqrt{2\pi}} \int_{-\infty}^\infty dp \, e^{-\mathrm{i}\hat{H}(p)t} e^{\mathrm{i}px}\ip{p}{\psi(0)}
\end{aligned} 
\end{align}
where we have taken $\hbar= 1$ in natural unit. For the non-Abelian SOC case, the Hamiltonian is explicitly 
\begin{align}
\hat{H}(p)=\begin{pmatrix}
2J\cos(\theta-pa)-\omega_z & i\omega_y\\
-i\omega_y   &   2J\cos(\theta + pa)+\omega_z
\end{pmatrix} \,.
\end{align}
The unitary transformation exponential matrix is explicitly computed to be:
\begin{align} \label{eq:Evolution}
e^{-\mathrm{i}\hat{H}(p)t} = e^{-\mathrm{i}2Jt\cos pa \cos\theta} \begin{pmatrix}
\cos g(p)t+\frac{\mathrm{i}(\omega_z -2J\sin pa \sin \theta) \sin g(p)t}{g(p)} & \frac{\omega_y \sin g(p)t}{g(p)}\\
 -\frac{\omega_y \sin g(p)t}{g(p)} & \cos g(p)t-\frac{\mathrm{i}(\omega_z -2J\sin pa \sin \theta) \sin g(p)t}{f(p)}
\end{pmatrix}
\end{align}
where 
\begin{align}
g(p) = \sqrt{\omega_y^2 + \omega_z^2 +J^2 -J^2\cos 2pa -2J\sin pa \,(J\cos 2\theta \sin pa + 2\omega_z \sin\theta )}
\end{align}
for $a=1$. It follows that the time-dependent position expectation value is
\begin{align}
\expval{x(t)} = \int_{x_\mathrm{min}}^{x_\mathrm{max}} dx \,x|\psi(x,t)|^2 = \frac{1}{2\pi}\int_{x_\mathrm{min}}^{x_\mathrm{max}} dx\, x\bigg\vert \int_{-\infty}^\infty dp \, e^{-\mathrm{i}\hat{H}(p)t} e^{\mathrm{i}px}\ip{p}{\psi(0)}\bigg\vert^2 \,,
\end{align}
which can be evaluated numerically. The result of $|\psi(x,t)|^2$ under $\sigma_x$, excitation is shown in Fig.~\ref{fig:psisquare_Allbasis}, which is in good agreement with the result achieved by the electromagnetic calculation (in particular Fig. ~\ref{fig:FullEMSimulation_vertical}(e) in the $\sigma_x$ basis) in the main text.

\section{Semi-analytical center-of-mass calculation under Heisenberg picture}\label{Hsbg}

We consider the situation with $\omega_z=0$ and employ the Heisenberg equation of motion $d\hat{v}/dt = \mathrm{i}[\hat{H}_{\mathrm{YM}}, \hat{v}]\,,$ where $\hat{H}_{\mathrm{YM}}$ is an analogous Yang-Mills Hamiltonian, given as the Taylor expansion of Eq.(10) in the main text up to second order of $k$, which gives a Hamiltonian in the continuum limit 
\begin{align}
\hat{H}_{\mathrm{YM}} \approx   2J-J(k-\theta\sigma_z)^2 -\omega_y \sigma_y \,,
\end{align}
where we have used the fact that Eq.(10) can be written as $H(k) = 2J\cos(  k -\theta \sigma_z) -\omega_y \sigma_y $. Therefore we have the analogous kinetic term as $\frac{\hat{P}^2}{2m} \leftrightarrow -(2J)\frac{1}{2}(k-\theta\sigma_z)^2$ with $\hat{P}= k-\theta\sigma_z \equiv k-A$ and for $J<0$. Thus we identify $1/m = -2J$
for $J<0$. Since $[\hat{p}, A] =0$ for constant gauge field, the analogous Yang-Mills Hamiltonian can be written down as
\begin{align} \label{eq:HYM}
\hat{H}_{\mathrm{YM}} =\frac{\hat{p}^2 + \theta^2 -2\hat{p}\theta\sigma_z}{2m} - \omega_y \sigma_y +2J\,.
\end{align}
Substituting Eq.~\eqref{eq:HYM} back into the Heisenberg equation of motion, we obtain
\begin{align}
\frac{d\hat{v}}{dt} = \mathrm{i}\left[ \hat{H}_{\mathrm{YM}}, \frac{\hat{P}-\hat{A}}{m}\right]=\frac{\mathrm{i}}{m}\left[-\omega_y \sigma_y , \hat{P} -\theta \sigma_z\right] =-\frac{2}{m}\omega_y \theta \sigma_x \,.
\end{align}
Comparing this result back to the expression of non-Abelian electric field $\mathrm{i}[V,A^i]$, rearranging the terms we finally derive the semi-classical Lorentz electric force law in operator form for non-Abelian electric field in one dimension,
\begin{align}
m\frac{d\hat{v}}{dt} = - e\hat{E} \,.
\end{align}

Using the dynamics equation of the Heisenberg picture, we can recover the the position, velocity and acceleration operators as a function of time,
\begin{align} \label{eq:HeisenberPic}
\begin{aligned}
\hat{x}(t) &= \exp\big(\mathrm{i}\hat{H}_{\mathrm{YM}}t \big) \hat{x}\exp\big(-\mathrm{i}\hat{H}_{\mathrm{YM}}t \big) \\
\hat{v}(t) &= \exp\big(\mathrm{i}\hat{H}_{\mathrm{YM}}t \big) \frac{\hat{p}-\theta \sigma_z}{m}\exp\big(-\mathrm{i}\hat{H} _{\mathrm{YM}}t \big) \\
\hat{a}(t) &= \exp\big(\mathrm{i}\hat{H}_{\mathrm{YM}}t \big) \frac{-2eV\theta \sigma_x}{m}\exp\big(-\mathrm{i}\hat{H}_{\mathrm{YM}}t \big) 
\end{aligned} \,.
\end{align}

By explicit computation, we can obtain analytical results for the position, velocity, and acceleration operators $\hat{x}(t),\hat{v}(t)$ and $\hat{a}(t)$ respectively. To simulate how the dynamics of a Gaussian wave packet evolves under such non-Abelian electric field, we evaluate the expectation value of position, velocity, and acceleration of the wave packet $\matrixel{\alpha}{\hat{x}(t)}{\alpha}$, $\matrixel{\alpha}{\hat{v}(t)}{\alpha}$ and $\matrixel{\alpha}{\hat{a}(t)}{\alpha}$ along the $\sigma_x$, $\sigma_y$ and $\sigma_z$ basis. 

\subsection{Acceleration}

Let us evaluate the acceleration first. Explicitly, using the identity of $\{\sigma_i , \sigma_j \}=2\delta_{ij}$, we get
\begin{align}
\begin{aligned}
\hat{a}(t) &= \exp\bigg[ \mathrm{i} \bigg(\frac{\hat{p}^2 + \theta^2 -2\hat{p}\theta\sigma_z}{2m} - \omega\sigma_y  +2J \bigg)t  \bigg]\frac{2eV\theta \sigma_x}{m} \exp\bigg[ -\mathrm{i} \bigg(\frac{\hat{p}^2 + \theta^2 -2\hat{p}\theta\sigma_z}{2m} - \omega\sigma_y  +2J\bigg)t  \bigg]\\
&=-\frac{2eV\theta }{m} \exp\bigg[\mathrm{i}\bigg(-\frac{\hat{p}\theta\sigma_z}{m} - \omega_y\sigma_y \bigg)t\bigg]\sigma_x\exp\bigg[\mathrm{i}\bigg(\frac{\hat{p}\theta\sigma_z}{m} + \omega_y\sigma_y \bigg)t\bigg] \\
&=-\frac{2eV\theta \sigma_x}{m} \exp\bigg[2\mathrm{i}\bigg(\frac{\hat{p}\theta\sigma_z}{m} +\omega_y \sigma_y\bigg)t \bigg] \,.
\end{aligned}
\end{align}
To evaluate the exponential terms, we first need to utilize a modified identity of 
\begin{align}
\exp\bigg(-\mathrm{i}\frac{\theta}{2}\pmb{\mathrm{\sigma}}\cdot \pmb{\mathrm{n}}  \bigg)= \cos\frac{\theta}{2}\sigma_0 -\mathrm{i}\pmb{\mathrm{n}}\cdot\pmb{\mathrm{\sigma}} \sin\frac{\theta}{2}\,,
\end{align}
which uses the fact of
$\pmb{\mathrm{n}}=(\sin\theta\cos\phi, \sin\theta\sin\phi, \cos\theta)$ with $|\pmb{\mathrm{n}}|^2 =1$ and 
\begin{align} 
(\pmb{\mathrm{\sigma}}\cdot \pmb{\mathrm{n}})^n =
\begin{cases}
1 & \text{for}\,\,n\,\,\text{is even} \\
\pmb{\mathrm{\sigma}}\cdot \pmb{\mathrm{n}} & \text{for}\,\,n\,\,\text{is odd}
\end{cases} \,.
\end{align} 
Here we need to use the extended identity:
\begin{align}
\exp\bigg(-\mathrm{i}\frac{\theta}{2}\pmb{\mathrm{\sigma}}\cdot \pmb{\mathrm{a}}  \bigg)= \cos\bigg(\frac{|\pmb{\mathrm{a}}|\theta}{2}\bigg)\sigma_0 -\mathrm{i}\frac{\pmb{\mathrm{\sigma}}\cdot \pmb{\mathrm{a}}}{|\pmb{\mathrm{a}}|} \sin\bigg(\frac{|\pmb{\mathrm{a}}|\theta}{2}\bigg)\,.
\end{align}
Notice the identity of $(\pmb{\mathrm{\sigma}}\cdot \pmb{\mathrm{a}})(\pmb{\mathrm{\sigma}}\cdot \pmb{\mathrm{b}}) = \pmb{\mathrm{a}}\cdot \pmb{\mathrm{b}} + \mathrm{i}\pmb{\mathrm{\sigma}}\cdot (\pmb{\mathrm{a}}\times \pmb{\mathrm{b}}) $. This will give us $(\pmb{\mathrm{\sigma}} \cdot\pmb{\mathrm{a}})^2 =|\pmb{\mathrm{a}}|^2$. Therefore, we further have
\begin{align} 
(\pmb{\mathrm{\sigma}}\cdot \pmb{\mathrm{a}})^n =
\begin{cases}
|\pmb{\mathrm{a}}|^{n} & \text{for}\,\,n\,\,\text{is even} \\
(\pmb{\mathrm{\sigma}}\cdot \pmb{\mathrm{a}}) |\pmb{\mathrm{a}}|^{n-1} & \text{for}\,\,n\,\,\text{is odd}
\end{cases} \,.
\end{align} 
For our case, we are interested in the following expression\begin{equation}
\mathrm{e}^{2\mathrm{i}\theta (a_x \sigma_x + a_y\sigma_y + a_z\sigma_z)} = \cos(2\theta |\pmb{\mathrm{a}}|)\sigma_0 + \mathrm{i}\frac{\pmb{\mathrm{\sigma}}\cdot \pmb{\mathrm{a}}}{|\pmb{\mathrm{a}}|} \sin(2\theta |\pmb{\mathrm{a}}|) \,.
\end{equation}
Then we pick $a_x =0,\, a_y = \frac{\omega_y}{\theta},\,\text{and}\, a_z = \frac{\hat{p}}{m}$. In addition, $|\pmb{\mathrm{a}}| =\sqrt{a_x^2 + a_y^2 + a_z^2} = \sqrt{\big(\frac{\hat{p}}{m}\big)^2 + \big(\frac{\omega_y}{\theta}\big)^2 } $.
Therefore, we have
\begin{align}
\begin{aligned}
\hat{a}(t) &=-\frac{2eV\theta \sigma_x}{m} \exp\bigg[2\mathrm{i}\bigg(\frac{\hat{p}\theta\sigma_z}{m} +\omega_y \sigma_y\bigg)t \bigg] \\
&=-\frac{2eV\theta \sigma_x}{m}\bigg[\cos(2\theta |\pmb{\mathrm{a}}| t )\sigma_0 + \mathrm{i}\sin(2\theta |\pmb{\mathrm{a}}|t) \bigg(\frac{ \frac{\hat{p}}{m}\sigma_z + \frac{\omega_y}{\theta}\sigma_y}{|\pmb{\mathrm{a}}|} \bigg)  \bigg] \,.
\end{aligned}
\end{align}
Finally, we obtain the complicated expression for the acceleration operator in the Heisenberg picture,
\begin{align}
\boxed{
\begin{aligned}
\hat{a}(t) &= -\frac{2eV\theta}{m}\Bigg[\cos\bigg(2\theta t\sqrt{\bigg(\frac{\hat{p}}{m} \bigg)^2 + \left(\frac{\omega_y}{\theta} \right)^2 }  \bigg) \sigma_x +  \sin\bigg(2\theta t\sqrt{\bigg(\frac{\hat{p}}{m} \bigg)^2 + \left(\frac{\omega_y}{\theta} \right)^2 }\bigg) \frac{\hat{p}/m}{\sqrt{\big(\frac{\hat{p}}{m} \big)^2 + \left(\frac{\omega_y}{\theta} \right)^2 }}     \sigma_y \\
&\quad -\sin\bigg(2\theta t\sqrt{\bigg(\frac{\hat{p}}{m} \bigg)^2 + \left(\frac{\omega_y}{\theta} \right)^2 }\bigg) \frac{\omega_y /\theta} {\sqrt{\big(\frac{\hat{p}}{m} \big)^2 + \left(\frac{\omega_y}{\theta} \right)^2 }} \sigma_z \Bigg] 
\end{aligned}}\,.
\end{align}
Next, we calculate the expectation value of the acceleration operator. First, considering the expansion of the acceleration operator in the momentum eigenbasis
\begin{align}
\hat{a}(t) = \iint dp\,dp^\prime | p^\prime\rangle \langle p^\prime| \hat{a}(\hat{p}, t) |p\rangle \langle p | \,.
\end{align}
Then, the expectation value of the acceleration operator is 
\begin{align}
\langle \alpha|\hat{a}(t)|\alpha\rangle = \int dp \psi^\dagger (p) \hat{a}(p,t) \psi(p) \,.
\end{align}
We are interested in how the Gaussian wave packet evolves over time. Consider the direction along the $\sigma_x$ basis, then we pick the normalized wave function in momentum space as
\begin{align}
\psi(p) = \frac{1}{\sqrt{2}} \bigg(\frac{2\gamma}{\pi} \bigg)^{1/4} \mathrm{e}^{-\gamma p^2}\begin{pmatrix}
1\\
1
\end{pmatrix} \,.
\end{align}
Then we get
\begin{align}
\langle \alpha|\hat{a}(t)|\alpha\rangle=-\frac{2eV\theta}{m}\sqrt{\frac{2\gamma}{\pi}}\int_{-\infty}^\infty\cos\bigg(2\theta t\sqrt{\bigg(\frac{p}{m} \bigg)^2 + \left(\frac{\omega_y}{\theta} \right)^2 }  \bigg) \mathrm{e}^{-2\gamma p^2} dp
\end{align}
which cannot be done analytically but numerically. The result is shown in Fig.~\ref{fig:PositionVelocityAcceleration_3Dallbasis}(c). Along the $\sigma_y$ basis, we pick the normalized wave function in momentum space as
\begin{align}
\psi(p) = \frac{1}{\sqrt{2}} \bigg(\frac{2\gamma}{\pi} \bigg)^{1/4} \mathrm{e}^{-\gamma p^2} \begin{pmatrix}
1\\
\mathrm{i}
\end{pmatrix} \,.
\end{align}
Then we get
\begin{align}
\langle \alpha|\hat{a}(t)|\alpha\rangle=-\frac{2eV\theta}{m}\sqrt{\frac{2\gamma}{\pi}}\int_{-\infty}^\infty \frac{p/m}{\sqrt{\big(\frac{p}{m} \big)^2 + \left(\frac{\omega_y}{\theta} \right)^2 }} \sin\bigg(2\theta t\sqrt{\bigg(\frac{p}{m} \bigg)^2 + \left(\frac{\omega_y}{\theta} \right)^2 }\bigg) \mathrm{e}^{-2\gamma p^2} dp = 0
\end{align}
as it is an odd function. Finally, along the $\sigma_z$ basis, we have 
\begin{align}
\psi(p) =\bigg(\frac{2\gamma}{\pi}\bigg)^{1/4} \mathrm{e}^{-\gamma p^2} \begin{pmatrix}
1\\
0
\end{pmatrix} \,.
\end{align}
Then we get
\begin{align}
\langle \alpha|\hat{a}(t)|\alpha\rangle=\frac{2eV\theta}{m}\sqrt{\frac{2\gamma}{\pi}}\int_{-\infty}^\infty \frac{\omega_y/\theta}{\sqrt{\big(\frac{p}{m} \big)^2 + \left(\frac{\omega_y}{\theta} \right)^2 }} \sin\bigg(2\theta t\sqrt{\bigg(\frac{p}{m} \bigg)^2 + \left(\frac{\omega_y}{\theta} \right)^2 }\bigg) \mathrm{e}^{-2\gamma p^2} dp 
\end{align}
which cannot be done analytically but numerically. The result is shown in Fig.~\ref{fig:PositionVelocityAcceleration_3Dallbasis}(f).

\subsection{Velocity}

Next, we calculate the velocity operator in the Heisenberg picture. We have
\begin{align}
\begin{aligned}
\hat{v}(t) &= \exp\bigg[ \mathrm{i} \bigg(\frac{\hat{p}^2 + \theta^2 -2\hat{p}\theta\sigma_z}{2m} - \omega\sigma_y +2J  \bigg)t  \bigg]\bigg(\frac{\hat{p}-\theta \sigma_z}{m} \bigg) \exp\bigg[ -\mathrm{i} \bigg(\frac{\hat{p}^2 + \theta^2 -2\hat{p}\theta\sigma_z}{2m} - \omega\sigma_y  +2J\bigg)t \bigg] \,.\\
\end{aligned}
\end{align}
Notice that for the first term,
\begin{align}
\begin{aligned}
\hat{v}(t) &= \exp\bigg[ \mathrm{i} \bigg(\frac{\hat{p}^2 + \theta^2 -2\hat{p}\theta\sigma_z}{2m} - \omega\sigma_y + 2J  \bigg)t  \bigg]\bigg(\frac{\hat{p}}{m}\bigg) \exp\bigg[ -\mathrm{i} \bigg(\frac{\hat{p}^2 + \theta^2 -2\hat{p}\theta\sigma_z}{2m} - \omega\sigma_y + 2J  \bigg)t \bigg] = \frac{\hat{p}}{m}\\
\end{aligned}
\end{align}
as $\hat{p}$ commutes with $\exp(\mathrm{i}\hat{H}_{\mathrm{YM}}t)$. Then it remains to calculate 
\begin{align}
\begin{aligned}
\hat{v}(t) &=\frac{\hat{p}}{m}-\frac{\theta}{m} \exp\bigg[ \mathrm{i} \bigg(-\frac{\hat{p}\theta\sigma_z}{m} - \omega_y \sigma_y\bigg)t \bigg] \sigma_z \exp\bigg[ \mathrm{i} \bigg(\frac{\hat{p}\theta\sigma_z}{m} + \omega_y \sigma_y\bigg)t \bigg] \\
&= \frac{\hat{p}}{m}-\frac{\theta}{m} \bigg[\cos(\theta t |\pmb{\mathrm{a}}|)\sigma_0 + \mathrm{i}\frac{-\frac{\hat{p}}{m}\sigma_z - \frac{\omega_y}{\theta}\sigma_y}{|\pmb{\mathrm{a}}|} \sin(\theta t |\pmb{\mathrm{a}}|)\bigg]\sigma_z \bigg[\cos(\theta t |\pmb{\mathrm{a}}|)\sigma_0 + \mathrm{i}\frac{\frac{\hat{p}}{m}\sigma_z + \frac{\omega_y}{\theta}\sigma_y}{|\pmb{\mathrm{a}}|} \sin(\theta t |\pmb{\mathrm{a}}|)\bigg]\\
&=\frac{\hat{p}}{m}- \frac{\theta\sigma_z}{m} \bigg[\cos(\theta t |\pmb{\mathrm{a}}|)\sigma_0 + \mathrm{i}\frac{-\frac{\hat{p}}{m}\sigma_z + \frac{\omega_y}{\theta}\sigma_y}{|\pmb{\mathrm{a}}|} \sin(\theta t |\pmb{\mathrm{a}}|)\bigg] \bigg[\cos(\theta t |\pmb{\mathrm{a}}|)\sigma_0 + \mathrm{i}\frac{\frac{\hat{p}}{m}\sigma_z + \frac{\omega_y}{\theta}\sigma_y}{|\pmb{\mathrm{a}}|} \sin(\theta t |\pmb{\mathrm{a}}|)\bigg] \\
&=\frac{\hat{p}}{m}- \frac{\theta\sigma_z}{m} \bigg[\cos^2(\theta t |\pmb{\mathrm{a}}|)\sigma_0 + \mathrm{i}\bigg(\frac{\frac{\hat{p}}{m}\sigma_z +\frac{\omega_y}{\theta}\sigma_y}{|\pmb{\mathrm{a}}|}\bigg) \cos(\theta t |\pmb{\mathrm{a}}|)\sin(\theta t |\pmb{\mathrm{a}}|)\\
&\quad\quad\quad\quad\quad
 + \mathrm{i}\bigg(\frac{-\frac{\hat{p}}{m}\sigma_z +\frac{\omega_y}{\theta}\sigma_y}{|\pmb{\mathrm{a}}|}\bigg) \cos(\theta t |\pmb{\mathrm{a}}|)\sin(\theta t |\pmb{\mathrm{a}}|)   + \frac{\big(\frac{\hat{p}}{m}\sigma_z  -\frac{\omega_y}{\theta}\sigma_y \big) \big(\frac{\hat{p}}{m}\sigma_z + \frac{\omega_y}{\theta}\sigma_y \big)  }{|\pmb{\mathrm{a}}|^2} \sin^2 (\theta t |\pmb{\mathrm{a}}| ) \bigg] \,.\\
\end{aligned}
\end{align}
To evaluate the last term, we consider the identity, $(\alpha \sigma_z - \beta \sigma_y)(\alpha\sigma_z +\beta \sigma_y) =(\alpha^2 -\beta^2)\sigma_0 -2\mathrm{i}\alpha\beta \sigma_x $. Then it follows that
\begin{align}
\begin{aligned}
\hat{v}(t) &= \frac{\hat{p}}{m}- \frac{\theta\sigma_z}{m} \bigg[ \cos^2 (\theta t |\pmb{\mathrm{a}}|)\sigma_0 +\frac{2\mathrm{i} \omega_y / \theta}{|\pmb{\mathrm{a}}|}\sin(\theta t |\pmb{\mathrm{a}}|)\cos(\theta t |\pmb{\mathrm{a}}|)\sigma_y \\
 &\quad\quad\quad\quad\quad \bigg[ \frac{\big(\frac{\hat{p}}{m})^2 - \big(\frac{\omega_y}{\theta} \big)^2}{\big(\frac{\hat{p}}{m})^2 + \big(\frac{\omega_y}{\theta} \big)^2}\bigg]\sigma_0 \sin^2(\theta t |\pmb{\mathrm{a}}| ) - \frac{2\mathrm{i}\big(\frac{\hat{p}}{m} \big)\big(\frac{\omega_y}{\theta}\big)\sigma_x}{|\pmb{\mathrm{a}}|^2}  \sin^2 (\theta t |\pmb{\mathrm{a}}|) \bigg] \\
&= \frac{\hat{p}}{m} - \frac{\omega_y /m}{|\pmb{\mathrm{a}}|}\sin(2\theta t |\pmb{\mathrm{a}}|)\sigma_x -\frac{2\big(\frac{\hat{p}}{m} \big)\big(\frac{\omega_y}{m}\big)}{|\pmb{\mathrm{a}}|} \sin^2(\theta t |\pmb{\mathrm{a}}|)\sigma_y \\
&\quad\quad\quad\quad\quad-\frac{\theta}{m}\bigg[ \cos^2 (\theta t |\pmb{\mathrm{a}}|) +  \bigg[ \frac{\big(\frac{\hat{p}}{m})^2 - \big(\frac{\omega_y}{\theta} \big)^2}{\big(\frac{\hat{p}}{m})^2 + \big(\frac{\omega_y}{\theta} \big)^2}\bigg]\sin^2(\theta t |\pmb{\mathrm{a}}| ) \bigg]\sigma_z \,.
\end{aligned}
\end{align}
Explicitly, we obtain the velocity operator as
\begin{align}
\boxed{
\begin{aligned}
\hat{v}(\hat{p},t) &=\frac{\hat{p}}{m} - \frac{\omega_y /m}{\sqrt{\left(\frac{\hat{p}}{m} \right)^2 + \left(\frac{\omega_y}{\theta} \right)^2 }}\sin\bigg(2\theta t \sqrt{\left(\frac{\hat{p}}{m} \right)^2 + \left(\frac{\omega_y}{\theta} \right)^2 }\bigg)\sigma_x -\frac{2\big(\frac{\hat{p}}{m} \big)\big(\frac{\omega_y}{m}\big)}{\sqrt{\big(\frac{\hat{p}}{m} \big)^2 + \left(\frac{\omega_y}{\theta} \right)^2 }} \sin^2 \bigg(\theta t \sqrt{\left(\frac{\hat{p}}{m} \right)^2 + \left(\frac{\omega_y}{\theta} \right)^2 }\bigg)\sigma_y \\
&\quad\quad\quad-\frac{\theta}{m}\bigg[ \cos^2 \bigg(\theta t \sqrt{\left(\frac{\hat{p}}{m} \right)^2 + \left(\frac{\omega_y}{\theta} \right)^2 }\bigg) +  \bigg[ \frac{\big(\frac{\hat{p}}{m})^2 - \big(\frac{\omega_y}{\theta} \big)^2}{\big(\frac{\hat{p}}{m})^2 + \big(\frac{\omega_y}{\theta} \big)^2}\bigg]\sin^2 \bigg(\theta t \sqrt{\left(\frac{\hat{p}}{m} \right)^2 + \left(\frac{\omega_y}{\theta} \right)^2 } \bigg) \bigg]\sigma_z 
\end{aligned}
}\,.
\end{align}
Then the expectation value of the velocity is
\begin{align}
\begin{aligned}
\langle \alpha|\hat{v}(t)|\alpha\rangle  
&= \int dp \psi^\dagger (p) \hat{v}(p,t) \psi(p) \,.
\end{aligned}
\end{align}
Notice that the first term $\frac{\hat{p}}{m}$ vanishes as the integral is odd,
 i.e.
 \begin{align}
 \frac{1}{m}\sqrt{\frac{2\gamma}{\pi}}\int_{-\infty}^{\infty} dp \, p\mathrm{e}^{-2\gamma p^2} = 0 \,. 
 \end{align}
Along the $\sigma_x$ basis, we have the expectation value as
\begin{align}
\langle \alpha|\hat{v}(t)|\alpha\rangle  = \sqrt{\frac{2\gamma}{\pi}}\int_{-\infty}^{\infty}\frac{-\omega_y /m}{\sqrt{\left(\frac{p}{m} \right)^2 + \left(\frac{\omega_y}{\theta} \right)^2 }}\sin\bigg(2\theta t \sqrt{\left(\frac{p}{m} \right)^2 + \left(\frac{\omega_y}{\theta} \right)^2 }\bigg) \mathrm{e}^{-2\gamma p^2} dp \,.
\end{align}
The result is shown in Fig.~\ref{fig:PositionVelocityAcceleration_3Dallbasis}(b).  Along the $\sigma_y$ basis, we have the expectation value as
\begin{align}
\langle \alpha|\hat{v}(t)|\alpha\rangle =\sqrt{\frac{2\gamma}{\pi}}\int_{-\infty}^{\infty}\frac{-2\big(\frac{p}{m} \big)\big(\frac{\omega_y}{m}\big)}{\sqrt{\big(\frac{p}{m} \big)^2 + \left(\frac{\omega_y}{\theta} \right)^2 }} \sin^2 \bigg(\theta t \sqrt{\left(\frac{p}{m} \right)^2 + \left(\frac{\omega_y}{\theta} \right)^2 }\bigg) \mathrm{e}^{-2\gamma p^2} dp =0 \,.
\end{align}
Along the $\sigma_z$ basis, we have the expectation value as
\begin{align}
\begin{aligned}
&\quad\langle \alpha|\hat{v}(t)|\alpha\rangle \\
&= -\frac{\theta}{m}\sqrt{\frac{2\gamma}{\pi}}\int_{-\infty}^{\infty} \bigg[ \cos^2 \bigg(\theta t \sqrt{\left(\frac{p}{m} \right)^2 + \left(\frac{\omega_y}{\theta} \right)^2 }\bigg) +  \bigg[ \frac{\big(\frac{p}{m})^2 - \big(\frac{\omega_y}{\theta} \big)^2}{\big(\frac{p}{m})^2 + \big(\frac{\omega_y}{\theta} \big)^2}\bigg]\sin^2 \bigg(\theta t \sqrt{\left(\frac{p}{m} \right)^2 + \left(\frac{\omega_y}{\theta} \right)^2 } \bigg) \bigg] \mathrm{e}^{-2\gamma p^2} dp \,.
\end{aligned}
\end{align}
The result is shown in Fig.~\ref{fig:PositionVelocityAcceleration_3Dallbasis}(e).

\subsection{Position}

Finally we calculate how the position operator evolve in time by evaluating the Heisenberg position operator. First, by definition, it is given by $\hat{x}(t) = \mathrm{e}^{\mathrm{i}\hat{H}(\hat{p})t} \hat{x}\mathrm{e}^{-\mathrm{i}\hat{H}(\hat{p})t} $. The position operator in momentum space representation is $\hat{x} = \mathrm{i}\hbar \frac{\partial}{\partial p}\,.$ However, there is a problem as $p$ is a variable in the derivative, but we have $\hat{p}$ operator in the Hamiltonian. Here, we will rigorously, by first calculate the amplitude $\langle \alpha | \hat{x}(t) | \alpha\rangle$. By inserting four completeness relation in momentum space, we have
\begin{align}
\begin{aligned}
&\quad\langle \alpha | \hat{x} (t) | \alpha\rangle \\
& = \langle\alpha|  \mathrm{e}^{\mathrm{i}\hat{H}(\hat{p})t} \hat{x}\mathrm{e}^{-\mathrm{i}\hat{H}(\hat{p})t} |\alpha\rangle \\
&=\iiiint dp dp^{\prime} dp^{\prime\prime} dp^{\prime\prime\prime} \langle \alpha | p \rangle \langle p | \mathrm{e}^{\mathrm{i}\hat{H}(\hat{p})t} |p^\prime \rangle \langle p^\prime | \hat{x} | p^{\prime\prime}\rangle \langle p^{\prime\prime} | \mathrm{e}^{-\mathrm{i}\hat{H}(\hat{p})t} |p^{\prime\prime\prime}\rangle \langle p^{\prime\prime\prime} | \alpha \rangle \\
&=\iiiint dp dp^{\prime} dp^{\prime\prime} dp^{\prime\prime\prime} \psi^*(p) \mathrm{e}^{\mathrm{i}\hat{H}(p)t}\langle p | p^\prime \rangle \langle p^\prime | \hat{x} | p^{\prime\prime}\rangle \mathrm{e}^{-\mathrm{i}\hat{H}(p^{\prime\prime})t} \langle p^{\prime\prime} | p^{\prime\prime\prime}\rangle \psi(p^{\prime\prime\prime}) \\ 
&=\iiiint dp dp^{\prime} dp^{\prime\prime} dp^{\prime\prime\prime} \psi^*(p) \mathrm{e}^{\mathrm{i}\hat{H}(p)t} \delta(p-p^\prime) \langle p^\prime | \hat{x} | p^{\prime\prime}\rangle \mathrm{e}^{-\mathrm{i}\hat{H}(p^{\prime\prime})t} \delta (p^{\prime\prime} - p^{\prime\prime\prime}) \psi(p^{\prime\prime\prime}) \\ 
&=\iint dp dp^{\prime\prime} \psi^*(p) \mathrm{e}^{\mathrm{i}\hat{H}(p)t} \langle p^\prime | \hat{x} | p^{\prime\prime}\rangle \mathrm{e}^{-\mathrm{i}\hat{H}(p^{\prime\prime})t}\psi(p^{\prime\prime}) \,.
\end{aligned}
\end{align}
Next we need to evaluate the matrix element $\langle p^\prime | \hat{x} | p^{\prime\prime}\rangle$, which is the position operator in momentum basis. We need to use the following identity:
\begin{align}
\langle p | \hat{x} | \alpha \rangle = \mathrm{i}\hbar \frac{\partial}{\partial p} \langle p| \alpha \rangle \,.
\end{align}
Then put $|\alpha \rangle = | p^{\prime\prime} \rangle$, then we have
\begin{align}
\langle p | \hat{x} | p^{\prime\prime} \rangle = \mathrm{i}\hbar \frac{\partial}{\partial p} \langle p| p^{\prime\prime} \rangle = \mathrm{i}\hbar \frac{\partial}{\partial p}\delta(p-p^{\prime\prime})\,.
\end{align}
Now we get
\begin{align}
\begin{aligned}
&\quad\langle \alpha | \hat{x} (t) | \alpha\rangle\\
&=\iint dp dp^{\prime\prime} \psi^*(p) \mathrm{e}^{\mathrm{i}\hat{H}(p)t} \mathrm{i}\hbar \frac{\partial}{\partial p}\delta(p-p^{\prime\prime}) \mathrm{e}^{-\mathrm{i}\hat{H}(p^{\prime\prime})t}\psi(p^{\prime\prime}) \\
&= \int dp \psi^*(p) \mathrm{e}^{\mathrm{i}\hat{H}(p)t}\mathrm{i}\hbar \frac{\partial}{\partial p}\mathrm{e}^{-\mathrm{i}\hat{H}(p)t}\psi(p) \,.
\end{aligned}
\end{align}
Since $\hat{H}^\prime (p)$ does not commute with $\hat{H}(p)$, we need to use a theorem for the identity of
\begin{align}
\frac{d}{dt}\mathrm{e}^{\hat{X}(t)} = \int_0^1 \mathrm{e}^{\beta \hat X(t)} \frac{d\hat{X}(t)}{dt} \mathrm{e}^{(1-\beta)\hat{X}(t)} d\beta \,.
\end{align}
Therefore we have
\begin{align}
\frac{\partial}{\partial p} \mathrm{e}^{-\mathrm{i}\hat{H}(p)t} =\int_0^1 \mathrm{e}^{-\mathrm{i}\beta t \hat{H}(p)} \bigg( \frac{p}{m} -\frac{\theta\sigma_z}{m} \bigg)t \mathrm{e}^{-\mathrm{i}(1-\beta)t \hat{H}(p)} d\beta \,.
\end{align}
Then we have
\begingroup
\allowdisplaybreaks
\begin{align}
&\quad\langle \alpha | \hat{x} (t) | \alpha\rangle  \nonumber\\
&=\int dp \psi^*(p) \mathrm{e}^{\mathrm{i}\hat{H}(p)t} \mathrm{i}\hbar \frac{\partial}{\partial p}\mathrm{e}^{-\mathrm{i}\hat{H}(p)t}\psi(p)  \nonumber\\
&=\int_0^1 \int_{-\infty}^{\infty} dp d\beta \psi^*(p) \mathrm{e}^{\mathrm{i}\hat{H}(p)t}  \bigg[ \mathrm{e}^{-\mathrm{i}\beta t \hat{H}(p)} \bigg( \frac{p}{m} -\frac{\theta\sigma_z}{m} \bigg)t \mathrm{e}^{-\mathrm{i}(1-\beta)t \hat{H}(p)} \bigg] \psi(p) + \mathrm{i} \int_{-\infty}^{\infty} dp \psi^*(p)\psi^{\prime}(p)   \nonumber\\
&=\int_0^1 \int_{-\infty}^{\infty} dpd\beta \psi^*(p) \mathrm{e}^{\mathrm{i}(1-\beta)t \hat{H}(p)} \frac{pt}{m} \mathrm{e}^{-\mathrm{i}(1-\beta)t \hat{H}(p)}\psi(p) \nonumber \\
&\quad\quad\quad\quad\quad\quad\quad -\int_0^1 \int_{-\infty}^{\infty} dpd\beta \psi^*(p) \mathrm{e}^{\mathrm{i}(1-\beta)t \hat{H}(p)} \frac{\theta\sigma_z t}{m} \mathrm{e}^{-\mathrm{i}(1-\beta)t \hat{H}(p)}\psi(p) +\mathrm{i} \int_{-\infty}^{\infty} dp \psi^*(p)\psi^{\prime}(p)  \nonumber \\
&=\int_{-\infty}^{\infty}dp \bigg(\frac{pt}{m}\bigg)|\psi(p)|^2 -\int_0^1 \int_{-\infty}^{\infty} dpd\beta \psi^*(p) \mathrm{e}^{\mathrm{i}(1-\beta)t \hat{H}(p)} \frac{\theta\sigma_z t}{m} \mathrm{e}^{-\mathrm{i}(1-\beta)t \hat{H}(p)}\psi(p) +\mathrm{i} \int_{-\infty}^{\infty} dp \psi^*(p)\psi^{\prime}(p)   \nonumber \\
&=\int_{-\infty}^{\infty}dp \bigg(\frac{pt}{m}\bigg)|\psi(p)|^2 \nonumber \\
&\quad\quad\quad- \frac{\theta t}{m} \int_0^1 \int_{-\infty}^{\infty} dpd\beta  \psi^*(p)\exp\bigg[\mathrm{i}(1-\beta)t \bigg( -\frac{p\theta\sigma_z}{m}-\omega_y\sigma_y\bigg) \bigg] \sigma_z \exp\bigg[i(1-\beta)t \bigg( \frac{p\theta\sigma_z}{m}+\omega_y\sigma_y\bigg) \psi(p) \bigg]  \nonumber\\
&\quad\quad\quad  +i \int_{-\infty}^{\infty} dp \psi^*(p)\psi^{\prime}(p)   \nonumber \\
&=\int_{-\infty}^{\infty}dp \bigg(\frac{pt}{m}\bigg)|\psi(p)|^2-\frac{\theta t}{m}\int_0^1 \int_{-\infty}^\infty dpd\beta  \psi^*(p) \bigg[\cos(\theta t (1-\beta) |\pmb{\mathrm{a}}|)\sigma_0 + \mathrm{i}\frac{-\frac{p}{m}\sigma_z - \frac{\omega_y}{\theta}\sigma_y}{|\pmb{\mathrm{a}}|} \sin(\theta t(1-\beta) |\pmb{\mathrm{a}}|)\bigg]\sigma_z \nonumber\\
&\quad\quad\quad\quad\quad\quad \times \bigg[\cos(\theta t(1-\beta) |\pmb{\mathrm{a}}|)\sigma_0 + \mathrm{i}\frac{\frac{p}{m}\sigma_z + \frac{\omega_y}{\theta}\sigma_y}{|\pmb{\mathrm{a}}|} \sin(\theta t(1-\beta) |\pmb{\mathrm{a}}|)\bigg] \psi(p)  +\mathrm{i} \int_{-\infty}^{\infty} dp \psi^*(p)\psi^{\prime}(p) \nonumber \\
&=\int_{-\infty}^{\infty}dp \bigg(\frac{pt}{m}\bigg)|\psi(p)|^2-\frac{\theta t \sigma_z }{m}\int_0^1 \int_{-\infty}^\infty dpd\beta  \psi^*(p) \bigg[\cos(\theta t (1-\beta)|\pmb{\mathrm{a}}|)\sigma_0 + \mathrm{i}\frac{-\frac{p}{m}\sigma_z + \frac{\omega_y}{\theta}\sigma_y}{|\pmb{\mathrm{a}}|} \sin(\theta t (1-\beta)|\pmb{\mathrm{a}}|)\bigg] \nonumber\\
&\quad\quad\quad\quad\quad\quad \times \bigg[\cos(\theta t (1-\beta) |\pmb{\mathrm{a}}|)\sigma_0 + \mathrm{i}\frac{\frac{p}{m}\sigma_z + \frac{\omega_y}{\theta}\sigma_y}{|\pmb{\mathrm{a}}|} \sin(\theta t (1-\beta) |\pmb{\mathrm{a}}|)\bigg] \psi(p)  +\mathrm{i} \int_{-\infty}^{\infty} dp \psi^*(p)\psi^{\prime}(p) \nonumber\\
&=\int_{-\infty}^{\infty}dp \bigg(\frac{pt}{m}\bigg)|\psi(p)|^2-t\int_0^1 \int_{-\infty}^{\infty} dp d\beta \psi^*(p) \bigg[  \frac{\omega_y /m}{|\pmb{\mathrm{a}}|}\sin(2\theta t (1-\beta)|\pmb{\mathrm{a}}|)\sigma_x +\frac{2\big(\frac{p}{m} \big)\big(\frac{\omega_y}{m}\big)}{|\pmb{\mathrm{a}}|} \sin^2(\theta t (1-\beta) |\pmb{\mathrm{a}}|)\sigma_y  \nonumber\\
&\quad\quad\quad\quad\quad+\frac{\theta}{m}\bigg[ \cos^2 (\theta t(1-\beta) |\pmb{\mathrm{a}}|) +  \bigg[ \frac{\big(\frac{p}{m})^2 - \big(\frac{\omega_y}{\theta} \big)^2}{\big(\frac{p}{m})^2 + \big(\frac{\omega_y}{\theta} \big)^2}\bigg]\sin^2(\theta t (1-\beta) |\pmb{\mathrm{a}}| ) \bigg]\sigma_z \bigg] \psi(p)  + \int_{-\infty}^{\infty} dp \psi^*(p) \hat{x}\psi^{\prime}(p) \,.
\end{align}
\endgroup

\begin{figure}
    \centering
    \includegraphics[width=\linewidth]{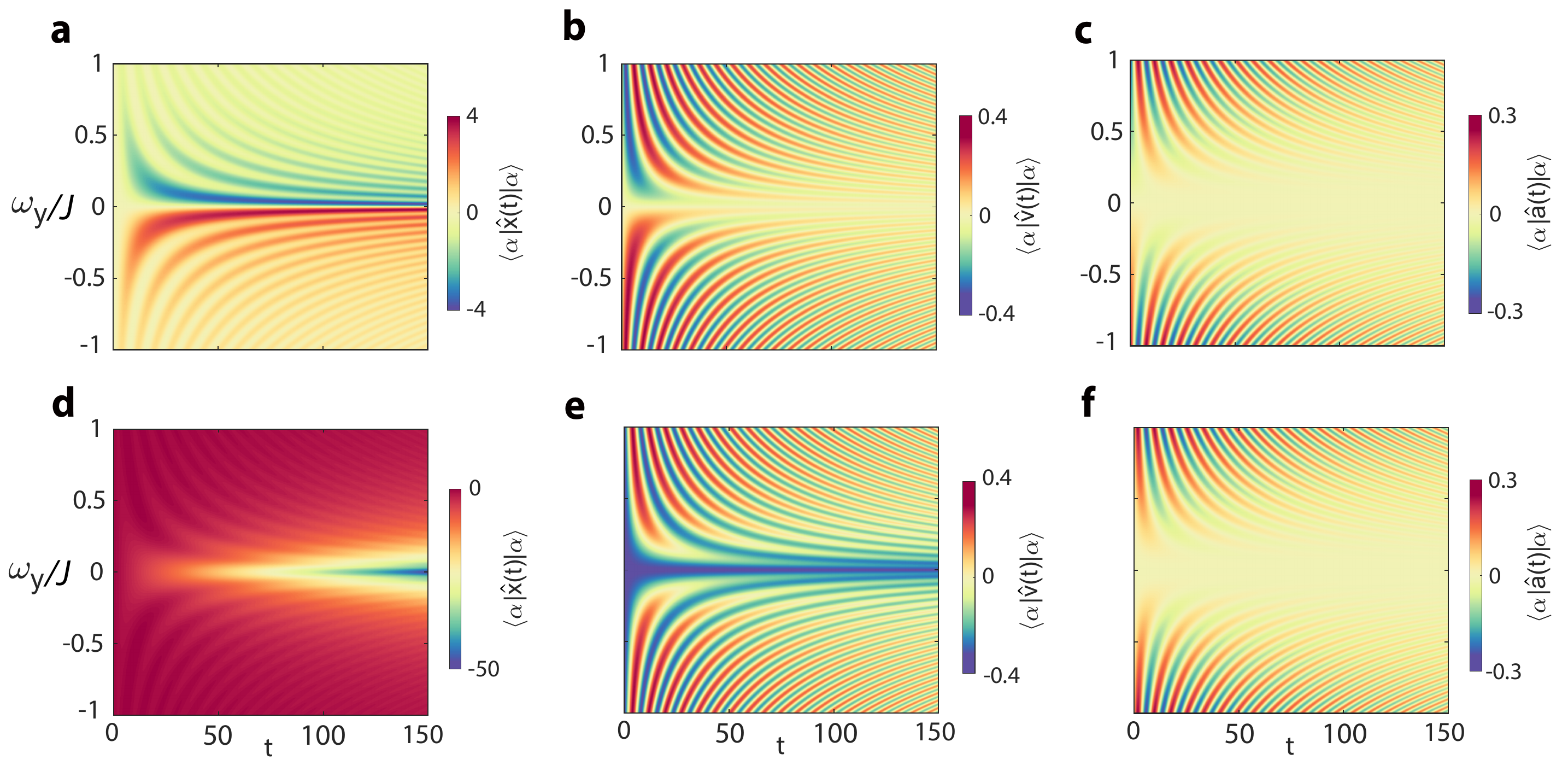}
    \centering
    \caption{\textbf{Acceleration, velocity, and position under synthetic non-Abelian electric field.}
        \textbf{a-c.} Input state as a Gaussian wave packet along the $\sigma_x$ basis.
        \textbf{d-f.} Input state as a Gaussian wave packet along the $\sigma_z$ basis.
        The results for the $\sigma_y$ basis are not shown because acceleration, velocity, and position are all zero. For all plots, $\omega_z =0$ and $\gamma=5$ are used.
        }
    \label{fig:PositionVelocityAcceleration_3Dallbasis}
\end{figure}

Then we need to evaluate the $\beta$ integrals. Notice that
\begin{align}
\int_0^1 d\beta \sin(2\theta t(1-\beta)|\pmb{\mathrm{a}}|  )  = \frac{\sin^2 (\theta t |\pmb{\mathrm{a}}|)}{\theta t |\pmb{\mathrm{a}}|}  \,,
\end{align}
\begin{align}
\int_0^1  d\beta\sin^2 (\theta t(1-\beta) |\pmb{\mathrm{a}}|) = \bigg( \frac{1}{2} - \frac{\sin(2\theta t |\pmb{\mathrm{a}}|)}{4\theta t |\pmb{\mathrm{a}}|} \bigg) \,,
\end{align}

\begin{align}
\int_0^1  d\beta\cos^2 (\theta t(1-\beta) |\pmb{\mathrm{a}}|)  = \bigg( \frac{1}{2} +\frac{\sin(2\theta t |\pmb{\mathrm{a}}|)}{4\theta t |\pmb{\mathrm{a}}|} \bigg) \,.
\end{align}
Therefore we have, 
\begin{align}
\begin{aligned}
&\quad \langle\alpha | \hat{x} (t) | \alpha\rangle\\
&=\int_{-\infty}^{\infty}dp \psi^* (p)\bigg(\hat{x} +\frac{pt}{m}\bigg)\psi(p) -t\int_{-\infty}^{\infty} dp  \psi^*(p)\bigg[ \frac{\omega_y /m}{|\pmb{\mathrm{a}}|} \frac{\sin^2 (\theta t |\pmb{\mathrm{a}}|)}{\theta t |\pmb{\mathrm{a}}|} \sigma_x +  \frac{2\big(\frac{p}{m} \big)\big(\frac{\omega_y}{m}\big)}{|\pmb{\mathrm{a}}|}  \bigg( \frac{1}{2} - \frac{\sin(2\theta t |\pmb{\mathrm{a}}|)}{4\theta t |\pmb{\mathrm{a}}|} \bigg) \\
&\quad\quad\quad\quad\quad\quad +\frac{\theta}{m}\bigg[ \bigg( \frac{1}{2} +\frac{\sin(2\theta t |\pmb{\mathrm{a}}|)}{4\theta t |\pmb{\mathrm{a}}|} \bigg) +  \bigg[ \frac{\big(\frac{p}{m})^2 - \big(\frac{\omega_y}{\theta} \big)^2}{\big(\frac{p}{m})^2 + \big(\frac{\omega_y}{\theta} \big)^2}\bigg] \bigg( \frac{1}{2} -\frac{\sin(2\theta t |\pmb{\mathrm{a}}|)}{4\theta t |\pmb{\mathrm{a}}|} \bigg)  \bigg]\sigma_z \bigg]  \psi(p)\\
&=\int_{-\infty}^{\infty}dp \psi^*(p) \bigg\{\bigg(\hat{x}+\frac{pt}{m}\bigg) -\bigg[ \frac{\omega_y /m}{|\pmb{\mathrm{a}}|} \frac{\sin^2 (\theta t |\pmb{\mathrm{a}}|)}{\theta  |\pmb{\mathrm{a}}|} \sigma_x +  \frac{2\big(\frac{p}{m} \big)\big(\frac{\omega_y}{m}\big)}{|\pmb{\mathrm{a}}|}  \bigg( \frac{t}{2} - \frac{\sin(2\theta t |\pmb{\mathrm{a}}|)}{4\theta  |\pmb{\mathrm{a}}|} \bigg)\sigma_y \\
&\quad\quad\quad\quad\quad\quad +\frac{\theta}{m}\bigg[ \bigg( \frac{t}{2} +\frac{\sin(2\theta t |\pmb{\mathrm{a}}|)}{4\theta  |\pmb{\mathrm{a}}|} \bigg) +  \bigg[ \frac{\big(\frac{p}{m})^2 - \big(\frac{\omega_y}{\theta} \big)^2}{\big(\frac{p}{m})^2 + \big(\frac{\omega_y}{\theta} \big)^2}\bigg] \bigg( \frac{t}{2} -\frac{\sin(2\theta t |\pmb{\mathrm{a}}|)}{4\theta  |\pmb{\mathrm{a}}|} \bigg)  \bigg]\sigma_z \bigg] \bigg\}  \psi(p) \,.\\
\end{aligned}
\end{align}
Therefore, in this way we can obtain the Heisenberg position operator as
\begin{align}
\begin{aligned}
\hat{x} (t) &= \hat{x}+ \frac{\hat{p}t}{m} - \frac{\omega_y /m}{|\pmb{\mathrm{a}}|} \frac{\sin^2 (\theta t |\pmb{\mathrm{a}}|)}{\theta  |\pmb{\mathrm{a}}|} \sigma_x -  \frac{2\big(\frac{\hat{p}}{m} \big)\big(\frac{\omega_y}{m}\big)}{|\pmb{\mathrm{a}}|}  \bigg( \frac{t}{2} - \frac{\sin(2\theta t |\pmb{\mathrm{a}}|)}{4\theta  |\pmb{\mathrm{a}}|} \bigg) \sigma_y \\
&\quad\quad\quad\quad\quad\quad -\frac{\theta}{m}\bigg[ \bigg( \frac{t}{2} +\frac{\sin(2\theta t |\pmb{\mathrm{a}}|)}{4\theta  |\pmb{\mathrm{a}}|} \bigg) +  \bigg[ \frac{\big(\frac{\hat{p}}{m})^2 - \big(\frac{\omega_y}{\theta} \big)^2}{\big(\frac{\hat{p}}{m})^2 + \big(\frac{\omega_y}{\theta} \big)^2}\bigg] \bigg( \frac{t}{2} -\frac{\sin(2\theta t |\pmb{\mathrm{a}}|)}{4\theta  |\pmb{\mathrm{a}}|} \bigg)  \bigg]\sigma_z  \,. \\
\end{aligned}
\end{align}
Explicitly,
\begin{align}
\boxed{
\begin{aligned}
\hat{x} (t) &= \hat{x} + \frac{\hat{p}t}{m} - \frac{\omega_y}{m} \frac{\sin^2 \big(\theta t \sqrt{\left(\frac{\hat{p}}{m} \right)^2 + \left(\frac{\omega_y}{\theta} \right)^2 }\big)}{\theta\big[  \left(\frac{\hat{p}}{m} \right)^2 + \left(\frac{\omega_y}{\theta} \right)^2 \big]} \sigma_x -  \frac{2\big(\frac{\hat{p}}{m} \big)\big(\frac{\omega_y}{m}\big)}{\sqrt{\left(\frac{\hat{p}}{m} \right)^2 + \left(\frac{\omega_y}{\theta} \right)^2 }}  \bigg( \frac{t}{2} - \frac{\sin\big(2\theta t \sqrt{\left(\frac{\hat{p}}{m} \right)^2 + \left(\frac{\omega_y}{\theta} \right)^2 }\big)}{4\theta  \sqrt{\left(\frac{\hat{p}}{m} \right)^2 + \left(\frac{\omega_y}{\theta} \right)^2 }} \bigg) \sigma_y \\
&\quad -\frac{\theta}{m}\bigg[ \bigg( \frac{t}{2} +\frac{\sin\big(2\theta t \sqrt{\left(\frac{\hat{p}}{m} \right)^2 + \left(\frac{\omega_y}{\theta} \right)^2 }\big)}{4\theta  \sqrt{\left(\frac{\hat{p}}{m} \right)^2 + \left(\frac{\omega_y}{\theta} \right)^2 }} \bigg) +  \bigg[ \frac{\big(\frac{\hat{p}}{m})^2 - \big(\frac{\omega_y}{\theta} \big)^2}{\big(\frac{\hat{p}}{m})^2 + \big(\frac{\omega_y}{\theta} \big)^2}\bigg] \bigg( \frac{t}{2} -\frac{\sin\big(2\theta t \sqrt{\left(\frac{\hat{p}}{m} \right)^2 + \left(\frac{\omega_y}{\theta} \right)^2 }\big)}{4\theta  \sqrt{\left(\frac{\hat{p}}{m} \right)^2 + \left(\frac{\omega_y}{\theta} \right)^2 }} \bigg)  \bigg]\sigma_z   \\
\end{aligned}
}\,.
\end{align}
The position operator contains a linear term in time in the $\sigma_0$ basis and a Zitterbewegung-like term in the $\sigma_x$ basis. Next, we compute the expectation value of the position operator. First for the first term, we have 
\begin{align}
\int_{-\infty}^{\infty} \mathrm{i}\frac{\partial}{\partial p} \mathrm{e}^{-2\gamma p^2} =0\,,
\end{align}
which vanishes as the integral is odd.
Then for second term,
\begin{align}
\int_{-\infty}^{\infty} dp \bigg( \frac{pt}{m}  \bigg) \mathrm{e}^{-2\gamma p^2} = 0 \,.
\end{align}
Along the $\sigma_x$ basis, 
\begin{align} \label{eq:xbasisHsbg}
 \langle\alpha | \hat{x} (t) | \alpha\rangle = - \frac{\omega_y}{m}\sqrt{\frac{2\gamma}{\pi}}\int_{-\infty}^{\infty}  \frac{\sin^2 \big(\theta t \sqrt{\left(\frac{p}{m} \right)^2 + \left(\frac{\omega_y}{\theta} \right)^2 }\big)}{\theta\big[  \left(\frac{p}{m} \right)^2 + \left(\frac{\omega_y}{\theta} \right)^2 \big]} \mathrm{e}^{-2\gamma p^2} dp \,.
\end{align}
The result is shown in the red curve in Fig. 4(e) in the main text as well as in Fig.~\ref{fig:PositionVelocityAcceleration_3Dallbasis}(a). In addition, we wish to find the asymptotic limit by evaluating the following limit of the integral
\begin{align}
L = \lim_{t\rightarrow \infty} -\frac{\omega_y}{m} \sqrt{\frac{2\gamma}{\pi}} \int_{-\infty}^{\infty}  \frac{\sin^2 \big(\theta t \sqrt{\left(\frac{p}{m} \right)^2 + \left(\frac{\omega_y}{\theta} \right)^2 }\big)}{\theta\big[  \left(\frac{p}{m} \right)^2 + \left(\frac{\omega_y}{\theta} \right)^2 \big]} \mathrm{e}^{-2\gamma p^2} dp \,.
\end{align}
We can separate the integral into a time-independent part and time-dependent part. Then we have
\begin{align}
\begin{aligned}
L &= \lim_{t\rightarrow \infty} -\frac{\omega_y}{m}\sqrt{\frac{2\gamma}{\pi}}\int_{-\infty}^{\infty}  \frac{\sin^2 \big(\theta t \sqrt{\left(\frac{p}{m} \right)^2 + \left(\frac{\omega_y}{\theta} \right)^2 }\big)}{\theta\big[  \left(\frac{p}{m} \right)^2 + \left(\frac{\omega_y}{\theta} \right)^2 \big]} \mathrm{e}^{-2\gamma p^2} dp \\
&=-\frac{\omega_y}{m}\sqrt{\frac{2\gamma}{\pi}}\lim_{t\rightarrow \infty} \int_{-\infty}^{\infty}\frac{1-\cos\big(2\theta t \sqrt{\left(\frac{p}{m} \right)^2 + \left(\frac{\omega_y}{\theta} \right)^2}\big)}{2 \theta \big[ \left(\frac{p}{m} \right)^2 + \left(\frac{\omega_y}{\theta} \right)^2 \big]} \mathrm{e}^{-2\gamma p^2} dp\\
&=-\frac{\omega_y}{m}\sqrt{\frac{2\gamma}{\pi}}\lim_{t\rightarrow \infty} \bigg[ \int_{-\infty}^{\infty}\frac{\mathrm{e}^{-2\gamma p^2}}{2\theta \big[ \left(\frac{p}{m} \right)^2 + \left(\frac{\omega_y}{\theta} \right)^2 \big]}dp - \int_{-\infty}^{\infty}\frac{\cos\big(2\theta t \sqrt{\left(\frac{p}{m} \right)^2 + \left(\frac{\omega_y}{\theta} \right)^2}\big)}{2\theta\big[ \left(\frac{p}{m} \right)^2 + \left(\frac{\omega_y}{\theta} \right)^2 \big]}\mathrm{e}^{-2\gamma p^2}dp \bigg] \,.
\end{aligned}
\end{align}
Next we define two integrals, the first one is
\begin{align}
I_1 =-\frac{\omega_y}{2m\theta}\sqrt{\frac{2\gamma}{\pi}}\lim_{t\rightarrow \infty}\int_{-\infty}^{\infty}\frac{\mathrm{e}^{-2\gamma p^2}}{ \big[ \left(\frac{p}{m} \right)^2 + \left(\frac{\omega_y}{\theta} \right)^2 \big]}dp =-\frac{\omega_y}{2m\theta}\sqrt{\frac{2\gamma}{\pi}}\int_{-\infty}^{\infty}\frac{\mathrm{e}^{-2\gamma p^2}}{ \big[ \left(\frac{p}{m} \right)^2 + \left(\frac{\omega_y}{\theta} \right)^2 \big]}dp \,,
\end{align}
which is independent of time, so the limit has no effect on this integral. We find that the analytical expression of $I_1$ is
\begin{align}
I_1 = -\frac{\pi}{2}\sqrt{\frac{2\gamma}{\pi}}\mathrm{sgn}\bigg(\frac{m\omega_y}{\theta}\bigg)  \exp\bigg[2\gamma\left(\frac{m\omega_y }{\theta}\right)^2 \bigg]\mathrm{erfc}\bigg( \sqrt{2\gamma}\bigg\vert \frac{m\omega_y }{\theta}  \bigg\vert \bigg) \,,
\end{align}
where $\mathrm{erfc}(z)$ is the complementary error function, its definition is given by
\begin{align}
\mathrm{erfc}(z) = \frac{2}{\sqrt{\pi}} \int_z^\infty \mathrm{e}^{-t^2} dt \,.
\end{align}
In fact $I_1$ gives the asymptote limit. We can in fact check our results. Using the parameters $\theta =0.4, \omega_y = 0.16, m= 1.25$ and $\gamma =5$, we obtain $I_1 = -0.865$, which can be verified in figure 4(e).

The second integral we define is dependent of time
\begin{align} \label{eq:789}
I_2 = \frac{\omega_y}{2m\theta}\sqrt{\frac{2\gamma}{\pi}} \lim_{t\rightarrow\infty}\int_{-\infty}^{\infty}\frac{\cos\big(2\theta t \sqrt{\left(\frac{p}{m} \right)^2 + \left(\frac{\omega_y}{\theta} \right)^2}\big)}{\big[ \left(\frac{p}{m} \right)^2 + \left(\frac{\omega_y}{\theta} \right)^2 \big]}\mathrm{e}^{-2\gamma p^2}dp \,.
\end{align}
However, this limit is very difficult to evaluate. First notice that in order to have $I_1$ be the asymptotic limit, we demand $I_2$ =0. Next we will try to prove that $I_2$ tends to zero when $t$ tends to infinity.

First, let $x=\frac{p}{m}$ and $a=\frac{\omega_y}{\theta}$, then the integral becomes
\begin{align}
I_2 = \frac{a}{2}\sqrt{\frac{2\gamma}{\pi}}\lim_{t\rightarrow\infty}\int_{-\infty}^{\infty} \frac{\cos(2\theta t \sqrt{x^2 + a^2})}{x^2 + a^2} \mathrm{e}^{-2\gamma m^2 x^2 }dx \,.
\end{align}
Then substitute $x=a \tan\phi =\frac{\omega_y}{\theta}\tan \phi$, the limit of integral becomes
\begin{align}
\begin{aligned}
I_2 &= \frac{a}{2}\sqrt{\frac{2\gamma}{\pi}}\lim_{t\rightarrow\infty}\int_{-\infty}^{\infty} \frac{\cos(2\theta t \sqrt{x^2 + a^2})}{x^2 + a^2} \mathrm{e}^{-2\gamma m^2 x^2 }dx \\
&= \frac{a}{2}\sqrt{\frac{2\gamma}{\pi}}\lim_{t\rightarrow\infty} \int_{-\pi/2}^{\pi/2} \frac{\cos(2\theta t a \sqrt{\tan^2 \phi +1} )}{a^2 (\tan^2 \phi +1)} \exp(-2\gamma m^2 a^2 \tan^2 \phi) a \sec^2 \phi \, d\phi\\
&=\frac{1}{2}\sqrt{\frac{2\gamma}{\pi}}\lim_{t\rightarrow\infty} \int_{-\pi/2}^{\pi/2} \cos(2\theta t a \sec\phi) \exp[-2\gamma m^2 a^2 (\sec^2 \phi -1)] a \sec^2 \phi \, d\phi \,.
\end{aligned}
\end{align}
Therefore finally we have
\begin{align}
I_2 =\lim_{t\rightarrow\infty} \frac{1}{2} \sqrt{\frac{2\gamma}{\pi}}\exp\bigg[2\gamma\left(\frac{\omega_y m}{\theta}\right)^2 \bigg]\int_{-\pi/2}^{\pi/2} \cos(2\omega_y t \sec\phi) \exp\bigg[ -2\gamma\bigg(\frac{m\omega_y}{\theta} \bigg)^2 \sec^2\phi  \bigg] d\phi \,.
\end{align}
This limit can be evaluated numerically, which can be found to be tending to zero when $t$ is very large. Therefore we can conclude that
\begin{align}
\lim_{t\rightarrow \infty} I_2 (t) = 0 \,.
\end{align}
Then finally we have
\begin{align}
L = \lim_{t\rightarrow \infty} [I_1 + I_2(t)] = I_1 \,.
\end{align}
Next, along $\sigma_y$,
\begin{align}
\langle\alpha | \hat{x}(t) | \alpha\rangle = \sqrt{\frac{2\gamma}{\pi}}\int_{-\infty}^{\infty} -  \frac{2\big(\frac{p}{m} \big)\big(\frac{\omega_y}{m}\big)}{\sqrt{\left(\frac{p}{m} \right)^2 + \left(\frac{\omega_y}{\theta} \right)^2 }}  \bigg( \frac{t}{2} - \frac{\sin\big(2\theta t \sqrt{\left(\frac{p}{m} \right)^2 + \left(\frac{\omega_y}{\theta} \right)^2 }\big)}{4\theta  \sqrt{\left(\frac{p}{m} \right)^2 + \left(\frac{\omega_y}{\theta} \right)^2 }} \bigg) \mathrm{e}^{-2\gamma p^2} dp =0
\end{align} 
as the function is odd. 

Finally along the $\sigma_z$ direction, 
\begin{align}
\begin{aligned}
&\quad\langle\alpha | \hat{x}(t) | \alpha\rangle \\
& = \int_{-\infty}^{\infty} -\frac{\theta}{m}\sqrt{\frac{2\gamma}{\pi}}\bigg[ \bigg( \frac{t}{2} +\frac{\sin\big(2\theta t \sqrt{\left(\frac{p}{m} \right)^2 + \left(\frac{\omega_y}{\theta} \right)^2 }\big)}{4\theta  \sqrt{\left(\frac{p}{m} \right)^2 + \left(\frac{\omega_y}{\theta} \right)^2 }} \bigg) +  \bigg[ \frac{\big(\frac{p}{m})^2 - \big(\frac{\omega_y}{\theta} \big)^2}{\big(\frac{p}{m})^2 + \big(\frac{\omega_y}{\theta} \big)^2}\bigg] \bigg( \frac{t}{2} -\frac{\sin\big(2\theta t \sqrt{\left(\frac{p}{m} \right)^2 + \left(\frac{\omega_y}{\theta} \right)^2 }\big)}{4\theta  \sqrt{\left(\frac{p}{m} \right)^2 + \left(\frac{\omega_y}{\theta} \right)^2 }} \bigg)  \bigg] \mathrm{e}^{-2\gamma p^2} dp \,.
\end{aligned} 
\end{align}

\providecommand{\noopsort}[1]{}\providecommand{\singleletter}[1]{#1}
%
